\documentclass[preprint,showpacs]{revtex4}
\usepackage{graphicx}
\usepackage{bm}
\usepackage{color}
\usepackage{amsmath}
\usepackage{natbib}

\begin{document}

\title{Many-particle quantum hydrodynamics: Basic principles and fundamental applications}

\author{P. A. Andreev}
\email{andreevpa@physics.msu.ru}
\author{L. S. Kuz'menkov}%
\email{lsk@phys.msu.ru}
\affiliation{Faculty of physics, M. V. Lomonosov Moscow State University, Moscow, Russian Federation.}

\date{\today}

\begin{abstract}
Basic principles and technics of many-particle quantum hydrodynamics are described in full details for the spinless case developed in L. S. Kuz'menkov and S. G. Maksimov, Theor. Math. Phys. $\textbf{118}$, 227 (1999). This method has been developed for different physical systems. Necessity of technical details of derivation are given in this paper.

We present a detailed derivation of the continuity, Euler, and energy balance equations from many particle Schrodinger equation. Interparticle interaction is explicitly considered as the Coulomb interaction. We show the QHD equations in a form suitable for low dimensional systems. The QHD equations in the cylindrical coordinates are presented as well. We present a method of non-linear Schrodinger equation derivation from the QHD equations. We describe a method of exchange interaction calculation for two- and three-dimensional quantum plasmas. We discuss explicit form and area of applicability of equations of state for 1D, 2D, and 3D degenerate electrons. We explicitly consider equations of state for degenerate electron gas in an external magnetic field. Considering QHD equations in cylindrical coordinates we focus our attention on the quantum part of the inertia forces and contribution of the quantum Bohm potential to the spectrum of quantum cylindrical waves.

We apply the QHD equations to small amplitude collective excitations in linear regime of QHD equations. We consider the Langmuir waves in 1D, 2D, and 3D quantum plasmas including the exchange Coulomb interaction. Considering quantum plasmas located in an external magnetic field we focus our attention on longitudinal waves propagating perpendicular to the external magnetic field. We present this consideration for 1D, 2D, and 3D plasmas. We discuss dependence of the exchange interaction on rate of polarization of spins in the external magnetic field. We include dependence of the pressure of degenerate electrons on the strength of the external magnetic field via the change of occupation of quantum states at application of an external magnetic field.

QHD equations in the spherical coordinates are derived. Corresponding inertia forces including the quantum part are obtained. Collective excitations of 2DEG on spherical surface are considered. Equation of state for degenerate 2DEG on a spherical surface is obtained. Spectrum of the collective excitations corresponding to Langmuir waves is calculated.

Many-particle QHD in the quasi-classic approximation is also derived. Comparison of the quasi-classic limit with the full scheme of the many-particle QHD is presented.
\end{abstract}






\maketitle


\tableofcontents

\section{\label{sec:INT} Introduction}

A long time has gone since pioneer papers of E. Madelung \cite{Madelung ZP 26} in 1926
and T. Takabayasi \cite{Takabayasi PTP 55} in 1955
considered hydrodynamic representation of basic equations of the quantum mechanics for one particle in an external electromagnetic field. This representation has interested recearchers \cite{Recami PRA 98},  \cite{Wong JMP 76}, \cite{Bialynicki PRD 71},  \cite{Sanin OS 96}. In 1964 \cite{Rand PF 64} it was applied to consider quantum plasmas including the quantum Bohm potential \cite{Bohm PR 52} in plasmas properties. The time has come for development of the many-particle quantum hydrodynamic (MPQHD). The crutial step in this field was made in 1999, when definition of the many-particle concentration as the quantum machanical average of the sum of Delta functions representing point-like particles \cite{MaksimovTMP 1999}. This appears to be the first and most important quantum variable for the QHD development. Here we have an analogy with the classical hydrodynamics, where derivation starts with the particle concentration definition \cite{Klimontovich book}, \cite{Weinberg book}.
To get continous functions in classic physics we have to make explicit averaging on the physically infinitezimal volume \cite{Drofa1996}, \cite{Kuz'menkov 12}.%
Similar scheme allows to derive relativistic kinetics with full electromagnetic interaction \cite{Kuz'menkov 91}.

A lot of application of both spinless quantum plasmas containing the quantum Bohm potential and spin-1/2 quantum plasmas having contribution of the force given by action of magnetic field on magnetic moments and the magnetic moment evolution equation. Some of these applications are presented in reviews Refs. \cite{Shukla RMP 11}, \cite{Shukla UFN 10}. Some of these applications are critically discussed in this paper. However, this paper is dedicated to description of method of derivation of the MPQHD equations. We want to show possibilities given by the MPQHD for development of different approximations for the quantum plasma description and description of other physical systems, such as quantum plasmas in 1999 \cite{MaksimovTMP 1999}, spin half quantum plasmas in 2001 \cite{MaksimovTMP 2001}, \cite{MaksimovTMP 2001 b}, \cite{Andreev Asenjo 13}, \cite{Trukhanova PTEP 13} including spin-spin interaction, explicit consideration of spin-current interaction in 2007 \cite{pavel} and spin-orbit interaction in 2009 \cite{Andreev IJMP B 12} in spin 1/2 quantum plasmas, the annihilation interaction in electron-positron plasmas \cite{Andreev annigilation}, semi-relativistic plasmas in 2012 \cite{Ivanov arxiv 2012}, \cite{Ivanov IJMP B 14}, ultracold quantum gases of neutral atoms and molecules in 2008 \cite{Andreev PRA08}, \cite{Andreev IJMP B 13}, systems of charged and neutral particles having the electric dipole moment in 2011 \cite{Andreev PRB 11} including polarized Bose-Einstein condensates in 2013 \cite{Andreev 2013 non-int GP}, \cite{Andreev EPJ D Pol}.

Most of interactions described here are semi-relativistic. Some relativistic and semi-relativistic effects in quantum plasmas were reviewed in Refs. \cite{Ivanov IJMP B 14}, \cite{Uzdensky arxiv review 14}.

A different approach to obtain QHD equation was suggested in Ref. \cite{Koide PRC 13}.

Increased interested revealed in 2000-2007, and this interest has not decreased in following years, in spite of  increasing interest to spin-1/2 quantum plasmas steadily developing in 2001-2007 \cite{MaksimovTMP 2001}, \cite{pavel}, \cite{Kuzmenkov RPJ 04}, \cite{Polyakov 79}, \cite{Andreev VestnMSU 2007}, \cite{Marklund PRL 07}, \cite{brodinMHD}. Some interesting application have been made in passing years \cite{Andreev Asenjo 13}, \cite{Andreev IJMP B 12}, \cite{Andreev annigilation}, \cite{Brodin PRL 08 a}, \cite{Misra JPP 10}, \cite{Mahajan PRL 11}, \cite{Mahajan PL A 13}, \cite{Andreev spin-up and spin-down 1405}, \cite{Asenjo kinetic 12}. One of first paper in huge number of papers was based on kinetic representation of the single particle Schrodinger equation by means of single particle Wigner function \cite{Haas PRE 00} (for review of Wigner formalism see paper \cite{Wigner PR 84}).

Derivations of MPQHD equations in spinless or spin-1/2 cases are a bit more complicate in compare with operation with single particle formalism. Hence a large part of this paper id dedicated to derivation of the MPQHD in simplest case of spinless quantum Coulomb plasmas. We consider general derivation. Next we focus attention on the self-consistent field approximation. This approach do not give closed set of equations. We need to do extra step to truncate QHD equations. We need to present pressure via the hydrodynamic variables, usually via the particle concentration. So, we pay a lot of attention to the equation of state and critically analyse equations of state have been used in literature. We present derivations of some important cases, which do not present in literature.

Processes involving many particles, atom and molecule structures, wave patterns, soliton and vortical structures, form of macroscopic objects, are revealed in three dimensional physical space, and are not very pronounce in the 3N dimensional configuration space. Particles involved in a many-particle process can also be involved in some other processes, such as thermal motion. Hence separate motion of each particle in each 3D cell of configuration space is rather complicated. Dealing with many-particle systems, quantum mechanics, as the classic mechanics, presents motion of particles in the configuration space. When we solve a two, or more, particle system motion we use our imagination to put all particles in the 3D physical space. There are mathematical methods describing systems in three dimensional space, such as hydrodynamics. The method of the MPQHD presents a mechanism to project coordinate all particles from 3N dimension configuration space to 3D physical space during derivation of equations of collective motion for application to coherent many-particle structures.

Some reasonable critics of application of the QHD based on single particle Schrodinger or Pauli equations exist in literature \cite{Krishnaswami PRL and arXiv 14}. This paper is dedicated to the QHD. Thus we need to say some words to compare our method with suggestion and criticisms of Refs. \cite{Krishnaswami PRL and arXiv 14}. The many-particle wave function gives microscopic picture of quantum motion. At transition to statistical description, such as the Fermi distribution function, we introduce extra averaging suitable for huge particle number system. So, we lose microscopic picture of motion. The MPQHD applies the many-particle wave function and extract interaction from it by construction of collective variables directly connected to microscopic evolution. Thus we have obtained microscopic description in terms of collective variables, so sometimes it looks like macroscopic description. Our method is not inferior to statistical description, being more general, since our methodincludes exact distribution of particles in quantum states via the many-particle wave function. We should also mention that the wave function has necessary symmetry properties relatively to permutation of particles. In case of non-interacting fermions. The many-particle wave function leads to Slatter determinant.


\section{\label{sec:Bas princ} Basic principles of QHD}

\subsection{\label{sec:level1}Continuity and Euler equations}

We have two starting points in our derivation. One of them is the choosing of proper many-particle quantum variable we will discuss below, and the second one is the many-particle Schr\"{o}dinger equation (MPSE) (or corresponding Pauli equation for spinning particles). In this case we have usual general form of the Schr\"{o}dinger equation
\begin{equation}\label{ChCoulQHD Schrodinger eq}\imath\hbar\partial_{t}\psi=\hat{H}\psi,\end{equation}
but explicit form of the Hamiltonian
\begin{equation}\label{ChCoulQHD Hamiltonian} \hat{H}=\sum_{i}\Biggl(\frac{1}{2m_{i}}\textbf{D}_{i}^{2}+e_{i}\varphi_{i,ext}\Biggr)+\frac{1}{2}\sum_{i,j\neq i}\Biggl(e_{i}e_{j}G_{ij}\Biggr),\end{equation}
shows that we have deal with the many-particle system. Information on the number of particles in the system is contained in the wave function $\Psi$. We can get it by looking on the arguments of wave function. For one particle $\Psi$ depends on three space variables $\textbf{r}$ and time $t$: $\Psi=\Psi(\textbf{r},t)$. In our case, when we study a many-particle system, the wave function depends on 3N space variables (three coordinates for each of N particles) and time $t$: $\Psi=\Psi(R,t)$, where $R=[\textbf{r}_{1}, ..., \textbf{r}_{N}]$. For system of non-interacting particles the many-particle wave function can be presented as the product of the one-particle wave functions $\Psi(R,t)=\prod_{i=1}^{N}\psi(\textbf{r}_{i},t)$. In this case the many-particle Schrodinger equation for quantum particles motion in an external field
\begin{equation}\label{ChCoulQHD} \imath\hbar\partial_{t}\psi(\textbf{r}_{i},t)=\biggl(\frac{1}{2m_{i}}\textbf{D}_{i}^{2}+e_{i}\varphi_{i, ext}\biggr)\psi(\textbf{r}_{i},t).\end{equation}
But we are interested in effects caused by interparticle interaction, the Coulomb interaction is under consideration in this paper. As we mention in the Introduction (\ref{sec:INT}) other interactions have been considered by means the MPQHD, but we focus our attention on simplest case to describe details of derivation and explain technique of derivation, including the exchange interaction.
The following designations are used in the Hamiltonian (\ref{ChCoulQHD Hamiltonian}): $D_{i}^{\alpha}=-\imath\hbar\partial_{i}^{\alpha}-e_{i}A_{i,ext}^{\alpha}/c$, $\partial_{i}^{\alpha}=\nabla_{i}^{\alpha}$ is the spatial derivatives,
$\varphi_{i,ext}=\varphi_{ext}(\textbf{r}_{i},t)$, $\textbf{A}_{i,ext}=\textbf{A}_{ext}(\textbf{r}_{i},t)$ the potentials of external electromagnetic field, and $G_{ij}=1/r_{ij}$ is the Green function of the Coulomb,
$e_{i}$ and $m_{i}$ are the charge and mass of particles, $\hbar$ is the reduced Plank constant, $c$ is the speed of light in vacuum. The first term gives sum of kinetic energies of all particles, it contains long derivative including action of the vortex part of the external electromagnetic field on particle charge. The second term describes potential energy of charges in the external electric field. The last term presents the Coulomb interaction between particles.

We have described the MPSE, we use in this paper, along with some general properties of the MPQHD method. Next, let us present and describe the particles concentration, which will allow to derive the set of MPQHD equations.

The first step in the construction of QHD apparatus is to determine the concentration of particles in the neighborhood of $\textbf{r}$ in a physical space. If we define the concentration of particles as quantum average \cite{Landau Vol 3} of the concentration operator in the coordinate representation $\hat{n}=\sum_{i}\delta(\textbf{r}-\textbf{r}_{i})$ we obtain:
\begin{equation}\label{ChCoulQHD def density}n(\textbf{r},t)=\int dR_{N}\sum_{i}\delta(\textbf{r}-\textbf{r}_{i})\psi^{*}(R,t)\psi(R,t),\end{equation}
where $dR_{N}=\prod_{p=1}^{N}d\textbf{r}_{p}$ is an element of $3N$ volume \cite{MaksimovTMP 1999}, \cite{MaksimovTMP 2001}, \cite{pavel}, \cite{Andreev PRB 11}. To be short we will put away the subindex $N$ at $dR$ in most formulas, where it is not necessarily to show subindex $N$ explicitly.

Quantum mechanics is formulated for model point size particles. Consequently hydrodynamic equations, which follow from quantum mechanics or classic mechanics \cite{Klimontovich book}, \cite{Drofa1996}, \cite{Kuz'menkov 12}, are obtained for point-like particles. An attempt to create hydrodynamics of finite size particles is presented in Ref. \cite{Andreev 1401 finite ions}.

The particle concentration $n$ is the result of quantum mechanical averaging \cite{Landau Vol 3} of the corresponding operator on many-particle wave function, so formula (\ref{ChCoulQHD def density}) can be written in a compact form
\begin{equation}\label{ChCoulQHD def density with brackets} n(\textbf{r},t)=\langle \Psi^{*}\hat{n}\Psi\rangle=\langle \Psi^{*}\sum_{i}\delta(\textbf{r}-\textbf{r}_{i})\Psi\rangle.\end{equation}

In statistical physics we do averaging on time or on ensemble to loose unappropriately complicate microscopic description. Quantum mechanics gives different meaning of averaging. If we need to obtain most precise value of a physical quantity in a quantum state the nature can provide we should find the average value of the corresponding operator of the physical quantity on the wave function of the quantum state. If the quantum state under discussion is an eigen-state of the operator of the physical quantity then the average value of the operator coincide with the eigen-value of the operator.

In this subsection we mainly follow Ref. \cite{MaksimovTMP 1999} introducing extra details.

Considering time derivative of the particle concentration we have that time derivative acts on wave functions being under the integral in formula (\ref{ChCoulQHD def density}). Using the Schr\"{o}dinger equation we replace time derivatives of wave functions with the Hamiltonian acting on wave functions, so we find
\begin{equation}\label{ChCoulQHD cont eq calcul 1}\partial_{t}n=\frac{\imath}{\hbar}\int dR_{N}\sum_{i}\delta(\textbf{r}-\textbf{r}_{i})\biggl((\hat{H}\psi)^{*}(R,t)\psi(R,t)-\psi^{*}(R,t)\hat{H}\psi(R,t)\biggr).\end{equation}
From this formula we can see that terms in the Hamiltonian (\ref{ChCoulQHD Hamiltonian}) do not depend on the momentum operator give no contribution in $\partial_{t}n$. Thus, we see that the first term in Hamiltonian (\ref{ChCoulQHD Hamiltonian}) only gives a contribution in the particle concentration evolution. Let us explicitly present we need to provide to obtain the continuity equation
$$\partial_{t}n=\frac{\imath}{\hbar}\int dR_{N}\sum_{i}\delta(\textbf{r}-\textbf{r}_{i})\frac{1}{2m_{i}}\biggl[(\textbf{D}_{i}^{2}\psi)^{*}(R,t)\psi(R,t)-\psi^{*}(R,t)\textbf{D}_{i}^{2}\psi(R,t)\biggr]$$
$$=\frac{\imath}{\hbar}\int dR_{N}\sum_{i}\delta(\textbf{r}-\textbf{r}_{i})\frac{1}{2m_{i}}$$
$$\biggl[-\frac{\hbar}{\imath}\nabla_{i}(\textbf{D}_{i}\psi)^{*}(R,t)\psi(R,t)+(\textbf{D}_{i}\psi)^{*}(R,t)\textbf{D}_{i}\psi(R,t)-\psi^{*}(R,t)\textbf{D}_{i}^{2}\psi(R,t)\biggr]$$
$$=\frac{\imath}{\hbar}\int dR_{N}\sum_{i}\delta(\textbf{r}-\textbf{r}_{i})\frac{1}{2m_{i}}\biggl[-\frac{\hbar}{\imath}\nabla_{i}(\textbf{D}_{i}\psi)^{*}(R,t)\psi(R,t)$$
\begin{equation}\label{ChCoulQHD cont eq calcul 2} -\frac{\hbar}{\imath}\nabla_{i}\biggl(\psi^{*}(R,t)\textbf{D}_{i}\psi(R,t)\biggr)+\psi^{*}(R,t)\textbf{D}_{i}^{2}\psi(R,t)-\psi^{*}(R,t)\textbf{D}_{i}^{2}\psi(R,t)\biggr].\end{equation}
During this calculation we kept in mind that $\textbf{D}_{i}$ is the sum of a derivative $\textbf{p}_{i}=-\imath\hbar\nabla_{i}$ and a function $-e_{i}\textbf{A}_{i}/c$, and we have used that $\nabla_{i}\delta(\textbf{r}-\textbf{r}_{i})=-\nabla\delta(\textbf{r}-\textbf{r}_{i})$, where $\nabla$ is the derivative on space variable $\textbf{r}$ making arithmetezation of the physical space.
Further calculations are
$$\partial_{t}n=-\int dR \sum_{i}\delta(\textbf{r}-\textbf{r}_{i})\frac{1}{2m_{i}}\nabla_{i}\biggl(\Psi^{*}\textbf{D}_{i}\Psi+c.c.\biggr)$$
$$=-\int dR \sum_{i}\nabla_{i}\Biggl(\delta(\textbf{r}-\textbf{r}_{i})\frac{1}{2m_{i}}\biggl(\Psi^{*}\textbf{D}_{i}\Psi+c.c.\biggr)\Biggr)+\int dR \sum_{i}\biggl(\nabla_{i}\delta(\textbf{r}-\textbf{r}_{i})\biggr)\frac{1}{2m_{i}}\biggl(\Psi^{*}\textbf{D}_{i}\Psi+c.c.\biggr)$$
\begin{equation}\label{ChCoulQHD cont eq calcul 3}=-\nabla\int dR \sum_{i}\delta(\textbf{r}-\textbf{r}_{i})\frac{1}{2m_{i}}\biggl(\Psi^{*}\textbf{D}_{i}\Psi+c.c.\biggr).\end{equation}
Considering integral $\int dR_{N}\nabla_{i}f(R)=\int dR_{N-1}d\mathbf{r}_{i}\nabla_{i}f(R)=0$ we have it equals to zero as an integral of the divergence over whole space. We put $\nabla$ outside of the integral since it can act on $\delta(\textbf{r}-\textbf{r}_{i})$ only.

Thus differentiation of the particle concentration $n(\textbf{r},t)$ with respect to time and applying of the Schr\"{o}dinger equation with Hamiltonian (\ref{ChCoulQHD Hamiltonian}) leads to the
continuity equation
\begin{equation}\label{ChCoulQHD continuity equation}\partial_{t}n(\textbf{r},t)+\nabla\textbf{j}(\textbf{r},t)=0\end{equation}
where we introduced the particle current
$$j^{\alpha}(\textbf{r},t)=\int dR\sum_{i}\delta(\textbf{r}-\textbf{r}_{i})\frac{1}{2m_{i}}\biggl(\psi^{*}(R,t)(D_{i}^{\alpha}\psi(R,t))$$
\begin{equation}\label{ChCoulQHD def of current of density}+(D_{i}^{\alpha}\psi(R,t))^{*}\psi(R,t)\biggr),\end{equation}
which is the momentum density either. We have not written down arguments of the wave functions in this definition (\ref{ChCoulQHD def of current of density}), and in large definitions of other hydrodynamic functions below.

The velocity of i-th particle $\textbf{v}_{i}(R,t)$ is determined by equation
\begin{equation}\label{ChCoulQHD vel of i part}\textbf{v}_{i}(R,t)=\frac{1}{m_{i}}\nabla_{i}S(R,t)-\frac{e_{i}}{m_{i}c}\textbf{A}_{i, ext}. \end{equation}
The $S(R,t)$ value in the formula (\ref{ChCoulQHD vel of i part}) represents the phase of the wave function
\begin{equation}\label{ChCoulQHD psi representation}\psi(R,t)=a(R,t) exp\biggl(\frac{\imath S(R,t)}{\hbar}\biggr),\end{equation}
and $a(R,t)$ is the amplitude of the many-particle wave function.
Velocity field $\textbf{v}(\textbf{r},t)$, which usually appears in hydrodynamic equations, is determined by formula
\begin{equation}\label{ChCoulQHD current to velocity}\textbf{j}(\textbf{r},t)=n(\textbf{r},t)\textbf{v}(\textbf{r},t).\end{equation}
We also can introduce quantum equivalent of the thermal velocity as the difference between the quantum velocity of a particle and the velocity field $\textbf{u}_{i}(\textbf{r},R,t)=\textbf{v}_{i}(R,t)-\textbf{v}(\textbf{r},t)$. We have obtained  the first equation of the chain of MPQHD equations. At next step we should derive equation for the momentum density evolution.

We differentiate the momentum density (\ref{ChCoulQHD def of current of density}) with respect to time and apply the many-particle Schrodinger equation (\ref{ChCoulQHD Schrodinger eq}) to time derivatives of the wave functions
$$ \partial_{t}\textbf{j}=-\int dR\sum_{i=1}^N
\delta (\textbf{r}-\textbf{r}_i)\frac{1}{2m_{i}}\frac{\imath}{\hbar}
\biggl( \psi^*\textbf{D}_i
\left(\hat H\psi\right)-\left(\hat H\psi\right)^*
\textbf{D}_i\psi$$
\begin{equation} \label{ChCoulQHD DtJ 01} +\textbf{D}_i^*\psi^*\left(\hat H\psi\right)
-\textbf{D}_i^*\left(\hat H\psi\right)^*\psi-
2i\hbar\frac{e_i}{c}\,\partial_{t}\textbf{A}_i\psi^*\psi\biggr).\end{equation}
At the next step we have
$$\partial_{t}\textbf{j}=-\int dR\sum_{i=1}^N\delta (\textbf{r}-\textbf{r}_i)\frac{1}{2m_{i}}\frac{\imath}{\hbar}
\sum_{k=1}^N\Biggl[ \psi^*\textbf{D}_i
\frac{\textbf{D}_{k}^{2}}{2m_{k}}\psi $$
$$-\Biggl(\frac{\textbf{D}_k^2}{2m_k}
\psi\biggr)^*\textbf{D}_i\psi+(\textbf{D}_i\psi)^*
\frac{\textbf{D}_{k}^{2}}{2m_{k}}\psi-\biggl(\textbf{D}_{i}
\frac{\textbf{D}_{k}^{2}}{2m_{k}}\psi\biggr)^*\psi$$
$$+\psi^* \left[ \textbf{D}_i, \varphi_{k} \right]
\psi-(\left[\textbf{D}_i, \varphi_k \right]\psi)^*\psi$$
$$+\frac{1}{2}\sum_{p\neq k}\biggl(\psi^* \left[ \textbf{D}_i, G_{pk} \right]
\psi-(\left[\textbf{D}_i, G_{pk} \right]\psi)^*\psi\biggr)\Biggr]$$
\begin{equation} \label{ChCoulQHD DtJ} -\frac{1}{c}e n(\textbf{r},t)\partial_{t}\textbf{A}(\textbf{r},t).\end{equation}

The following commutators have to be calculated to get the momentum balance equation
$$\left[ {D_i}_{\alpha}\frac{\textbf{D}_k^2}{2m_k}
\right]\psi= \frac{1}{2m_i}\delta_{ik}\frac{e_i}{c}\frac{\hbar}{i}
\left\{2\left(\frac{\partial {A_i}_{\alpha}}{\partial x_i^{\beta}}-
\frac{\partial {A_i}_{\beta}}{\partial x_i^{\alpha}}\right){D_i}_{\beta}+\right.$$
\begin{equation} \label{ChCoulQHD commut D2} \left.+\frac{\hbar}{i}\frac{\partial}{\partial x_i^{\beta}}
\left( \frac{\partial {A_i}_{\alpha}}{\partial x_i^{\beta}}-
\frac{\partial {A_i}_{\beta}}{\partial x_i^{\alpha}} \right) \right\}\psi ,\end{equation}
\begin{equation} \label{ChCoulQHD commut phi} \left[ \textbf{D}_i, \varphi_k
\right]\psi=\delta_{ik}\frac{\hbar}{i}(\nabla_i\varphi_i)\psi,\end{equation}
and
\begin{equation} \label{ChCoulQHD commut G} \left[ \textbf{D}_i, G_{kn}
\right]\psi=\biggl(\delta_{ik}\frac{\hbar}{i}(\nabla_i G_{in})+\delta_{in}\frac{\hbar}{i}(\nabla_i G_{ki})\biggr)\psi.\end{equation}

A momentum balance equation can be derived by differentiating the current density (\ref{ChCoulQHD def of current of density}) with respect to time:
\begin{equation} \label{ChCoulQHD bal of imp gen}\partial_{t}j^{\alpha}+\frac{1}{m}\partial^{\beta}\Pi^{\alpha\beta}=F^{\alpha},\end{equation}
where $F^{\alpha}(\textbf{r},t)$ is a force field and
$$\Pi^{\alpha\beta}=\int dR\sum_{i}\delta(\textbf{r}-\textbf{r}_{i})\times$$
$$\times\frac{1}{4m_{i}}\biggl(\psi^{*}(R,t)(\hat{D}^{\alpha}_{i}\hat{D}^{\beta}_{i}\psi)(R,t)+$$
\begin{equation} \label{ChCoulQHD Pi}+(\hat{D}^{\alpha}_{i}\psi)^{*}(R,t)(\hat{D}^{\beta}_{i}\psi)(R,t)+c.c.\biggr) \end{equation}
represents the momentum current density tensor or the momentum flux tensor. The four first terms in the right-hand side of formula (\ref{ChCoulQHD DtJ}) are splinched on two groups. One of them contains the first four terms having operator $\widehat{\textbf{D}}$ three times each. The second group contains other terms including commutators (\ref{ChCoulQHD commut D2}). The second group gives the force field $F^{\alpha}(\textbf{r},t)$. The first group appears as
$$\Im=-\int dR\sum_{i=1}^N\delta (\textbf{r}-\textbf{r}_i)\frac{1}{2m_{i}}\frac{\imath}{\hbar}
\sum_{k=1}^N\Biggl[ \psi^*\textbf{D}_i
\frac{\textbf{D}_{k}^{2}}{2m_{k}}\psi $$
$$-\Biggl(\frac{\textbf{D}_k^2}{2m_k}
\psi\biggr)^*\textbf{D}_i\psi+(\textbf{D}_i\psi)^*
\frac{\textbf{D}_{k}^{2}}{2m_{k}}\psi-\biggl(\textbf{D}_{i}
\frac{\textbf{D}_{k}^{2}}{2m_{k}}\psi\biggr)^*\psi\Biggr] .$$
Using manipulation analogous to presented in formula (\ref{ChCoulQHD cont eq calcul 2}) we get $\Im^{\alpha}=\partial_{\beta}\Pi^{\alpha\beta}/m$.

Let us now perform explicit separation of particles thermal movement with velocities $\textbf{u}_{i}(\textbf{r},R,t)$ and the collective movement of particles with velocity $\textbf{v}(\textbf{r},t)$ in equations of continuity (\ref{ChCoulQHD continuity equation}) and of the momentum balance (\ref{ChCoulQHD bal of imp gen}). Putting representation of the wave function (\ref{ChCoulQHD psi representation}) in formula (\ref{ChCoulQHD Pi}) we that in the tensor $\Pi^{\alpha\beta}(\textbf{r},t)$ we can extract three group of terms
\begin{equation}\label{ChCoulQHD} \Pi^{\alpha\beta}=mnv^{\alpha}v^{\beta}+p^{\alpha\beta}+T^{\alpha\beta},\end{equation}
the first of them is expressed via the velocity field; the second
\begin{equation}\label{ChCoulQHD pressure} p^{\alpha\beta}(\textbf{r},t)=\int dR\sum_{i=1}^{N}\delta(\textbf{r}-\textbf{r}_{i})a^{2}(R,t)m_{i}u^{\alpha}_{i}u^{\beta}_{i} \end{equation}
depends on the thermal velocities and has meaning of the kinetic pressure tensor. This tensor tends to zero by
letting $\textbf{u}_{i}\rightarrow 0$; and the last term
$$T^{\alpha\beta}(\textbf{r},t)=-\frac{\hbar^{2}}{2m}\times$$
\begin{equation}\label{ChCoulQHD Bom1} \times\int dR\sum_{i=1}^{N}\delta(\textbf{r}-\textbf{r}_{i})a^{2}(R,t)\partial_{i}^{\alpha}\partial_{i}^{\beta}\ln a(R,t) \end{equation}
is proportional to $\hbar^{2}$ and has a purely
quantum origin. $a(R,t)=\prod_{j=1}^{N}a_{j}(\textbf{r}_{j},t)$. $T^{\alpha\beta}=-\frac{\hbar^{2}}{2m}\sum_{i=1}^{N}\biggl(\frac{1}{2}\partial^{\alpha}\partial^{\beta}a_{i}^{2}(\textbf{r},t)-2(\partial^{\alpha}a_{i}(\textbf{r},t))\partial^{\beta}a_{i}(\textbf{r},t)\biggr)$$=-\frac{\hbar^{2}}{4m}n+\frac{\hbar^{2}}{m}\sum_{i=1}^{N}(\partial^{\alpha}a_{i})\partial^{\beta}a_{i}$. For the large system of noninteracting particles,
this tensor can be presented as
$$T^{\alpha\beta}(\textbf{r},t)=-\frac{\hbar^{2}}{4m}\partial^{\alpha}\partial^{\beta}n(\textbf{r},t)$$
\begin{equation}\label{ChCoulQHD Bom2}+\frac{\hbar^{2}}{4m}\frac{1}{n(\textbf{r},t)}(\partial^{\alpha}n(\textbf{r},t))(\partial^{\beta}n(\textbf{r},t))
.\end{equation}
This term is called the Bohm quantum potential. Divergence of this tensor in many papers is written as
\begin{equation}\label{ChCoulQHD}F_{Q}^{\alpha}=-\partial^{\beta}T^{\alpha\beta}=\frac{\hbar^{2}}{2m}n\partial^{\alpha}\frac{\triangle\sqrt{n}}{\sqrt{n}}\end{equation}
it can be rewritten as
\begin{equation}\label{ChCoulQHD}\textbf{F}_{Q}=\frac{\hbar^{2}}{2m}n\nabla\Biggl(\frac{\triangle n}{n}-\frac{(\nabla n)^{2}}{2n^{2}}\Biggr)\end{equation}

Putting the explicit form of momentum current density in the momentum balance equation we get the Euler equation for the velocity field evolution
$$mn(\partial_{t}+\textbf{v}\nabla)v^{\alpha}+\partial_{\beta}(p^{\alpha\beta}+T^{\alpha\beta})$$
$$=enE_{ext}^{\alpha}+en\varepsilon^{\alpha\beta\gamma}v^{\beta}B_{ext}^{\gamma}$$
\begin{equation}\label{ChCoulQHD bal imp eq with N2}-e^{2}\int
d\textbf{r}'(\partial^{\alpha}G(\textbf{r},\textbf{r}'))n_{2}(\textbf{r},\textbf{r}',t),\end{equation}
where
$\textbf{E}_{ext}=-\nabla\varphi_{ext}-\partial_{t}\textbf{A}_{ext}/c$, $\textbf{B}_{ext}=\nabla\times \textbf{A}_{ext}$ present an external electromagnetic field, and the two-particle concentration
\begin{equation}\label{ChCoulQHD n2 FT}n_{2}(\textbf{r},\textbf{r}',t)=\sum_{s}\int dR\sum_{i,k\neq i}\delta(\textbf{r}-\textbf{r}_{i})\delta(\textbf{r}'-\textbf{r}_{k})\psi^{*}(R,t)\psi(R,t). \end{equation}
At derivation of equation (\ref{ChCoulQHD bal imp eq with N2}) we have used that
$$\int dR\sum_{i=1}^{N}\delta(\textbf{r}-\textbf{r}_{i})\textbf{E}(\textbf{r}_{i},t)\psi^{*}(R,t)\widehat{\textbf{L}}\psi(R,t)$$
\begin{equation}\label{ChCoulQHD taking func out of int}=\textbf{E}(\textbf{r},t)\int dR\sum_{i=1}^{N}\delta(\textbf{r}-\textbf{r}_{i})\psi^{*}(R,t)\widehat{\textbf{L}}\psi(R,t),\end{equation}
where a function of $\textbf{r}_{i}$ was put out of the integral with replacing of argument $\textbf{r}_{i}$ with the argument $\textbf{r}$ due to the $\delta$ function being under the integral.

As the particles of the system under consideration interact via long-range forces the approximation of a self-consistent field is sufficient to analyze collective processes. With the use of this approximation two-particle functions in the momentum balance equation can be split into a product of single-particle functions
\begin{equation}\label{ChCoulQHD n2 long r SCField appr}
n_2(\textbf{r},\textbf{r}',t)\simeq n(\textbf{r},t)n(\textbf{r}',t).\end{equation}
That corresponds to neglecting the two-particle correction existing at general expansion of the two-particle concentration
\begin{equation}\label{ChCoulQHD n2 long r expansion}
n_2(\textbf{r},\textbf{r}',t)=n(\textbf{r},t)n(\textbf{r}',t)+g_{2}(\textbf{r},\textbf{r}',t).\end{equation}
Thus we neglect the quantum correlations $g_{2}$, which include the exchange corrections. QHD beyond the self-consistent field approach is considered below in Sections (\ref{sec:Quantum correlations}) and (\ref{sec:Exchange explicit}).

Let us consider formula (\ref{ChCoulQHD n2 long r expansion}) in more details accurately tracing all arguments of the many-particle wave function $\Psi(R,t)$. Two-particle concentration $n_{2}(\textbf{r},\textbf{r}',t)$ shows correlations between two groups of particles. Coordinate $\textbf{r}$ presents particles moving under action of the Coulomb electric field. Coordinate $\textbf{r}'$ is related to particles, which are sources of the electric field. In plasmas we have at least two species of particles: electrons and ions. If we consider evolution of electrons, then electrons feel action of other electrons and ions.
Thus the term in equation (\ref{ChCoulQHD bal imp eq with N2}) containing the Coulomb interaction is a pair of terms
$$\textbf{F}_{Coul}=-e^{2}\int d\textbf{r}' (\nabla G(\textbf{r},\textbf{r}'))n_{2}(\textbf{r},\textbf{r}',t)\rightarrow$$
\begin{equation}\label{ChCoulQHD F Coulomb in two terms} \rightarrow -q_{e}^{2}\int d\textbf{r}' (\nabla G(\textbf{r},\textbf{r}'))n_{2,ee}(\textbf{r},\textbf{r}',t) -q_{e}q_{i}\int d\textbf{r}' (\nabla G(\textbf{r},\textbf{r}'))n_{2,ei}(\textbf{r},\textbf{r}',t).\end{equation}
This pair of terms describes the electron-electron and electron-ion interactions correspondingly. Considering the self-consistent field approximation both for the electron-electron and electron-ion Coulomb interactions we can represent the two-particle concentrations entering formula (\ref{ChCoulQHD F Coulomb in two terms}) in the traditional form
\begin{equation}\label{ChCoulQHD} n_{2,ee}(\textbf{r},\textbf{r}',t)=n_{e}(\textbf{r},t)n_{e}(\textbf{r}',t)+g_{2,ee}(\textbf{r},\textbf{r}',t),\end{equation}
and
\begin{equation}\label{ChCoulQHD} n_{2,ei}(\textbf{r},\textbf{r}',t)=n_{e}(\textbf{r},t)n_{i}(\textbf{r}',t)+g_{2,ei}(\textbf{r},\textbf{r}',t).\end{equation}
Hence the self-consistent field Coulomb force in the Euler equation for electrons (\ref{ChCoulQHD bal imp eq with N2}) reappears as
\begin{equation}\label{ChCoulQHD} \textbf{F}_{Coul}(\textbf{r},t)=-q_{e}n_{e}(\textbf{r},t) \int d\textbf{r}' (\nabla G(\textbf{r},\textbf{r}'))(q_{e}n_{e}(\textbf{r}',t)+q_{i}n_{i}(\textbf{r}',t)).\end{equation}

Taken in the approximation of self-consistent field, the set of QHD equation, continuity equation and momentum balance equation has the following form:
\begin{equation}\label{ChCoulQHD cont eq inset}
\partial_{t}n+\nabla(n\textbf{v})=0, \end{equation}
and
$$mn(\partial_{t}+\textbf{v}\nabla)v^{\alpha}+\partial_{\beta}(p^{\alpha\beta}+T^{\alpha\beta})
$$
$$=enE_{ext}^{\alpha}+en\varepsilon^{\alpha\beta\gamma}v^{\beta}B_{ext}^{\gamma}$$
\begin{equation}\label{ChCoulQHD bal imp eq selfCons int}-e^{2}n\partial^{\alpha}\int
d\textbf{r}'G(\textbf{r},\textbf{r}')\biggl[n(\textbf{r}',t)-n_{0i}\biggr].\end{equation}

Note, that for 3D systems of particles, the momentum balance equation (\ref{ChCoulQHD bal imp eq selfCons int}) can be written down in terms of the electrical field created by charges of the systems:
$$mn(\partial_{t}+\textbf{v}\nabla)v^{\alpha}+\partial_{\beta}(p^{\alpha\beta}+T^{\alpha\beta})
$$
\begin{equation}\label{ChCoulQHD bal imp eq short}=enE^{\alpha}+en\varepsilon^{\alpha\beta\gamma}v^{\beta}B_{ext}^{\gamma},\end{equation}
where:
$\textbf{E}(\textbf{r},t)=\textbf{E}_{ext}(\textbf{r},t)+\textbf{E}_{int}(\textbf{r},t)$
and
$\textbf{E}_{int}(\textbf{r},t)$ is the internal electric field describing the interparticle Coulomb interaction
\begin{equation}\label{ChCoulQHD el field of charges expl}\textbf{E}_{int}(\textbf{r},t)=-e \nabla\int
d\textbf{r}'G(\textbf{r},\textbf{r}')n(\textbf{r}',t)\end{equation}
These variables meet equations
\begin{equation}\label{ChCoulQHD Max us div E} \nabla\textbf{E}(\textbf{r},t)=4\pi \sum_{i}e_{i}n_{i},\end{equation}
and
\begin{equation}\label{ChCoulQHD Max us rot E} \nabla\times\textbf{E}(\textbf{r},t)=0.\end{equation}

If we have several species of particles, system of electrons and ions, or system of electrons and positrons, for instance, we introduce the particle concentration of each species of particles and derive the continuity, momentum balance and other equations of the chain of the MPQHD equations for each species of particles. The right-hand side of these equations contains interaction between particles of the species and interaction between species.

The QHD equations for electron subsystem in electron-ion quantum plasmas
\begin{equation}\label{ChCoulQHD cont eq electrons}
\partial_{t}n_{e}+\nabla(n_{e}\textbf{v}_{e})=0, \end{equation}
and
$$m_{e}n_{e}(\partial_{t}+\textbf{v}_{e}\nabla)\textbf{v}_{e}+\nabla p_{e}-\frac{\hbar^{2}}{2m_{e}}n_{e}\nabla\Biggl(\frac{\triangle n_{e}}{n_{e}}-\frac{(\nabla n_{e})^{2}}{2n_{e}^{2}}\Biggr)$$
$$=q_{e}n_{e}\biggl(\textbf{E}+\frac{1}{c}[\textbf{v}_{e},\textbf{B}_{ext}]\biggr)-q_{e}^{2}n_{e}\nabla\int G(\textbf{r},\textbf{r}')n_{e}(\textbf{r}',t)d\textbf{r}'$$
\begin{equation}\label{ChCoulQHD Euler eq electrons} -q_{e}q_{i}n_{e}\nabla\int G(\textbf{r},\textbf{r}')n_{i}(\textbf{r}',t)d\textbf{r}',\end{equation}
\textit{and} the QHD equations for ions
\begin{equation}\label{ChCoulQHD cont eq ions}
\partial_{t}n_{i}+\nabla(n_{i}\textbf{v}_{i})=0, \end{equation}
$$m_{i}n_{i}(\partial_{t}+\textbf{v}_{i}\nabla)\textbf{v}_{i}+\nabla p_{i}-\frac{\hbar^{2}}{2m_{i}}n_{i}\nabla\Biggl(\frac{\triangle n_{i}}{n_{i}}-\frac{(\nabla n_{i})^{2}}{2n_{i}^{2}}\Biggr)$$
$$=q_{i}n_{i}\biggl(\textbf{E}+\frac{1}{c}[\textbf{v}_{i},\textbf{B}_{ext}]\biggr)-q_{i}^{2}n_{i}\nabla\int G(\textbf{r},\textbf{r}')n_{i}(\textbf{r}',t)d\textbf{r}'$$
\begin{equation}\label{ChCoulQHD Euler eq ions} -q_{e}q_{i}n_{i}\nabla\int G(\textbf{r},\textbf{r}')n_{e}(\textbf{r}',t)d\textbf{r}'.\end{equation}
Set of equations (\ref{ChCoulQHD cont eq electrons}), (\ref{ChCoulQHD Euler eq electrons}) and (\ref{ChCoulQHD cont eq ions}), (\ref{ChCoulQHD Euler eq ions}) are coupled to each other by means of the last terms in the Euler equations (\ref{ChCoulQHD Euler eq electrons}) and (\ref{ChCoulQHD Euler eq ions}). In equation set (\ref{ChCoulQHD cont eq electrons})-(\ref{ChCoulQHD Euler eq ions}) we assumed that thermal pressure is isotropic:
$p_{a}^{\alpha\beta}=p_{a}\delta^{\alpha\beta}$.

Presence of the short-range interaction between particles gives an additional force field $\textbf{F}_{s}$, which can be written as the divergence of the stress tensor $F_{s}^{\alpha}=-\partial^{\beta}\sigma^{\alpha\beta}$ (see Ref. \cite{Andreev PRA08} formula 10). So we can generalize the momentum current density $\Pi^{\alpha\beta}=mnv^{\alpha}v^{\beta}+p^{\alpha\beta}+T^{\alpha\beta}+\sigma^{\alpha\beta}$.

\subsection{QHD and quasi-classic approach of the quantum mechanics}

In general case the many-particle wave function $\Psi(R,t)$ is the complex function of 3N coordinates $R$ and time $t$. It can be presented as a composition of two real functions. Each of them is a function of 3N coordinates and time $t$ as well. For instance we can choose the following general exponential presentation of the complex many-particle wave function
\begin{equation}\label{ChCoulQHD WF via phase and amplitude} \Psi(R,t)=a(R,t)e^{\imath\frac{S(R,t)}{\hbar}}, \end{equation}
where $a(R,t)$ is the amplitude of the wave function, and $S(R,t)$ is the phase of the wave function. This presentation of the many-particle wave function is useful at introduction velocity field.

If we want to derive the quasi-classic approach of the quantum mechanics \cite{Landau Vol 3} we should represent the wave function in different rather specific form
\begin{equation}\label{ChCoulQHD representation via complex phase} \Psi(R,t)= e^{\imath\sigma(R,t)/\hbar},\end{equation}
where $\sigma(R,t)$ is a complex function, but $\sigma(R,t)$ appears in the form of expansion on degrees of the Planck constant $\hbar$. So we have
\begin{equation}\label{ChCoulQHD expantion of phase} \sigma(R,t) =\sigma_{0}(R,t)+\frac{\hbar}{\imath}\sigma_{1}(R,t) +\biggl(\frac{\hbar}{\imath}\biggr)^{2}\sigma_{2}(R,t)+... \end{equation}

Substituting expression (\ref{ChCoulQHD expantion of phase}) in the many-particle Schrodinger equation (\ref{ChCoulQHD Schrodinger eq}) we find a set of equations arising at different degrees of the Planck constant.

This approach is valuable if the de-Broglie wave lengths of particles are small in compare with characteristic length in the system.

In the case of the Coulomb interparticle interaction, which is an example of interaction having no dependence on the momentum, the substitution gives the following representation of the Schr\"{o}dinger equation
\begin{equation}\label{ChCoulQHD} -\partial_{t}\sigma=\sum_{i}\frac{1}{2m_{i}}\biggl((\nabla_{i}\sigma)^{2}-\imath\hbar\triangle\sigma\biggr)+\sum_{i,j\neq i}U_{ij}.\end{equation}

Now we can separate contribution of different orders of the quasi-classical approach. In the zeroth order on $\hbar$ we have
\begin{equation}\label{ChCoulQHD QCA zeroth order} \partial_{t}\sigma_{0} =\sum_{i}\frac{1}{2m_{i}}(\nabla_{i}\sigma_{0})^{2}+\sum_{i,j\neq i}U_{ij}.\end{equation}
In the first order on the Planck constant $\hbar$ one obtains
\begin{equation}\label{ChCoulQHD QCA first order} \partial_{t}\sigma_{1} =\sum_{i}\frac{1}{2m_{i}}\biggl((2\nabla_{i}\sigma_{0}\cdot\nabla_{i}\sigma_{1})+\triangle_{i}\sigma_{0}\biggr).\end{equation}
In the second order on the Planck constant $\hbar$ one finds
\begin{equation}\label{ChCoulQHD QCA second order} \partial_{t}\sigma_{2} =\sum_{i}\frac{1}{2m_{i}}\biggl((\nabla_{i}\sigma_{1})^{2}+2(\nabla_{i}\sigma_{0}\cdot\nabla_{i}\sigma_{2})+\triangle_{i}\sigma_{1}\biggr).\end{equation}

Having equations (\ref{ChCoulQHD QCA zeroth order})-(\ref{ChCoulQHD QCA second order}) we can derive QHD equations in the quasi-classic approach. Hence we need to define the particle concentration. Applying general definition of the particle concentration (\ref{ChCoulQHD def density with brackets}) and the expansion of the wave function in quasi-classic approach (\ref{ChCoulQHD representation via complex phase}), (\ref{ChCoulQHD expantion of phase}) we obtain
\begin{equation}\label{ChCoulQHD} n=\langle \Psi^{*}\hat{n}\Psi\rangle=\langle \hat{n}e^{\imath(\sigma-\sigma^{*})/\hbar}\rangle=\langle \hat{n}e^{2\sigma_{1}}\rangle.\end{equation}
This definition shows that in zeroth order on the Planck constant the time derivative of the particle concentration equals to zero $\partial_{t}n=0$. Evolution of concentration appears in the first order on $\hbar$. Correction to this evolution can be found in higher uneven orders on $\hbar$. Let us mention that time evolution of $n$ depends on time evolution of $\sigma_{1}$: $\partial_{t}n=2\langle \hat{n}e^{2\sigma_{1}}\partial_{t}\sigma_{1}\rangle$. Whereas $\partial_{t}\sigma_{1}$ depends on $\sigma_{0}(t)$, which depends on interaction.

Simple manipulation with the time derivative of the particle concentration leads to the continuity equation (\ref{ChCoulQHD continuity equation}) with the particle current
\begin{equation}\label{ChCoulQHD def current QCA} \textbf{j}=\frac{1}{m} \langle (\hat{n}\nabla\sigma_{0})e^{2\sigma_{1}}\rangle. \end{equation}

At derivation of the MPQHD equations, presented above, we have deal with the full wave function (\ref{ChCoulQHD WF via phase and amplitude}) without expansion (\ref{ChCoulQHD expantion of phase}). If quantum effects getting smaller, that corresponds decreasing of the de-Broglie wave length, the MPQHD is an equivalent of application of full series (\ref{ChCoulQHD expantion of phase}) instead of several term we consider in the quasi-classic approximation.

Time evolution of the particle current $\textbf{j}$ appears as simultaneous time evolution of $\sigma_{0}$ and $\sigma_{1}$. We see that the interaction is included in the "classical" approach, in the zeroth order on $\hbar$ (see equation (\ref{ChCoulQHD QCA zeroth order})). Let us derive the quasi-classic Euler equation. Differentiating function (\ref{ChCoulQHD def current QCA}) with respect to time and applying equation (\ref{ChCoulQHD QCA zeroth order}) and (\ref{ChCoulQHD QCA first order}) we find
\begin{equation}\label{ChCoulQHD} \partial_{t}j^{\alpha}+\partial_{\beta}\Pi^{\alpha\beta}=F^{\alpha},\end{equation}
where
\begin{equation}\label{ChCoulQHD} \textbf{F}=\frac{1}{m}\int d\textbf{r}' (\nabla U(\textbf{r},\textbf{r}')) n_{2}(\textbf{r},\textbf{r}',t) \end{equation}
is the interaction force field with the quasi-classic two-particle concentration
\begin{equation}\label{ChCoulQHD} n_{2}(\textbf{r},\textbf{r}',t)=-2\int dR\sum_{i,j\neq i}\delta(\textbf{r}-\textbf{r}_{i})\delta(\textbf{r}'-\textbf{r}_{j})e^{2\sigma_{1}(R,t)}, \end{equation}
and
\begin{equation}\label{ChCoulQHD} \Pi^{\alpha\beta}= \frac{1}{m^{2}}\langle\hat{n}(\partial^{\alpha}_{i}\sigma_{0})(\partial^{\beta}_{i}\sigma_{0})e^{2\sigma_{1}}\rangle\end{equation}
is the quasi-classical momentum current.

The MPQHD in the quasi-classic approach allows to introduce the velocity field $\textbf{v}(\textbf{r},t)$ and thermal velocities $\textbf{u}_{i}$. Introducing usual relation between the particle current $\textbf{j}$ and the velocity field $\textbf{v}=\textbf{j}/n$ we obtain definition for the thermal velocities $\textbf{u}_{i}=\nabla_{i}\sigma_{0}/m-\textbf{v}$ and their main property
\begin{equation}\label{ChCoulQHD thermal vel QCA aver} \langle \hat{n}\textbf{u}_{i} e^{2\sigma_{1}}\rangle=0.\end{equation}
Formula (\ref{ChCoulQHD thermal vel QCA aver}) shows that quantum mechanical average of thermal velocities of all particles in the system equals to zero in each point of space.

As the next step we obtain representation of the momentum current $\Pi^{\alpha\beta}$ via the velocity field and thermal velocities
\begin{equation}\label{ChCoulQHD Pi via v in QCA} \Pi^{\alpha\beta}=nv^{\alpha}v^{\beta}+p^{\alpha\beta},\end{equation}
where
\begin{equation}\label{ChCoulQHD} p^{\alpha\beta}=\langle\hat{n} u_{i}^{\alpha}u_{i}^{\beta}e^{2\sigma_{1}}\rangle \end{equation}
is the thermal pressure. Formula (\ref{ChCoulQHD Pi via v in QCA}) does not contain the quantum Bohm potential contribution.

We see that the MPQHD contracted in terms of collective variables suitable for description of many-particle physics does not reduce itself to the quasi-classic approach, but it keeps many traces of classical structure of equation due to many-particle systems description.

In this subsection $n$, $\textbf{j}$, $\Pi^{\alpha\beta}$ are quantities existing the first order on the Planck constant $\hbar$. Working in the first order of the quasi-classic theory we do not obtain quantum contributions, the famous quantum Bohm potential, in the Euler equation. If we want to derive the quantum Bohm potential we should extend our model and include contributions up to the third order by the Planck constant $\hbar$.

Explicit consideration of the MPQHD in terms of the quasi-classic approach of quantum mechanics shows that the MPQHD is not a quasi-classic theory. Even minimal coupling QHD model containing the continuity (\ref{ChCoulQHD cont eq inset}) and Euler (\ref{ChCoulQHD bal imp eq with N2}) equations, with no account of relativistic effects, exchange effects, and other quantum correlations, includes effects, which do not arise in the first order on the quasi-classic theory.

\subsection{Density matrix \textit{and} basic definitions and area applicability of MPQHD}

The many-particle wave function gives full description of physical systems. However, if we want to consider a part of the system interacting with surrounding particles, then this part will not be in a pure state allowing description in terms of the wave function. The density matrix is required for description of the part of the system \cite{Landau Vol 3}. Full system can also be described in terms of the density matrix $\varrho(R,R')$, but in this case the density matrix presents a description, which is an equivalent to description by means of the many-particle wave function. The density matrix of the full system is defined in terms of the many-particle wave function depending on coordinates of all particles
\begin{equation}\label{ChCoulQHD} \varrho(R,R')=\Psi(R,t)\Psi^{*}(R',t).\end{equation}

The density matrix of a part of the system containing particles with the set of coordinates $Q$ can also be defined via wave function of the full system . Let us present full set of coordinate $R$ as two parts: coordinates of the subsystem under consideration $Q$, and coordinates of other particles $P$. Hence the full set of coordinates $R$ can be written as $R=\{Q,P\}$.

The density matrix of subsystem appears as
\begin{equation}\label{ChCoulQHD} \varrho_{ss}=\int \Psi^{*}(Q,P,t)\Psi(Q,P,t) dP,\end{equation}
where $dP$ is the product of differentials of coordinates entering the set of coordinates $P$.

Considering closed system of particles we do not need to apply the density matrix, hence we keep applying the many-particle wave function for definition of hydrodynamic variables.

\section{\label{sec:Energy ev} Energy evolution}

Even in the self-consistent field approximation the set of continuity (\ref{ChCoulQHD cont eq inset}) and Euler equations (\ref{ChCoulQHD bal imp eq short}) is not closed, since it contains the thermal pressure. Usually the thermal pressure tensor is reduced to a scalar function $p^{\alpha\beta}=p\delta^{\alpha\beta}$, but we have to present an equation of state to get relation of pressure with the particle concentration, velocity field, and other parameters describing the system.

Let us consider some properties of the pressure when system is adiabatic. If we have a homogeneous ideal gas undergoing a reversible adiabatic process, when the pressure satisfy the equation
\begin{equation}\label{ChCoulQHD eq adiabate}pV^{\gamma}=const,\end{equation}
where $p$ is the pressure in the system, $V$ is the volume of the system, and $\gamma$ is the adiabatic index. For the point like particles having three degree of freedom $\gamma=5/3$, when for adiabatic excitations in plasmas $\gamma=3$ \cite{Kroll book}. It corresponds fast compression of medium at  wave propagation. Low frequency excitations, such as ion-sound, are better to be described by isothermal perturbations what gives $\gamma=1$. It is useful, for our purpose, to represent equation (\ref{ChCoulQHD eq adiabate}) via the particle concentration $n$. Since the particles number does not change we can divide it by the particles number $N$, and introducing the concentration $n=N/V$, we get
\begin{equation}\label{ChCoulQHD}\frac{p}{n^{\gamma}}=const.\end{equation}
We see that during the adiabatic process $p/n^{\gamma}$ keeps to be constant. Then, $\nabla(p/n^{\gamma})=0$ and
\begin{equation}\label{ChCoulQHD eq adiabate 2}n\nabla p-\gamma p\nabla n=0.\end{equation}
Making calculations in linear approximation on small perturbations $\delta n$, $\delta p$ around equilibrium state described by $n_{0}$, $p_{0}$ we can rewrite equation (\ref{ChCoulQHD eq adiabate 2}) as
$$\nabla\delta p=\gamma\frac{p_{0}}{n_{0}}\nabla\delta n.$$
Therefore we have that the gradient of the pressure perturbation is expressed via $\delta n$ and parameters describing the equilibrium state.

During derivation of the Euler equation (\ref{ChCoulQHD bal imp eq short}) we separated kinetic terms (we put them in the left-hand side) and terms describing interactions (in the right-hand side). Consequently we can conclude that the thermal pressure $p$ contains no trace of the interaction, so we can use equation of state suggested for the ideal gas
\begin{equation}\label{ChCoulQHD eq of state}p=nT.\end{equation}
Thus, for the pressure perturbation we find $\nabla\delta p=\gamma T_{0}\nabla\delta n$. This result will be used in the section (\ref{sec:Apps}) of this paper, where we present applications of the method under description for dispersion of waves in the quantum plasma.

Going up to nonlinear perturbations we should come back to equation (\ref{ChCoulQHD eq adiabate 2}). Putting equation of state in formula (\ref{ChCoulQHD eq adiabate 2}) we have $\nabla p=\gamma T\nabla n$ and we need to have an equation of the temperature evolution to get a closed set of equations. This is an example showing that we need to derive other equations of quantum hydrodynamics, in addition to the continuity and Euler equations, at least one equation more.

On the other hand, the local Maxwell distribution function in the classic physics
\begin{equation}\label{ChCoulQHD}f(\textbf{r},t)=n(\textbf{r},t)\biggl(\frac{m}{2\pi T(\textbf{r},t)}\biggr)^{3/2}\exp\biggl(-\frac{m\textbf{v}^{2}(\textbf{r},t)}{2T(\textbf{r},t)}\biggr),\end{equation}
and the Fermi distribution function in the quantum physics
\begin{equation}\label{ChCoulQHD}f(\textbf{r},t)=\frac{n(\textbf{r},t)}{\exp\biggl(\frac{\varepsilon(\textbf{r},t)-\mu(\textbf{r},t)}{T(\textbf{r},t)}\biggr)+1}\end{equation}
depend on the particle concentration $n(\textbf{r},t)$, velocity field $\textbf{v}(\textbf{r},t)$, and temperature field $T(\textbf{r},t)$. The last one is the part of the local kinetic energy $\varepsilon_{kin}(\textbf{r},t)$ associated with the thermal motion of particles (we will discuss it below).

The energy density can be defined in the MPQHD as the quantum mechanical average of the energy operator \cite{MaksimovTMP 1999}, \cite{MaksimovTMP 2001}
$${\cal E}(\textbf{r},t)={\cal E}_{\rm kin}(\textbf{r},t)+{\cal
E}_{\rm pot}(\textbf{r},t)$$
$$=\int dR\sum_{i=1}^N\delta(\textbf{r}-\textbf{r}_i)\frac{1}{4m_i }
\biggl[\psi^*(R,t)\textbf{D}_i^2\psi(R,t)+ \biggl(\left(\textbf{D}_i\right)^2\psi\biggr)^*(R,t)\psi(R,t)\biggr]$$
\begin{equation}\label{ChCoulQHD def.E}
+\frac{1}{2} \int dR  \sum_{i,j=1; j\neq i}^{N}
\delta (\textbf{r}-\textbf{r}_i) e_{i} e_j G_{ij}  \psi^{*}(R,t)\psi (R,t).
\end{equation}
This definition can be applied for derivation of the energy balance equation. Using definition (\ref{ChCoulQHD def.E}) we can also derive inner or thermal energy of the system. We do it below extracting the energy of the local centre mass flow. The thermal energy allows to get hydrodynamic definition of temperature. Hence we see that a lot of important application are related with the energy density (\ref{ChCoulQHD def.E}).

Total energy arises as integral of the energy density over whole space
$E=\displaystyle\int{\cal E}(\textbf{r},t)d \textbf{r}=\int dR\psi^*(R,t)\hat H\psi(R,t)$. This is the energy quantized in quantum mechanics.

Considering time evolution of the energy density ${\cal E}$ (\ref{ChCoulQHD def.E}) obeying the many-particle Schrodinger equation (\ref{ChCoulQHD Schrodinger eq}) we find the quantum energy balance equation
\begin{equation}\label{ChCoulQHD eq.balance E} \partial_{t}{\cal E}+\nabla  \textbf{Q}
=\textbf{j}_e \textbf{E}^{\rm ext}(\textbf{r},t)+{\cal A}(\textbf{r},t),
\end{equation}
with the kinetic energy current
$$\textbf{Q}_{\rm kin}(\textbf{r},t)={1\over 8} \int dR\sum_{i=1}^N \delta (\textbf{r}-\textbf{r}_i)
{1\over m_i^2} \biggl[\psi^*(R,t) \textbf{D}_i  \textbf{D}_i^2 \psi (R,t)$$
$$+(\textbf{D}_i \psi)^*(R,t)  \textbf{D}_i^2 \psi (R,t)+
\biggl(\textbf{D}_i^2\psi\biggr)^*(R,t)  \textbf{D}_i \psi (R,t)$$
\begin{equation}\label{ChCoulQHD def.Q_kin} +\biggl(\textbf{D}_i \textbf{D}_i^2\psi\biggr)^*(R,t)\psi (R,t)\biggr],\end{equation}
and the potential energy current
$$\textbf{Q}_{pot}(\textbf{r},t)={1\over 4}\int dR \mathop{{\sum}'}_{i,j=1}^N
e_{i} e_{j} G_{ij}{1\over m_i} \biggl[    \psi^*(R,t)   \textbf{D}_i\psi (R,t)$$
\begin{equation}\label{ChCoulQHD def.Q_pot}+(\textbf{D}_i\psi)^*(R,t)\psi (R,t)  \biggr].\end{equation}
Together they give full current of energy, which is also called the energy flux.

The scalar field of work arises in the energy balance equation. Its explicit form to be
$$ {\cal A}(\textbf{r},t)=-{1\over 4}  \int dR \mathop{{\sum}'}_{i,j=1}^N
\delta(\textbf{r}-\textbf{r}_i) e_i e_j \nabla_i G_{ij}\times$$
$$\times \Biggl[ {1\over m_i} \biggl(\psi^*(R,t)\textbf{D}_i \psi (R,t)+(\textbf{D}_i \psi)^*(R,t)\psi (R,t)\biggr)$$
\begin{equation}\label{ChCoulQHD def.A} +{1\over m_j}\biggl(\psi^{*}(R,t)
\textbf{D}_j \psi(R,t)+  (\textbf{D}_j \psi)^{*}(R,t)\psi (R,t)\biggr) \Biggr].\end{equation}

\section{\label{sec:Vel field in energy} Velocity field in the energy balance equation}

Part of the energy density is related to motion of the local centre of mass. Motion of particles relatively the local centre of mass is the thermal motion. We can separate energy density for these types of motion. Hence we present energy density as sum of two terms
\begin{equation}\label{ChCoulQHD relation of energy and thermal en} {\cal E}=\frac{1}{2}mn\textbf{v}^{2}+n\epsilon,\end{equation}
where $\epsilon$ is the thermal energy related to the temperature $\epsilon=\frac{3}{2}T$. To get equation for the thermal energy evolution, giving the equation for the quantum evolution of temperature, we should use substitution (\ref{ChCoulQHD relation of energy and thermal en}) \emph{and} the continuity and Euler equations to extract time evolution of the particle concentration $n$ and the velocity field $\textbf{v}$.

The energy current can be represented as follows
\begin{equation}\label{ChCoulQHD Q expansion as four terms}
Q^\alpha=v^\alpha\varepsilon+v^\beta (p^{\alpha\beta}+T^{\alpha\beta})+q^\alpha.
\end{equation}

Substituting formulae (\ref{ChCoulQHD relation of energy and thermal en}) and (\ref{ChCoulQHD Q expansion as four terms}) in the energy balance equation (\ref{ChCoulQHD eq.balance E}), and applying the continuity equation (\ref{ChCoulQHD cont eq inset}) and the Euler equation (\ref{ChCoulQHD bal imp eq with N2}) to time derivatives of the particle concentration $\partial_{t}n$ and the velocity field $\partial_{t}\textbf{v}$, we obtain equation for the "thermal energy" $\epsilon$ evolution
$$n \biggl[
\frac{\partial}{\partial t}+\textbf{v}\nabla
\biggr]\epsilon+p_{\alpha\beta}
\frac{\partial v_{\alpha}}{\partial x^{\beta}}+\nabla \textbf{q}$$
\begin{equation}\label{ChCoulQHD eq of thermal energy evol} =-{e^2\over 2}\int d \textbf{r}'
\left(\textbf{v}( \textbf{r}',t)- \textbf{v}( \textbf{r},t)
\right)  \nabla G(\textbf{r}-\textbf{r}') n_2(\textbf{r},\textbf{r}',t)+\alpha .\end{equation}
Above we put notion thermal energy in quotes since this thermal energy $\epsilon$ contains the thermal and quantum contributions, as we can see from the explicit form of the thermal energy $\epsilon$ presented below.

Explicit form of the thermal energy density can be written as
$$n(\textbf{r},t)\epsilon (\textbf{r},t)
=\int dR\sum_{i=1}^N \delta (\textbf{r}-\textbf{r}_i)a^2(R,t)
\left(\frac{m \textbf{u}_i^2}{2}-\frac{\hbar^2}{2m}
\frac{\Delta_ia}{a}\right)$$
\begin{equation}\label{ChCoulQHD}+{1\over 2}\int d\textbf{r}'e^2 G(\textbf{r}-\textbf{r}')n_2(\textbf{r},\textbf{r}',t).\end{equation}
The explicit form of $\epsilon (\textbf{r},t)$ arises in terms of amplitude of the many-particle wave function $a=a(R,t)$, the thermal velocities $\textbf{u}_i$, and the two-particle concentration $n_2(\textbf{r},\textbf{r}',t)$ giving energy of interaction. We also have quantum contribution in the energy density presented by the term proportional $\hbar^{2}$.

The part of the energy current related to the local centre of mass can be written explicitly
$$q_{\alpha}(\textbf{r},t)=\int dR\sum_{i=1}^N \delta (\textbf{r}-\textbf{r}_i)
a^2(R,t)\Biggl[ {u_i}_{\alpha}\left(
\frac{m \textbf{u}_i^2}{2}-\frac{\hbar^2}{2m}
\frac{\Delta_ia}{a}\right)$$
$$-\frac{\hbar^2}{2m}\partial_{i}^{\alpha}\left(\textbf{u}_{i}
\nabla_{i}\ln a\right)-\frac{\hbar^2}{4m}
\partial_{i}^{\alpha}\partial_{i}^{\beta}u_i^{\beta} \Biggr]-\frac{\hbar^2}{4m}(\partial_{\beta}n)\partial_{\alpha}v^{\beta}-\frac{\hbar^2}{4m}n\partial^{\alpha}(\nabla \textbf{v})$$
\begin{equation}\label{ChCoulQHD q explicit}+{1\over 2}\int d\textbf{r}'e^2 G(\textbf{r}-\textbf{r}')
\int dR \mathop{{\sum}'}\limits_{i,j=1}^N
\delta (\textbf{r}-\textbf{r}_i)\delta (\textbf{r}'-\textbf{r}_j)
a^2(R,t){u_i}_{\alpha}.\end{equation}
This formula presents two contributions: the thermal energy current and the quantum energy current. The last one is similar to the quantum Bohm potential in the Euler equation.
As we see from formula (\ref{ChCoulQHD q explicit}) the thermal energy current consists of two parts. The first part is the first term in formula (\ref{ChCoulQHD q explicit}), which is proportional to ${u_i}_{\alpha} \textbf{u}_i^2$. So this is the the thermal current of kinetic energy. The second part of the thermal energy current is the last term of formula (\ref{ChCoulQHD q explicit}) containing the Green function of the Coulomb interaction. Hence this is the the thermal current of potential energy. The second-sixth terms of formula (\ref{ChCoulQHD q explicit}) give the quantum energy current. The quantum terms in $q_{\alpha}(\textbf{r},t)$ have different forms. Some of the related to motion of centre of mass, when this motion is related to de-Brougle wave nature of particles. These are the fifth and sixth terms, which do not contain thermal velocities. The second-fourth terms are quantum-thermal terms since they are contain the Plank constant and the thermal velocities. Due to this description we can present $\textbf{q}$ as sum of three terms $\textbf{q}=\textbf{q}_{cl}+\textbf{q}_{\hbar}+\textbf{q}_{th\hbar}$, which are classic, quantum and thermal-quantum parts described above. Let us separately present the quantum part of $\textbf{q}$
\begin{equation}\label{ChCoulQHD}
q^\alpha_{\hbar}= -\frac{\hbar^2}{4m}(\partial_{\beta}n)\partial_{\alpha}v^{\beta}-\frac{\hbar^2}{4m}n\partial^{\alpha}(\nabla \textbf{v}).
\end{equation}

The work field on thermal velocities enters equation (\ref{ChCoulQHD eq of thermal energy evol}), it has the following form
$$\alpha(\textbf{r},t)=-{1\over 2} \int d\textbf{r}' e^2
\nabla G(\textbf{r}-\textbf{r}') \int dR \mathop{{\sum}'}_{i,j=1}^N
\delta (\textbf{r}-\textbf{r}_i) \delta (\textbf{r}'-\textbf{r}_j) a^2(R,t)\times$$
\begin{equation}\label{ChCoulQHD} \times\biggl(
\textbf{u}_i(\textbf{r},R,t)+\textbf{u}_j(\textbf{r}',R,t)
\biggr).\end{equation}

The energy balance equation can be derived from the single-particle Schrodinger equation, but the single-particle energy evolution equation will be the consequences of the continuity and Euler equations, since the single-particle energy density is a combination of the concentration and the velocity field.
Consideration of the many-particle system reveals in independent energy evolution equation and a set of other functions, such as the pressure, energy current, etc.

\section{\label{sec:NLSE} NLSE for quantum plasmas}

"One-particle" non-linear Schrodinger equations (NLSEs) appear to be very useful tool in some areas of the condensed matter physics. One of the most well-known NLSEs is the Gross-Pitaevskii equation describing dynamical properties of the inhomogeneous non-ideal Bose-Einstein condensates of neutral particles with the short-range interaction \cite{L.P.Pitaevskii RMP 99}.

It is essential to mention that a derivation of the Gross-Pitaevskii equation was performed by means of the MPQHD method \cite{Andreev PRA08}. Consideration of the exchange interaction plays crucial role in this derivation.

In the MPQHD method, the NLSE arises as a representation of truncated set of the QHD equations down to the continuity and Euler equations. Ultracold non-ideal neutral fermions \cite{Giorgini RMP 08} were also considered by the MPQHD method \cite{Andreev PRA08}.

Considering the eddy-free motion
$\textbf{v}(\textbf{r},t)=\nabla\phi(\textbf{r},t)$ and the
barotropicity condition for the isotropic thermal pressure
\begin{equation}\label{ChCoulQHD barotropicity condition}\frac{\nabla p(\textbf{r},t)}{mn(\textbf{r},t)}=\nabla \mu(\textbf{r},t),\end{equation}
where $\mu(\textbf{r},t)$ is the chemical
potential, we see that the momentum balance equation (\ref{ChCoulQHD bal imp eq short})
has the Cauchy integral. In this case the momentum balance equation
can be associated with an equivalent
one-particle non-linear Schr\"{o}dinger equation (NLSE) for an effective wave
function $\Phi(\textbf{r},t)$. In literature this wave function is called the order parameter or the wave function in the medium. We define this function as follows
\begin{equation}\label{ChCoulQHD WF in medium}
\Phi(\textbf{r},t)=\sqrt{n(\textbf{r},t)}\exp\biggl(\frac{\imath}{\hbar}m\phi(\textbf{r},t)\biggr)
.\end{equation}
We see from definition that $\Phi(\textbf{r},t)$ describe many particle behavior in terms of one complex wave function.
Differentiating it with respect to time and using equations
(\ref{ChCoulQHD cont eq inset}), (\ref{ChCoulQHD bal imp eq short}) we find the non-linear
one-particle Schr\"{o}dinger equation
\begin{equation}\label{ChCoulQHD NLSE} \imath\hbar\partial_{t}\Phi(\textbf{r},t)=\Biggl(-\frac{\hbar^{2}\nabla^{2}}{2m}+\mu(\textbf{r},t)+q\widetilde{\phi}(\textbf{r},t)\Biggr)\Phi(\textbf{r},t),
\end{equation}
where the electric field $\textbf{E}=-\nabla\widetilde{\phi}$ obeys the Maxwell equations $\textmd{curl} \textbf{E}=0$ and $\textmd{div} \textbf{E}=4\pi (-e)(\Phi^{*}\Phi-n_{i})$. Hence equation (\ref{ChCoulQHD NLSE}) is a non-linear equation, since the electric field $E(\textbf{r},t)$ is a function of the wave function in the medium $\Phi(\textbf{r},t)$. We can represent equation (\ref{ChCoulQHD NLSE}) as a integral equation
$$\imath\hbar\partial_{t}\Phi(\textbf{r},t)=\Biggl(-\frac{\hbar^{2}\nabla^{2}}{2m}+\mu(\textbf{r},t)$$
\begin{equation}\label{ChCoulQHD NLSE int} +e^{2}\int
d\textbf{r}'G(\textbf{r},\textbf{r}')\biggl[n(\textbf{r}',t)-n_{0i}\biggr]\Biggr)\Phi(\textbf{r},t),
\end{equation}

The NLSE is an alternative to the couple of the continuity and Euler equations for potential velocity field.

Integral NLSE is useful for low dimensional plasmas. Corresponding examples are presented below after derivation of the exchange Coulomb potential for 1D, 2D, and 3D plasmas. Hence we will present more general NLSEs with the exchange interaction derived by means of the MPQHD method.

NLSEs have been also applied for studying of ultracold neutral fully polarised spinning particles and particles having electric dipole moment \cite{Yi PRA 00}-\cite{Stamper-Kurn RMP 13}. The first principles derivation of these model and some generalisations was performed by means of the MPQHD method \cite{Andreev 2013 non-int GP}, \cite{Andreev EPJ D Pol}.

\section{\label{sec:Equations of state} Equations of state in 3D, 2D and 1D degenerate electron gas located in external magnetic field}

Explicit form of equation of state is an important part of the truncation procedure in quantum and classic hydrodynamics. Above we have discussed equation of state for classic plasmas. Here we pay attention to equation of state for 3D, 2D, and 1D quantum plasmas of degenerate electrons. Especially we focus our attention on electron gas located in external magnetic field.  In spite the fact that we do not consider the spin evolution of electron, the equilibrium spin gives contribution in plasma dynamics. This contribution appears due to modification of equation of state in presence of the external magnetic field.

Before we start description of equations of state for degenerate electron gas we should make a note that contribution of 3D Fermi pressure in spectrums of plasmas differs from similar contribution given by physical kinetics. This different reveals in different numerical coefficients in from of the square of the Fermi pressure.  Application of the Fermi pressure in the Euler equation gives $1/3$, then kinetics leads to $3/5$, it was shown by Vlasov in 1938 \cite{Landau Vol 10} (see problem in section 40), \cite{Rukhadze book 9} (see section 19).

Different form of symmetric perturbations can propagate in three dimensional plasmas. Most symmetric of them are the plane, cylindric and spherical waves. These three examples are kind of "one dimensional" perturbations in 3D plasmas, since they depend on one coordinate: $x$ of $x$, $y$, $z$ for  plane waves, $\rho$ of $\rho$, $\varphi$, $z$ for cylindric waves, $r$ of $r$, $\theta$, $\varphi$ for spherical waves.

Since we have 3D mediums we can apply the Maxwell equations in the usual form (\ref{ChCoulQHD Max us div E}) and (\ref{ChCoulQHD Max us rot E}) along with the equation of state for three dimensional electron gas (for degenerate electron gas see formulae (\ref{ChCoulQHD EqState un Pol}) below). In three dimensional case plasmas characterise by the "three dimensional" concentrations of particles $[n_{3D}]=cm^{-3}$.

In two dimensional plasmas, for instance plane plasmas, two-dimensional electron gas (2DEG), carbon nano-tubes as cylindrical two dimensional plasmas, and fullerene molecules as spherical two dimensional plasmas. we have two type of symmetrical perturbations. They are the circular perturbations, which are 2D analog of cylindrical waves, and "plane" waves. The wave front of the 2D plane waves is a straight line. These are examples of "one dimensional" perturbations in two dimensional plasmas.

In 2D plasmas the particle concentration $n_{2D}$ is the number of particles on the unit of surface $[n_{2D}]=cm^{-2}$.

In 2D plasmas we can use the usual form of the Maxwell equation, but they contain the Dirac delta function. If we consider the Poisson equation for plane plasmas we can write $\nabla \textbf{E}=4\pi q_{e}(n_{e,2D}-n_{i,2D})\delta(z)$, where we have assumed that the plane is located at $z=0$. However the integral form is rather useful, especially when we want to consider nano-tubes and fullerene molecules.

Consideration of 2D plasmas require to be careful with equations of field. But it also required to choose another equation of state (explicit form can be found below (\ref{ChCoulQHD eq State 2D unPol})).

We can also have one dimensional plasmas, which are string-like objects. In simplest case it is a straight line of electrons, where we can observe longitudinal waves of particles number or density. In this case the particle concentration $[n_{1D}]=cm^{-1}$ is the number of particles on the unit of length. In this case we can also dial with the usual form of the Maxwell equations $\nabla \textbf{E}=4\pi q_{e}(n_{e,1D}(x)-n_{i,1D}(x))\delta(y)\delta(z)$, but it is advisable to apply the integral form (see formula (\ref{ChCoulQHD F coul int 1D}) below). Equation of state of 1D electron gas differs from  equations of states for 2D and 3D electron gas. Explicit for of equation of sate for degenerate 1D electron gas we present by formula (\ref{ChCoulQHD eq State 1D unPol}).

Unfortunately, we should mention that there is mess in literature on this subject. Some times authors simultaneously apply  equation of state and equation of field corresponding to plasmas of different dimensional structure. There are a lot examples of such confusion. We describe here just some of them to illustrate our statement (see subsection (\ref{subsec:Modified eq St})).  Or we can also find misapplication of the exchange potential, when the exchange potential derived for 3D plasmas is applied for 2D case \cite{Khan JAP 14}. In section (\ref{sec:Exchange explicit}) we consider exchange interaction for 2D and 3D electron gas, so readers can see difference between them.

\subsection{Equation of state in 3D degenerate electron gas}

We need to get a closed set of equations, so we should use an equation of state for the pressure of electrons $p_{e}$ via hydrodynamic variables, namely via the particle concentration $n_{e}$. We consider degenerate electrons. Hence, in non-relativistic case and in absence of any external field, we have
\begin{equation}\label{ChCoulQHD EqState un Pol} p_{unpol}=p_{Fe,3D}=\frac{(3\pi^{2})^{2/3}}{5}\frac{\hbar^{2}}{m_{e}}n_{e}^{5/3}.\end{equation}
From this equation of state we find $\frac{\partial p_{Fe}}{\partial n_{e}}=\frac{(6\pi^{2})^{2/3}}{3}\frac{\hbar^{2}}{m_{e}}n_{e}^{2/3}$ giving contribution in the Euler equation via $\nabla p_{Fe}=\frac{\partial p_{Fe}}{\partial n_{e}}\nabla n_{e}$. At derivation of the Fermi pressure one assumes that two particles with different spin directions could occupy a quantum state $n_{0\uparrow}= n_{0\downarrow}$. Formula (\ref{ChCoulQHD EqState un Pol}) and method derivation of this formula can be found in most textbooks on statistical physics, we present reference on the Landau Course of Theoretical Physics Ref. \cite{Landau v5 eq st} (see section 57).

Considering quantum spin 1/2 plasmas researchers usually apply equation of state for unpolarized electrons (\ref{ChCoulQHD EqState un Pol})
see Refs. \cite{Landau v5 eq st}-\cite{Mushtaq PP 10}.  Hence, being focused on the spin contribution in dynamics of small amplitude perturbations authors have not included change of equilibrium properties. However equilibrium quantities, such as derivative of the pressure on the particle concentration, appears as coefficients in equations describing perturbation evolution.  Consequently, change equation of state by spins in an uniform external magnetic field affects dynamical properties, such as the spectrum of collective excitations, of quantum plasmas. Explicit form equation of state has more dramatic influence on nonlinear properties of physical systems.

Including different occupation of quantum states by spin-up and spin-down electrons located in magnetic field we obtain the following equation of state of 3D degenerate electrons in an external uniform magnetic field
$$p_{pol,3D}=\biggl[\frac{(6\pi^{2})^{\frac{2}{3}}}{5}\frac{\hbar^{2}}{m}\biggl(n_{(av)}+\frac{\Delta n}{2}\biggr)^{\frac{5}{3}}$$ \begin{equation}\label{ChCoulQHD eq State single Fl polarised 3D} +\frac{(6\pi^{2})^{\frac{2}{3}}}{5}\frac{\hbar^{2}}{m}\biggl(n_{(av)}-\frac{\Delta n}{2}\biggr)^{\frac{5}{3}}\biggr],\end{equation}
where $n_{\uparrow}=n_{(av)}-\frac{\Delta n}{2}$, $n_{\downarrow}=n_{(av)}+\frac{\Delta n}{2}$, $n=n_{\uparrow}+n_{\downarrow}=2n_{(av)}$, $\Delta n=n_{3D}\tanh(\mu B_{0}/E_{Fe,3D})$, $E_{Fe,3D}=(3\pi^{2}n)^{\frac{2}{3}}\hbar^{2}/(2m)$ is the Fermi energy, the Fermi velocity $v_{Fe,3D}=(3\pi^{2}n)^{\frac{1}{3}}\hbar/m$ will be applied over the paper. If the external magnetic field approach zero value $B_{0}\rightarrow 0$, thereat $\Delta n\rightarrow0$ and pressure of polarised electrons (\ref{ChCoulQHD eq State single Fl polarised 3D}) move into usual Fermi pressure of degenerate electrons (\ref{ChCoulQHD EqState un Pol}) $p_{pol,3D}\rightarrow p_{unpol}=p_{Fe,3D}$.

We keep in mind that $n_{\uparrow} \leq n_{\downarrow}$. Since preferable direction of magnetic moments in external magnetic field coincides with direction of the external magnetic field and electrons have negative charge therefore spins have opposite preferable direction. Consequently concentration of spin-down electrons is larger than the concentration of the spin-up electrons.

Let us consider limit of the small external magnetic field then $\Delta n\ll n_{(av)}$. In this limit we can make expansion of the equation of state of spin polarised electrons on small parameter $\Delta n/ n_{(av)}$ and obtain
\begin{equation}\label{ChCoulQHD eq State single Fl polarised 3D expansion} p_{pol,3D}=\frac{(3\pi^{2})^{\frac{2}{3}}}{5}\frac{\hbar^{2}}{m}\biggl[1+\frac{5}{9}\biggl(\frac{\Delta n}{n}\biggr)^{2}\biggr],\end{equation}
where the first term presents the Fermi pressure and the second term describes correction caused by the magnetic field.

Formula (\ref{ChCoulQHD eq State single Fl polarised 3D}) presents pressure as sun of two contributions: spin-up and spin-down electrons. Explicit consideration of separate evolution of spin-up and spin-down electrons in the external magnetic field employ each of terms in formula (\ref{ChCoulQHD eq State single Fl polarised 3D}) separately. Difference of these terms leads to existence on new longitudinal waves in plasmas \cite{Andreev spin-up and spin-down 1405}, \cite{Andreev spin-up and spin-down p2}.

\subsection{Equation of state in 2D degenerate electron gas}

Equation of state for degenerate 2D Fermi gas is
\begin{equation}\label{ChCoulQHD eq State 2D unPol} p_{2D,unpol}=\frac{\pi\hbar^{2}n_{2D}^{2}}{2m}.\end{equation}
This is a nonrelativistic equation of state in absence of external fields.

The gradient of pressure existing in the Euler equation appears as
\begin{equation}\label{ChCoulQHD eq State 2D unPol Grad} \nabla p_{2D,unpol}=\frac{\pi\hbar^{2}n_{2D}}{m}\nabla n_{2D}=\frac{1}{2}mv_{Fe,2D}^{2}\nabla n_{2D},\end{equation}
where $v_{Fe,2D}=\sqrt{2\pi n_{2D}}\hbar/m$ is non equilibrium Fermi velocity.

Considering the 2DEG in an external magnetic field we find modified equation of state
$$p_{pol,2D}=\frac{\pi\hbar^{2}}{m}\biggl[\biggl(n_{(av)}+\frac{\Delta n}{2}\biggr)^{2}$$
\begin{equation}\label{ChCoulQHD eq State single Fl polarised 2D} +\biggl(n_{(av)}-\frac{\Delta n}{2}\biggr)^{2}\biggr],\end{equation}
where $\Delta n=n_{2D}\tanh(\mu B_{0}/E_{Fe,2D})$, $E_{Fe,2D}=\pi n_{2D}\hbar^{2}/m$ is the Fermi energy of two dimensional ideal gas.

Opening brackets we can easily represent in a form explicitly showing contribution of the magnetic field
\begin{equation}\label{ChCoulQHD eq State single Fl polarised 2D expansion} p_{pol,2D}=\frac{\pi\hbar^{2}}{2m}\Biggl(1+\biggl(\frac{\Delta n}{n}\biggr)^{2}\Biggr),\end{equation}
Due to the of the particle concentration of spin-up and spin-down electrons in the equation of state (\ref{ChCoulQHD eq State single Fl polarised 2D}) we do not need to consider limit of small magnetic field as we have done for three dimensional case (\ref{ChCoulQHD eq State single Fl polarised 3D expansion}).

Application of equation (\ref{ChCoulQHD eq State single Fl polarised 2D expansion}) to the spectrum of collective excitations in magnetised 2D quantum plasmas is presented below in Section (\ref{sec:Apps}).

We would like to stress attention that we have considered a plane like 2DEG. Other example of 2DEG, 2DEGs on a sphere surface, will be considered in section (\ref{sec:QHD spheric}).

\subsection{Equation of state in 1D degenerate electron gas}

Equation of state for degenerate 1D Fermi gas is
\begin{equation}\label{ChCoulQHD eq State 1D unPol} p_{1D,unpol}=p_{Fe,1D}=\frac{\pi^{2}}{6}\frac{\hbar^{2}n_{1D}^{3}}{m}.\end{equation}

One dimensional Fermi velocity appears as
\begin{equation}\label{ChCoulQHD vel Fermi 1D} v_{Fe,1D}=\pi n_{1D}\hbar/(2m).\end{equation}

The one dimensional Fermi pressure can be rewritten in terms of the one dimensional Fermi velocity
\begin{equation}\label{ChCoulQHD eq State 1D unPol Other forms} p_{1D,unpol}=\frac{2}{3}mv_{Fe,1D}^{2}\cdot n_{1D},\end{equation}
where we have used non equilibrium one dimensional Fermi velocity.

In paper \cite{Haas PP 03} authors presented the equation of state of 1D degenerate electron gas in a different form. It is rather unusual that three dimensional particle concentration is applied for pressure of one dimensional plasmas.

In this section we have shown that if we do not explicitly consider spin dynamics in quantum plasmas, the equilibrium spin gives contribution in the plasma properties via equation of state.

Equation of state of the one dimensional in magnetic fields differs from the 1D Fermi pressure presented by formula (\ref{ChCoulQHD eq State 1D unPol}). Spin-up and spin-down electrons have different different occupation of the energy levels that reveals in modification of equation of state:
$$p_{pol,1D}=\frac{\pi^{2}}{6}\frac{4\hbar^{2}}{m}\biggl[\biggl(n_{(av)1D}+\frac{\Delta n_{1D}}{2}\biggr)^{3}$$
\begin{equation}\label{ChCoulQHD eq State single Fl polarised 1D} +\biggl(n_{(av)1D}-\frac{\Delta n_{1D}}{2}\biggr)^{3}\biggr].\end{equation}

We can open round brackets in formula (\ref{ChCoulQHD eq State single Fl polarised 1D}) and obtain another representation of the equation of state of 1D degenerate electrons in magnetic field
\begin{equation}\label{ChCoulQHD eq State single Fl polarised 1D expansion} p_{pol,1D}=\frac{\pi^{2}}{6}\frac{\hbar^{2}n_{1D}^{3}}{m}\biggl[1+3\biggl(\frac{\Delta n_{1D}}{n_{1D}}\biggr)^{2}\biggr].\end{equation}
Formula (\ref{ChCoulQHD eq State single Fl polarised 1D expansion}) explicitly shows shift of pressure caused by the magnetic field.

\subsection{\label{subsec:Modified eq St} Modified equations of state}

It is well-known that application of the Fermi pressure in three dimensional plasmas gives different contribution in compare with results of physical kinetics.
Thus, there is a necessity to modify Fermi pressure to have agreement with the kinetics.
However, to best of our knowledge there is no modification of equation of state giving good behavior in all regimes: equilibrium, linear perturbation and non-linear perturbations.

We can find a lot of application of the unmodified Fermi pressure (\ref{ChCoulQHD EqState un Pol}) (here we have some examples \cite{unmodified pressure beg}-\cite{unmodified pressure end}), or
equivalent representation $P=mv_{Fe,3D}^{2}n^{\frac{5}{3}}/(5n_{0}^{\frac{2}{3}})$ \cite{represented Fermi 3D beg}-\cite{represented Fermi 3D end}, where the Fermi velocity $v_{Fe,3D}$ is the function of the equilibrium concentration $n_{0}$. This representation a bit unusual, since it applies the the equilibrium concentration $n_{0}$, which does not exist in general equations we need to truncate. Application of the equilibrium concentration $n_{0}$ is also unnecessary in this case, since we can obtain the same result without it, applying Fermi pressure in the traditional form (\ref{ChCoulQHD EqState un Pol}).

At consideration of degenerate relativistic plasmas corresponding relativistic equation of state has been applied \cite{relativistic   pressure beg}-\cite{relativistic   pressure end}. Relativistic exchange interaction presented in the form of extra pressure can also be found in some of these papers (see for instance \cite{Rel exch 1}, \cite{Rel exch 2}, \cite{relativistic   pressure end}) In non-relativistic limit it gives the Fermi pressure (\ref{ChCoulQHD EqState un Pol}) (see for instance Ref. \cite{relativistic   pressure beg}).

Sometimes authors use mixed equations of state, when pressure is the sum of classic thermal pressure and the pressure of degenerate electron gas \cite{mixed eq state 1}, \cite{mixed eq state 2}, \cite{Sharma PP 14} .

There are different modifications of equation of state. A rather unusual modification has been applied in recent years \cite{mod over 3 beg}-\cite{mod over 3 end}, it has the following form
\begin{equation}\label{ChCoulQHD P m over 3} P_{m}=\frac{1}{3}\frac{mV_{F}^{2}}{n_{0}^{2}}n^{3}, \end{equation}
with $V_{F}$ is the Fermi velocity. Not all mentioned papers give explicit form of the Fermi velocity to make it clear, is it the 3D Fermi velocity? Judging by the explicit form of the Langmuir frequency, they consider 3D concentration, so we can conclude that all of them use 3D Fermi velocity.

We should mention that equation (\ref{ChCoulQHD P m over 3}) contains two different concentrations $n$ and $n_{0}$, where $n_{0}$ is the equilibrium concentration and $n$ is the full concentration. So, if we have a small amplitude perturbation, even non-linear, we can present full concentration as $n=n_{0}+\delta n$, where $\delta n$ is the perturbation. Let us note that this simple representation of concentration can be unapplicable in regime of strong nonlinearities.

If we consider equilibrium value of (\ref{ChCoulQHD P m over 3}) we find $P_{m}=mV_{F}^{2}n_{0}/3=\frac{5}{3}p_{Fe,3D}$. We can also consider linearised gradient of pressure (\ref{ChCoulQHD P m over 3}) $\nabla P_{m}=mV_{F}^{2} \nabla \delta n$. It is three times bigger than contribution of the Fermi pressure, and it is $5/3$ times bigger than similar terms given by kinetics.

This modification has been applied for dust \cite{mod over 3  For dust} and for relativistic plasmas \cite{mod over 3 in Rel plasmas}.

There is an another modification of equation of state for 3D plasmas having some advantages \cite{good modif beg}-\cite{good modif end}. Its explicit form is
\begin{equation}\label{ChCoulQHD P m over 5} \tilde{P}_{m}={\color{blue}\frac{1}{5}}\frac{mV_{F}^{2}}{n_{0}^{2}}n^{3}, \end{equation}
with $V_{F}=v_{Fe,3D}=(3\pi^{2}n_{3D})^{\frac{1}{3}}\hbar/m$ is the 3D Fermi velocity.

Let us consider equilibrium value of pressure (\ref{ChCoulQHD P m over 5}). Putting $n=n_{0}$ we obtain $\tilde{P}_{m}=mV_{F}^{2}n/5=p_{Fe,3D}$. 3D Fermi pressure is the most reasonable result for the equilibrium pressure. Next we consider linearised gradient of the pressure, since we have this quantity in the Euler equation. It appears as $\nabla\tilde{P}_{m}=\frac{3}{5}mV_{F}^{2}\nabla \delta n$ that corresponds to linear results of kinetics.

On the first look equations (\ref{ChCoulQHD P m over 3}) and (\ref{ChCoulQHD P m over 5}) are similar to 1D equation of state (\ref{ChCoulQHD eq State 1D unPol}), since it is proportional to the third degree of the particle concentration. Some time authors refer to equations (\ref{ChCoulQHD P m over 3}) and (\ref{ChCoulQHD P m over 5}) as 1D equations of state, but they apply 3D concentration and the Poisson equation, which is a signature of 3D plasmas.

We have described equilibrium and linear properties of equations (\ref{ChCoulQHD P m over 3}) and (\ref{ChCoulQHD P m over 5}). Let us describe their nonlinear properties. As we have mentioned equations (\ref{ChCoulQHD P m over 3}) and (\ref{ChCoulQHD P m over 5}) hardly able to be applied to one dimensional plasmas. They could be applied for 3D mediums. Let us repeat that equation (\ref{ChCoulQHD P m over 3}) fails to describe equilibrium and linear behavior, whereas equation (\ref{ChCoulQHD P m over 5}) describes these regimes even better than the Fermi pressure. The Fermi pressure gives nonlinearity of $5/3$ degree, which is an non-integer number, while equations (\ref{ChCoulQHD P m over 3}) and (\ref{ChCoulQHD P m over 5}) presents non-linearities of the third degree, which is rather different and integer. At consideration of problems with small non-linearities, different degree of nonlinearity reveals in different coefficients. However, it also reveals in different higher order of nonlinearity. A term proportional to $n^{3}$ can give nonlinearities up to the third order, than expansion of $(n_{0}+\delta n)^{5/3}$ gives a series of nonlinear terms with infinite number of terms. In case of strong nonlinearity $n^{3}$ and $n^{5/3}$ support different nonlinear structure, whereas we can expect only one of them exist in degenerate plasmas.

In reviews \cite{Shukla UFN 10} and \cite{Shukla RMP 11}, P. K. Shukla, B. Eliasson present a formula for pressure for different number of degrees of freedom $D$ in the system:
\begin{equation}\label{ChCoulQHD P m to D} \textrm{P}_{m}=\frac{mv_{Fe}^{2}n_{0}}{3}\biggl(\frac{n}{n_{0}}\biggr)^{\frac{D+2}{D}},\end{equation}
see \cite{Shukla UFN 10} (formula 49) and \cite{Shukla RMP 11} (formula 29). We should mention that judging by \cite{Shukla RMP 11} (formula 29), we see that formula (\ref{ChCoulQHD P m to D}) contains the 3D Fermi velocity $v_{Fe}=v_{Fe,3D}$.

Substituting $D=3$ and $n=n_{0}$ in formula (\ref{ChCoulQHD P m to D}) we find $\textrm{P}_{m}=\frac{5}{3}P_{Fe,3D}$.
Substituting $D=1$ and $n=n_{0}$ in formula (\ref{ChCoulQHD P m to D}) we obtain $\textrm{P}_{m}=P_{m}$, where $P_{m}$ is presented by formula (\ref{ChCoulQHD P m over 3}). Taking into account discussion presented above, we see that formula (\ref{ChCoulQHD P m to D}) is not suitable for 3D and 1D plasmas.

Formula was a key formula in Ref. \cite{Eq of state Strange diff D}.

In \cite{Shukla RMP 11} (formula 29) we can find reference on paper Manfredi and Haas \cite{Manfredi PRB 2001}. Manfredi and Haas \cite{Manfredi PRB 2001} have formula (\ref{ChCoulQHD P m over 3}), which corresponds to formula (\ref{ChCoulQHD P m to D}) at $D=1$, but in their paper $v_{Fe}=\pi\hbar n_{0}/(2m)=v_{Fe,1D}$ (see formula (\ref{ChCoulQHD vel Fermi 1D})). But, judging on other equations in Ref. \cite{Manfredi PRB 2001}, authors have deal with 3D particle concentration. Hence dimension of velocity is not $cm/s$, otherwise other equations (see for instance formulae (2), (28), (29) of Ref. \cite{Manfredi PRB 2001}) are incorrect. Thus we can conclude that analysis presented in Ref. \cite{Manfredi PRB 2001} is inconsistent and it is not suitable for 1D systems.

Paper \cite{Khan PP 08 DIAW} consider 3D electrodynamics, judging by equation (5) requiring 3D concentration, and 2D equation of state requiring 2D particle concentration. In Ref. \cite{Akbari Moghanjoughi PP 10 RaOC} a modification of formula (\ref{ChCoulQHD P m to D}) is presented $1/3\rightarrow1/(D+2)$. It improves coefficient in this formula, but it still contain some confusing information about the Fermi velocity containing in the formula (\ref{ChCoulQHD P m to D}). Whole structure and dependence on concentration in formula (3) of Ref. \cite{Akbari Moghanjoughi PP 10 RaOC} is uncorrect for low dimensional systems. The Poisson equation in set (1) should explicitly contain delta function, otherwise it is suitable for three dimensional mediums only. We find rather clear situation with equation of state in Ref. \cite{Akbari Moghanjoughi PRAMANA 11}, but equation of state looks to be three dimensional, as in previous paper.
Authors of Ref. \cite{Fathalian PP 10} consider plasmas on a cylinder, a specific example of two dimensional systems, but they apply 3D equation of state. They also do not include quantum part of the inertia force existing in the hydrodynamic equations in curvilinear coordinates. The quantum inertia forces reveals as an addition to the quantum Bohm potential (see sections \ref{sec:QHD cylindr} and \ref{sec:QHD spheric}).
Judging by equation (1) of Ref. \cite{Wang PLA 08 SaCS} and notations to it we conclude that a two dimensional system is under consideration. They apply equation of state for "2D Fermi plasma follow the pressure law" $p\sim n^{2}$, but they have 3D Langmuir frequency and 3D Poisson equation as a part of set (1).
General structure of equation of state in Ref. \cite{Ali NJP 08 PSoNE}, \cite{Sahu ASS 13 KPsiqp},  authors consider 2D system of particles, whereas they apply the 3D Poisson equation.

\section{\label{sec:Representation} Representation of hydrodynamic variables in terms of occupation numbers}

Being described by the many-particle wave function N indistinguishable quantum particles occupy N different quantum states. Evolution of the wave function is related to migration of particles between different states. This migration leads to change of the occupation numbers. If we want to follow this picture we can get the wave function in representation of occupation  numbers instead of the coordinate representation, which we have used through this paper. At transition to language of the occupation numbers, which will be very useful for calculation of two-particle concentration, the explicit form of definitions of QHD variables changes as well \cite{MaksimovTMP 1999}, \cite{Andreev PRA08}:
\begin{equation}\label{ChCoulQHD}n(\textbf{r})=\sum_f n_f \varphi_f^*(\textbf{r}) \varphi_f(\textbf{r}),\end{equation}
\begin{equation}\label{ChCoulQHD}mn(\textbf{r})\textbf{v}(\textbf{r})={1\over 2 }\sum_f n_f
\biggl(\varphi_f^*(\textbf{r})\textbf{D}\varphi_f(\textbf{r})+
(\textbf{D}\varphi_f)^*(\textbf{r}) \varphi_f(\textbf{r})\biggr),\end{equation}
$$\Pi_{\alpha\beta}(\textbf{r})={1\over 4m}\sum_f n_f
\biggl(\varphi_f^*(\textbf{r}) D_{\beta} D_{\alpha} \varphi_f(\textbf{r})+
(D_{\beta}\varphi_f)^*(\textbf{r}) D_{\alpha}\varphi_f(\textbf{r})$$
\begin{equation}\label{ChCoulQHD} +(D_{\alpha}\varphi_f)^*(\textbf{r}) D_{\beta}\varphi_f(\textbf{r})+
(D_{\beta} D_{\alpha} \varphi_f)^*(\textbf{r}) \varphi_f(\textbf{r})\biggr),\end{equation}
\begin{equation}\label{ChCoulQHD} {\cal E}_{\rm kin}(\textbf{r})={1\over 4m}\sum_f n_f
\biggl(\varphi_f^*(\textbf{r}) \textbf{D}^2\varphi_f(\textbf{r})
+(\textbf{D}^2 \varphi_f)^*(\textbf{r}) \varphi_f(\textbf{r})
\biggr),\end{equation}
and
$$\textbf{Q}_{\rm kin}(\textbf{r})={1\over 8m^2}\sum_f n_f
\biggl( \varphi^*(\textbf{r})\textbf{D}
\textbf{D}^2\varphi_f(\textbf{r})+(\textbf{D}\varphi)^*(\textbf{r})
\textbf{D}^2\varphi_f(\textbf{r})$$
\begin{equation}\label{ChCoulQHD}+(\textbf{D}^2\varphi)^*(\textbf{r})
\textbf{D} \varphi_f(\textbf{r})+
(\textbf{D}\textbf{D}^2
\varphi)^*(\textbf{r}) \varphi_f(\textbf{r})\biggr),\end{equation}
where $\varphi_f(\textbf{r})$ are wave functions of occupied and unoccupied states, $n_f$ are the occupation numbers, for fermions $n_f=0,1$, and $f$ is the set of "quantum numbers" describing each state. For instance, if we consider a system of noninteracting quantum particles, then each state can be described by the momentum $\textbf{p}$ and the spin projection $\sigma$ we have $f=\{\textbf{p},\sigma\}$.

\section{\label{sec:Quantum correlations}  Quantum correlations and exchange interaction}

The Coulomb exchange interaction in electron gas has been studied with 50-ies of XX century \cite{Nozieres PR 58}, \cite{Kanazawa PTP 60}. After derivation of an explicit form of the exchange force field we will compare it with the earlier results. We consider now the method of derivation of the exchange force field in terms of the MPQHD developed in 1999 (see ref. \cite{MaksimovTMP 1999}). In other words, knowing that it is a potential force field, we can reformulate that we present the method of derivation potential.

We have also used notion of the two-particle concentration, which general definition is
\begin{equation}\label{ChCoulQHD two-part conc def}n_{2}(\textbf{r},\textbf{r}',t)=\int dR\sum_{i,j\neq i}\delta(\textbf{r}-\textbf{r}_{i})\delta(\textbf{r}'-\textbf{r}_{j})\psi^{*}(R,t)\psi(R,t). \end{equation}

Function (\ref{ChCoulQHD two-part conc def}) has appeared at derivation of the Euler equation (\ref{ChCoulQHD bal imp eq with N2}), in the term describing the interparticle Coulomb interaction. In previous sections we have considered the self-consistent field approximation, which is suitable for the long-range interactions. In those cases we have presented the two-particle concentration $n_{2}(\textbf{r},\textbf{r}',t)$ as a product of the one-particle concentrations $n_{2}(\textbf{r},\textbf{r}',t)=n(\textbf{r},t)n(\textbf{r}',t)$. In this section we are going beyond of the self-consistent field approximation.

General form of the two-particle concentration of fermions containing the exchange correlation was derived in Ref. \cite{MaksimovTMP 1999} in 1999. Here we present details of this derivation. We also present further analysis of the general formula mentioned above. We calculate it for weakly interacting 1D, 2D, and 3D electron gases. We obtain the exchange potential via the particle concentration and the fundamental physical constants. Thus we find closed sets of the MPQHD equations containing contribution of the exchange interaction.

Definition (\ref{ChCoulQHD two-part conc def}) can be rewritten in more useful form
\begin{equation}\label{ChCoulQHD two-part occup number}n_2(\textbf{r},\textbf{r}',t)=N(N-1)
\int dR_{N-2}\langle n_1,n_2,\ldots |\textbf{r},\textbf{r}',R_{N-2},t\rangle \langle\textbf{r},\textbf{r}',R_{N-2},t |n_1,n_2,\ldots\rangle ,\end{equation}
where $\langle n_1,n_2,\ldots |\textbf{r},\textbf{r}',R_{N-2},t\rangle$ is the N-particle wave function in representation of the occupation numbers, and $dR_{N-2}=\displaystyle\prod\limits_{k=3}^{N}d\textbf{r}_k$.

For further transformation we need to extract evolution of particles related to arguments $\textbf{r}$ and $\textbf{r}'$. For this purpose, we consider expansion of the wave function
$\langle\textbf{r},\textbf{r}',R_{N-2},t |n_1,n_2\ldots\rangle$
\cite{Schweber}. In the case of fermions, making expansion on one of arguments, we find
$$\langle \textbf{r}, \textbf{r}', R_{N-2},t |n_1, n_2 \ldots
\rangle=\sum_f \sqrt{\frac{n_f}{N}} \: \langle\textbf{r},t | f\rangle \:
\langle \textbf{r}', R_{N-2},t |n_1, \ldots (n_f-1),\ldots \rangle,$$
where we have that particle in  an
arbitrary quantum state $f$ gives dependence on the coordinate $\textbf{r}$, and all particles
alternately make contribution in $\langle\textbf{r},\textbf{r}',R_{N-2},t |n_1,n_2\ldots\rangle$ via $\: \langle\textbf{r},t | f\rangle \:$ due to summation on all states.

Making the second expansion of the wave function, including symmetry of the wave function due to permutation of arguments, we obtain different formulas for the fermions and bosons. For the Fermi particles we have
$$\langle \textbf{r}, \textbf{r}', R_{N-2},t |n_1, n_2 \ldots \rangle=$$
$$=\sum_f \sum_{f'<f} \sqrt{\frac{n_f}{N}} \sqrt{\frac{n_{f'}}{N-1}}
(-1)^{\sum\limits_{f'\le g<f} n_g }
\: \left(\: \langle \textbf{r},t | f\rangle \: \langle\textbf{r}',t | f'\rangle-
\langle\textbf{r}',t | f\rangle \: \langle\textbf{r},t | f'\rangle\:\right)\times $$
\begin{equation}\label{ChCoulQHD}\times
\langle R_{N-2},t |n_1, \ldots (n_{f'}-1),\ldots (n_f-1), \ldots \rangle.
\end{equation}
where $\langle \textbf{r},t|f\rangle =\varphi_f(\textbf{r},t)$ are the single-particle wave functions.

Calculation of the two-particle concentration requires integration of the product of two wave functions in formula (\ref{ChCoulQHD two-part occup number})
$$\langle n_1, \ldots (n_{f'}-1), \ldots (n_f-1), \ldots |
n_1, \ldots (n_{q'}-1),\ldots (n_q-1),\ldots \rangle$$
\begin{equation}\label{ChCoulQHD product of wf}=\delta (f-q) \delta (f'-q')-\delta
(f-q')\delta (f'-q).\end{equation}
Formula (\ref{ChCoulQHD product of wf}) explicitly reveals symmetry of fermion wave function. The second term gives exchange term, and, consequently, contribution of the exchange interaction. Using these formulas, after some calculations, we find following result for the two particle concentration
\begin{equation}\label{ChCoulQHD n2 long r}
n_2(\textbf{r},\textbf{r}',t)=n(\textbf{r},t)n(\textbf{r}',t)-|\rho(\textbf{r},\textbf{r}',t)|^{2},\end{equation}
where $n_{g}$ is a number of particles in the quantum state $\varphi_{g}$, with a set of quantum numbers $g$,
\begin{equation}\label{ChCoulQHD nvarphi}
n(\textbf{r},t)=\sum_{g}n_{g}\varphi_{g}^{*}(\textbf{r},t)\varphi_{g}(\textbf{r},t)\end{equation}
is the particle concentration in terms of the arbitrary single-particle wave functions $\varphi_{g}(\textbf{r},t)$,
\begin{equation}\label{ChCoulQHD rhovarphi}\rho(\textbf{r},\textbf{r}',t)=\sum_{g}n_{g}\varphi_{g}^{*}(\textbf{r},t)\varphi_{g}(\textbf{r}',t)\end{equation}
is the macroscopic density matrix.

Formula (\ref{ChCoulQHD n2 long r}) is obtained for symmetric spin part of wave function and anti-symmetric coordinate part of the wave function. That corresponds to parallel spin orientation.

If reader are interested in calculation of $n_2(\textbf{r},\textbf{r}',t)$ for Bose particles, they can find it in Refs.
\cite{MaksimovTMP 1999}, \cite{Andreev PRA08}. Paper \cite{Andreev PRA08} is dedicated to systems of neutral particles with the short-range interaction, but analog of the general formula (\ref{ChCoulQHD n2 long r}) is presented there by formula (25).

\section{\label{sec:Exchange explicit}  Explicit form of the force field for the Coulomb exchange interaction}

A recent review of results for the Coulomb exchange interaction obtained earlier in terms of other methods is presented in Ref. \cite{Andreev 1403 exchange}.

In the previous section we have obtained general representation of the two-particle concentration as sum of two parts: the first part is the product of one particle concentrations and and the second one is the correlation function (\ref{ChCoulQHD n2 long r expansion}). The first term of this sum leads to the self-consistent field approximation in hydrodynamic equations. Generalization of the self-consistent field approximation may be reached by account of the exchange interaction hidden in the correlation function $g_{2}(\textbf{r},\textbf{r}',t)\equiv-\rho(\textbf{r},\textbf{r}',t)$.

Below we derive contribution of the exchange interaction in correlation function $g_{2}$ for weakly interaction charged particles. Hence we choose plane de Broglie waves for one-particle wave functions of each particle:
\begin{equation}\label{ChCoulQHD} \varphi=\frac{1}{\sqrt{V}}\exp\biggl(\frac{\imath}{\hbar}(\textbf{k}\textbf{r}-\varepsilon t)\biggr),\end{equation}
with $\textbf{p}=\hbar\textbf{k}$ and $\varepsilon=\hbar\omega$.

Writing correlation function in terms of plane waves we find the following formula
\begin{equation}\label{ChCoulQHD} g_{2}=-\frac{1}{V^{2}}\sum_{\textbf{p},\textbf{p}'}\sum_{\sigma,\sigma'}n_{p,\sigma}n_{p',\sigma'} \exp\biggl(\frac{\imath}{\hbar}(\textbf{p}-\textbf{p}')(\textbf{r}-\textbf{r}')\biggr).\end{equation}

Momentum $\textbf{p}$ and spin projection $\sigma$ are good quantum numbers for weakly interacting particles, thus we have chosen $g=\{\textbf{p},\sigma\}$.

\subsection{\label{sec:level1} Coulomb exchange interaction in three dimensional medium}

In this paper we do not consider spin evolution. However we should describe distribution of spin directions to calculate correlations. In this section we assume that all particles have parallel spin being at zero temperature occupying all states below Fermi level.

As a consequence of all particle being in the same spin state $\sigma'=\sigma$ we have one term of sum on the spin only. So the correlation function to be
\begin{equation}\label{ChCoulQHD} g_{2}=-\frac{1}{V^{2}}\sum_{\textbf{p},\textbf{p}'}n_{p,\sigma}n_{p',\sigma}\exp\biggl(\frac{\imath}{\hbar}(\textbf{p}-\textbf{p}')(\textbf{r}-\textbf{r}')\biggr)\end{equation}
Let us to calculate energy density for the exchange interaction as an intermediate step of getting of corresponding force field
$$\varepsilon_{C,3D}=q^{2}\int G(\textbf{r},\textbf{r}')g_{2}(\textbf{r},\textbf{r}',t) d\textbf{r}'$$
\begin{equation}\label{ChCoulQHD}=-\frac{q^{2}}{(2\pi\hbar)^{6}}\int d\textbf{p}d\textbf{p}'n_{\textbf{p}}n_{\textbf{p}'}\int d\textbf{r}'\frac{1}{\mid \textbf{r}-\textbf{r}'\mid}\exp\biggl(\frac{\imath}{\hbar}(\textbf{r}'-\textbf{r})(\textbf{p}'-\textbf{p})\biggr),\end{equation}
where $d\textbf{p}=dp_{x}dp_{y}dp_{z}$ and $d\textbf{r}=dx dy dz$.

In the following we use $n_{\textbf{p}}=n_{\textbf{p}'}=1$ and perform the Fourier transformation of the Coulomb potential
$$\varepsilon_{C,3D}=-\frac{q^{2}}{(2\pi\hbar)^{6}}\int d\textbf{p}d\textbf{p}'G_{3D}(\textbf{p}-\textbf{p}'),$$
where $G_{3D}(\textbf{p}-\textbf{p}')=4\pi\hbar^{2}/\mid \textbf{p}-\textbf{p}'\mid^{2}$ is the three dimensional Fourier image of the Green function of the Coulomb interaction. After some calculation we get explicit form of $G(\textbf{p}-\textbf{p}')$, so we have
$$\varepsilon_{C}=-\frac{4\pi q^{2}\hbar^{2}}{(2\pi\hbar)^{6}}\int d\textbf{p}d\textbf{p}'\frac{1}{\mid \textbf{p}-\textbf{p}'\mid^{2}},$$
where both integrals are taken over inner part of the Fermi sphere, which has radius $p_{F}$.

Further straightforward calculations give
\begin{equation}\label{ChCoulQHD} \varepsilon_{C,3D}=-\frac{4\pi q^{2}}{(2\pi\hbar)^{4}}\tilde{p}_{F,3D}^{4}.\end{equation}

Using explicit form of the Fermi momentum $\tilde{p}_{F,3D}=\sqrt[3]{3\pi^{2}n_{3D}}\hbar$ we find final form of the energy density of exchange interaction
\begin{equation}\label{ChCoulQHD varepsilon C 3D explicit} \varepsilon_{C,3D}=-\frac{3q^{2}}{4\pi}\sqrt[3]{3\pi^{2}}n_{3D}^{4/3}.\end{equation}
In this subsection we consider three dimensional plasmas, hence the particle concentration is the number of particles in cubic centimeter $[n]=cm^{-3}$. We put tilde above $p$ for the Fermi momentum to distinguish it from the Fermi pressure.

For partially polarised fermions we can obtain generalisation of formula (\ref{ChCoulQHD varepsilon C 3D explicit}) in terms of particle number density of spin-up and spin-down electrons
\begin{equation}\label{ChCoulQHD varepsilon C 3D explicit up and down} \varepsilon_{C,3D}=-\frac{3q^{2}}{4\pi}\sqrt[3]{3\pi^{2}}(n_{\downarrow,3D}^{4/3}-n_{\uparrow,3D}^{4/3}). \end{equation}
Different form of the exchange correlation energy and corresponding force field for partially polarised fermions can be found in Ref. \cite{Andreev 1403 exchange}.

Result (\ref{ChCoulQHD varepsilon C 3D explicit}) allows us to find the force field of the Coulomb exchange interaction
\begin{equation}\label{ChCoulQHD F C exchange 3D} \textbf{F}_{C,3D}=-\nabla\varepsilon_{C,3D}=q^{2}\sqrt[3]{\frac{3}{\pi}}\sqrt[3]{n_{3D}}\nabla n_{3D}. \end{equation}

Formula (\ref{ChCoulQHD F C exchange 3D}) is derived for fully polarised system of spin-1/2 charged particles. Let us note that equation of state for fully polarized spin 1/2 particles differs from the Fermi pressure obtained for unpolarised fermions. The equation of state for fully polarized spin 1/2 particles arises as $p_{3D}=\sqrt[3]{2}p_{Fe,3D}=(6\pi^{2})^{\frac{2}{3}}\hbar^{2}n_{3D}^{\frac{5}{3}}/(5m)$. General formulae for the force field of the Coulomb exchange interaction and equation of state appearing for partially polarised spin-1/2 fermions are presented and discussed in Ref. \cite{Andreev 1403 exchange}. Here we have briefly shown a method of derivation of the exchange quantum correlations for fully polarised fermions.

Formula (\ref{ChCoulQHD F C exchange 3D}) shows an attractive force between electrons.

Different technics were applied for study of the exchange interaction in the 3D electron gas: the Green function method \cite{Kanazawa PTP 60}, \cite{Nozieres PR 58}, \cite{DuBois AnnP 59}, two-particle density matrix and quantum kinetics \cite{Burt PR 62}, \cite{Roos PR 61}, \cite{Ter Haar RPP 61}, density functional theory \cite{Hedin JP C 71}, \cite{Brey PRB 90}.

Dependence on spin polarisation was not discussed by those authors. Comparison of their and our results shows that their results correspond to fully polarised electrons. Here, following paper \cite{Andreev 1403 exchange}, we present dependence of the exchange Coulomb interaction on soin polarization of electrons. Similar consideration can be performed by methods applied in Refs. \cite{Kanazawa PTP 60}, \cite{Nozieres PR 58}, \cite{DuBois AnnP 59}, \cite{Burt PR 62}, \cite{Roos PR 61}, \cite{Ter Haar RPP 61}, \cite{Hedin JP C 71}.

Exchange interaction in 2DEG was considered in 1983 in Ref.
\cite{Datta JAP 83}. We will present QHD description of the Coulomb exchange interaction in the next subsection following Ref. \cite{Andreev 1403 exchange}, where we will also compare our results with result of Ref. \cite{Datta JAP 83}.

NLSE with the Coulomb exchange interaction
$$\imath\hbar\partial_{t}\Phi(\textbf{r},t)=\Biggl(-\frac{\hbar^{2}\nabla^{2}}{2m}+\vartheta_{3D}\frac{(3\pi^{2})^{2/3}\hbar^{2}n_{e}^{2/3}}{2m_{e}}$$
\begin{equation}\label{ChCoulQHD NLSE int 3D with exchange} -3\sqrt[3]{\frac{3}{\pi}}\zeta_{3D}q^{2}n^{1/3}+e^{2}\int
d\textbf{r}'G(\textbf{r},\textbf{r}')\biggl[n(\textbf{r}',t)-n_{0i}\biggr]\Biggr)\Phi(\textbf{r},t),
\end{equation}
where coefficients $\vartheta_{3D}$ and $\zeta_{3D}$ describe rate of spin polarisation of three dimensional electron gas, where
\begin{equation}\label{ChCoulQHD} \zeta_{3D}=(1+\eta)^{4/3}-(1-\eta)^{4/3},\end{equation}
and
\begin{equation}\label{ChCoulQHD} \vartheta_{3D}=\frac{1}{2}[(1+\eta)^{5/3}+(1-\eta)^{5/3}],\end{equation}
where $\eta=\Delta n/n_{0}$, with $\Delta n=n_{\downarrow}-n_{\uparrow}$, $n=n_{\downarrow}+n_{\uparrow}$, with $n_{\uparrow}$ and $n_{\downarrow}$ are concentrations of spin-up and spin-down fermions.

Results of Ref. \cite{Andreev 1403 exchange} described in this subsection were applied to quantum magnetized plasmas in Ref. \cite{Trukhanova 1405 exchange}. Other approaches have been developed as well (see for instance \cite{Zamanian PRE 13}).

\subsection{\label{sec:level1} Coulomb exchange interaction in two dimensional medium}

In this subsection we present main steps of the force field derivation for the Coulomb exchange interaction in two dimensional electron gas located in a plane. As in the previous subsection we focus our attention on fully polarised electron gas. This example reveals constitutive picture of the exchange interaction. The exchange interaction has maximal strength in case of fully polarised fermions. We will see that in 2D plasmas, as in 3D plasmas, the Coulomb exchange interaction appears as an extra attractive force field in the Euler equation. By the way we notice that equation of state for fully polarised spin-1/2 fermions has the following form $p_{2D}=2p_{Fe(2D)}=\pi\hbar^{2}n^{2}_{2D}/m$.

Here $\textbf{r}={x,y}$. In contrast with previous subsection, in this subsection we consider two dimensional plasmas, hence the particle concentration is the number of particles in quadratic centimeter $[n]=cm^{-2}$. In other words it is number of particles on unit of surface area.

Energy of 2DEG corresponding to the Coulomb exchange correlations has the following form
$$\varepsilon_{C,2D}=q^{2}\int G(\textbf{r},\textbf{r}')g_{2}(\textbf{r},\textbf{r}',t) d\textbf{r}'$$
\begin{equation}\label{ChCoulQHD} =-\frac{q^{2}}{(2\pi\hbar)^{4}}\int d\textbf{p}d\textbf{p}'n_{\textbf{p}}n_{\textbf{p}'}\int d\textbf{r}'\frac{1}{\mid \textbf{r}-\textbf{r}'\mid}\exp\biggl(\frac{\imath}{\hbar}(\textbf{r}'-\textbf{r})(\textbf{p}'-\textbf{p})\biggr),\end{equation}
with $n_{\textbf{p}}=n_{\textbf{p}'}=1$ is the occupation numbers of fully polarised spin-1/2 fermions, $d\textbf{p}=dp_{x}dp_{y}$ is the element of volume in 2D momentum space.

As the first step we evaluate the integral on $d\textbf{r}'$. So we perform the two dimensional Fourier transformation of the Coulomb potential and find the Fourier image of the Green function of the Coulomb interaction $G_{2D}(\mid \textbf{p}-\textbf{p}'\mid)=2\pi\hbar/\mid \textbf{p}-\textbf{p}'\mid$.

The explicit form of the exchange correlation energy in term of two dimensional concentration of particles (density of particle number) is
\begin{equation}\label{ChCoulQHD} \varepsilon_{C,2D}=-\frac{8q^{2} \textrm{arsh}1}{\pi^{3}\hbar^{3}}\tilde{p}_{F,2D}^{3},\end{equation}
with the two dimensional Fermi momentum $\tilde{p}_{F,2D}=\sqrt{2\pi n}\hbar$.

After straightforward calculations we obtain \cite{Andreev 1403 exchange}
\begin{equation}\label{ChCoulQHD} \varepsilon_{C,2D}=-2^{3/2} \sqrt{2\pi}\frac{16 \textrm{arsh}1}{\pi^{2}}q_{a}^{2} n_{2D}^{\frac{3}{2}}.\end{equation}

\begin{equation}\label{ChCoulQHD F C exchange 2D} \textbf{F}_{C,2D}=-\nabla\varepsilon_{C}=2^{3/2} \sqrt{2\pi}\frac{24 \textrm{arsh}1}{\pi^{2}}q_{a}^{2}\sqrt{n_{2D}}\nabla n_{2D}, \end{equation}
where we have $24 \textrm{arsh}1=21.153$.

Assuming that the static dielectric constant of surrounding medium equals to one we can write down potential energy of exchange interaction obtained in Ref. \cite{Datta JAP 83} (see formula (15) and below)
$U_{ex}=-\frac{q^{2}}{2k_{Fe}}n$ with $k_{Fe}=\sqrt{2\pi n}$.

From formulas (\ref{ChCoulQHD F C exchange 3D}) and (\ref{ChCoulQHD F C exchange 2D}) we see that in linear approximation the force fields of Coulomb exchange interaction in three- and two-dimensional quantum plasmas are proportional to perturbations of the particle concentration. Hence they give shift of the thermal pressure or Fermi pressure for degenerate electrons. $\textbf{F}_{C}$ presented by formulas (\ref{ChCoulQHD F C exchange 3D}), for 3D plasmas, and (\ref{ChCoulQHD F C exchange 2D}), for 2D plasmas, give additional attraction decreasing contribution of the pressure.

Recent application of the 2D exchange potential in the formed obtained in Ref. \cite{Datta JAP 83} can be found in Ref. \cite{Akbari PP 14}.

NLSE for 2D quantum electron gas with the Coulomb exchange interaction
$$\imath\hbar\partial_{t}\Phi_{2D}(\textbf{r},t)=\Biggl(-\frac{\hbar^{2}\nabla^{2}}{2m}+\vartheta_{2D}\frac{\pi\hbar^{2}}{m_{e}}n$$
\begin{equation}\label{ChCoulQHD NLSE int 2D with exchange} -\frac{2\beta \sqrt{2\pi}}{\pi^{2}}\zeta_{2D}q_{e}^{2}\sqrt{n} +e^{2}\int
d\textbf{r}'G(\textbf{r},\textbf{r}')\biggl[n(\textbf{r}',t)-n_{0i}\biggr]\Biggr)\Phi_{2D}(\textbf{r},t),
\end{equation}
where $\beta\equiv 24 \textrm{arsh}1=21.153$ and coefficients $\vartheta_{2D}$ and $\zeta_{2D}$ describe rate of spin polarisation of 2DEG. These coefficients have the following form
\begin{equation}\label{ChCoulQHD} \zeta_{2D}=(1+\eta)^{3/2}-(1-\eta)^{3/2},\end{equation}
and
\begin{equation}\label{ChCoulQHD} \vartheta_{2D}=1+\eta^{2},\end{equation}
where $\eta=\Delta n/n_{0}$, with $\Delta n=n_{\downarrow}-n_{\uparrow}$, $n=n_{\downarrow}+n_{\uparrow}$, with $n_{\uparrow}$ and $n_{\downarrow}$ are concentrations of spin-up and spin-down fermions.

\section{\label{sec:Apps} Applications}

As an application of the quantum hydrodynamics we consider dispersion of the Langmuir waves in the quantum plasmas. We described one-, two-, and three dimensional quantum plasmas.

\subsection{\label{sec:level1} Three dimensional quantum plasma waves}

Let us consider small amplitude high frequency collective excitations having form of the plane wave in three dimensional quantum plasmas.
\begin{equation}\label{ChCoulQHD} -\imath\omega\delta n+\imath n_{0} \textbf{k}\delta \textbf{v}=0, \end{equation}
\begin{equation}\label{ChCoulQHD} -\imath\omega mn_{0}\delta \textbf{v}+\imath \textbf{k} mU^{2}\delta n+\frac{\imath\hbar^{2}k^{2}\textbf{k}}{4m}\delta n=en_{0}\delta \textbf{E}, \end{equation}
and
\begin{equation}\label{ChCoulQHD Puasson lin} \imath \textbf{k} \textbf{E}=4\pi e\delta n,\end{equation}
where $U^{2}=\gamma T_{0}/m$ for classic electron gas and $U^{2}=v_{Fe,3D}^{2}/3$ for quantum degenerate electron gas.

In the last equation we included that electrons and ions keep together, but we consider motion of electrons only.
Consequently, in the right-hand side of equation (\ref{ChCoulQHD Puasson lin}) we have for the charge density $\rho=\sum_{i}e_{i}n_{i}=e_{e}n_{e}+e_{i}n_{i}=e_{e}n_{0e}+e_{e}\delta n_{e}+e_{i}n_{0i}=e_{e}\delta n_{e}$. We included that equilibrium concentrations of electrons and ions equal to each other, $e_{e}$, $e_{i}$ are charges of electrons and ions having different sign.
Considering small perturbations of the equilibrium state described by $n_{0}$, and $\textbf{v}_{0}=0$. We have an equal concentrations of electrons and ions, since we assume that plasma is quasi neutral. Assuming that perturbations are monochromatic
\begin{equation}\label{ChCoulQHD} \delta n=N_{A} e^{-\imath\omega t+\imath \textbf{k} \textbf{r}},\end{equation}
\begin{equation}\label{ChCoulQHD} \delta \textbf{v}=\textbf{V}_{A} e^{-\imath\omega t+\imath \textbf{k} \textbf{r}},\end{equation}
and
\begin{equation}\label{ChCoulQHD} \delta \textbf{E}=\textbf{E}_{A} e^{-\imath\omega t+\imath \textbf{k} \textbf{r}},\end{equation}
we get a set of linear algebraic equations relatively to $N_{A}$ and $V_{A}$. Condition of existence of nonzero solutions for amplitudes of perturbations gives us a dispersion equation. As the result of calculations we find the following dispersion dependence
\begin{equation}\label{ChCoulQHD spectrum Langm 3D} \omega^{2}=\omega_{Le,3D}^2+U^{2}k^2+
\frac{\hbar^2k^4}{4m^2}, \end{equation}
with the three dimensional Langmuir frequency
\begin{equation}\label{ChCoulQHD Langmuir freq 3D}\omega_{Le,3D}^{2}=\frac{4\pi e^2n_0}{m}.\end{equation}
The spectrum (\ref{ChCoulQHD spectrum Langm 3D}) is obtained from equations (\ref{ChCoulQHD cont eq inset}) and (\ref{ChCoulQHD bal imp eq short}).

\subsection{\label{sec:level1} Spectrums of 3D plasmas with the Coulomb exchange interaction}

If 3D electron gas is fully polarised, then its spectrum differs from unpolarised one. First of all contribution of the exchange interaction appears in the spectrum. The second change is due to modification of equation of state, which significantly depends on occupation of quantum states by electrons. In our case it multiplies by $\sqrt[3]{4}$. Consequently we find
\begin{equation}\label{ChCoulQHD spectrum Langm 3D} \omega^{2}=\omega_{Le,3D}^2+\biggl(\frac{\sqrt[3]{4}}{3}v_{Fe,3D}^{2}-\sqrt[3]{\frac{3}{4\pi}}\frac{1}{\pi\sqrt[3]{n_{0}^{2}}}\omega_{Le,3D}^2\biggr)k^2+
\frac{\hbar^2k^4}{4m^2}. \end{equation}

We have included the exchange interaction existing in the fully polarised system of particles, corresponding change of the pressure contribution is included as well. However we have not considered any external magnetic field, which could cause different occupation of spin-up and spin-down states. We have assumed that nature of polarisation is related to the equilibrium Coulomb exchange interaction between atom bound electrons. This interaction is the mechanism of spin polarization of ferromagnetic and other ferrite materials. The equilibrium Coulomb exchange interaction can lead to partial spin polarization. General formula including partial polarization of electrons has been obtained in Ref. \cite{Andreev 1403 exchange}.

\subsection{\label{sec:level1} Two dimensional quantum plasma waves with the Coulomb exchange interaction}

The second example of the MPQHD application is the dispersion of waves in quantum 2DEG. This example reveals an important feature of the MPQHD that this method can be used for low dimensional physical systems. Considering a low dimensional system we can not use the Maxwell equation, since it contains the Dirac delta function $n_{3D}=n_{2D}\delta(z)$, where $n_{3D}$ is the three dimensional particle concentration measured in cm$^{-3}$, $n_{2D}$ is the two dimensional particle concentration measured in cm$^{-2}$, and $\delta(z)$ shows that 2DEG is located in the XoY plane. Nevertheless we can easily use integral form of the QHD equations, which we had before introducing of the electric field caused by the charges. An analog of the Fourier transformed Poisson equation for 2DEG has been used \cite{Krasheninnikov JETP 80}. However it has meaning in the linear approximation. If we want to study nonlinear properties of 2DEG we need to use integral QHD equations or corresponding kinetic equations \cite{Andreev kinetics 12}, \cite{Andreev kinetics 13}.

Let us ones again explicitly present set of the QHD equations suitable for consideration of low dimensional systems of electrons and ions. This set consists of
the continuity equation for electrons
\begin{equation}\label{ChCoulQHD continuity equation el 2 for 2D}\partial_{t}n_{e}+\nabla (n_{e}\textbf{v}_{e})=0,
\end{equation}
the Euler equation (momentum balance equation) for electrons in the integral form
$$m_{e}n_{e}(\partial_{t}+\textbf{v}_{e}\nabla)\textbf{v}_{e}+\nabla p_{e}$$
$$=q_{e}n_{e}(\textbf{r},t)\Biggl(-q_{e}\nabla\int
d\textbf{r}'G(\textbf{r},\textbf{r}')n_{e}(\textbf{r}',t)$$
\begin{equation}\label{ChCoulQHD Euler el 2 int for 2D} -q_{i}\nabla\int
d\textbf{r}'G(\textbf{r},\textbf{r}')n_{i}(\textbf{r}',t)\Biggr),
\end{equation}
the continuity equation for ions
\begin{equation}\label{ChCoulQHD continuity equation ion 2 for 2D}\partial_{t}n_{i}+\nabla (n_{i}\textbf{v}_{i})=0,
\end{equation}
and the Euler equation for ions in the integral form
$$m_{i}n_{i}(\partial_{t}+\textbf{v}_{i}\nabla)\textbf{v}_{i}+\nabla p_{i}$$
$$=q_{i}n_{i}(\textbf{r},t)\Biggl(-q_{e}\nabla\int
d\textbf{r}'G(\textbf{r},\textbf{r}')n_{e}(\textbf{r}',t)$$
\begin{equation}\label{ChCoulQHD Euler ion 2 int for 2D} -q_{i}\nabla\int
d\textbf{r}'G(\textbf{r},\textbf{r}')n_{i}(\textbf{r}',t)\Biggr),
\end{equation}
where
\begin{equation}\label{ChCoulQHD Colomb Green func in sect 2D} G(\textbf{r},\textbf{r}')=\frac{1}{\mid\textbf{r}-\textbf{r}'\mid}
\end{equation}
is the Green function for Coulomb interaction. Integrals in equations (\ref{ChCoulQHD Euler el 2 int for 2D}) and (\ref{ChCoulQHD Euler ion 2 int for 2D}) are over whole space. Thus a point $\textbf{r}-\textbf{r}'$ is also included. It corresponds to point like particles. For consideration of finite radius of ions we need to restrict area of integration taking integral over whole space except a sphere of radius $r_{0}=2r_{i}$, where $r_{i}$ is a radius of ion (see Ref. \cite{Andreev 1401 finite ions}.

Before we present the set of two dimensional QHD equations we need to consider the last term in formula (\ref{ChCoulQHD bal imp eq selfCons int}) in the linear approximation.

$$-e^{2}n(\textbf{r},t)\nabla\int G(\mid \textbf{r}-\textbf{r}'\mid)n(\textbf{r}',t)d\textbf{r}'$$
$$=-e^{2}(n_{0}+\delta n)\nabla\int G(\mid \textbf{r}-\textbf{r}'\mid)(n_{0}+\delta n(\textbf{r}',t))d\textbf{r}'$$
$$=-e^{2}n_{0}^{2}\nabla\int G(\mid \textbf{r}-\textbf{r}'\mid)d\textbf{r}'$$
$$-e^{2}n_{0}\delta n(\textbf{r},t)\nabla\int G(\mid \textbf{r}-\textbf{r}'\mid)d\textbf{r}'$$
\begin{equation}\label{ChCoulQHD Coulomb int term calc 1} -e^{2}n_{0}\nabla\int G(\mid \textbf{r}-\textbf{r}'\mid)\delta n(\textbf{r}',t)d\textbf{r}', \end{equation}
where
\begin{equation}\label{ChCoulQHD grad of int of the Green func}\nabla\int G(\mid \textbf{r}-\textbf{r}'\mid)d\textbf{r}'=\nabla\int \frac{1}{\mid \textbf{r}-\textbf{r}'\mid}d\textbf{r}'=\nabla\int \frac{1}{\mid \xi\mid}d\xi=\nabla const=0,\end{equation}
and $\xi=\textbf{r}'-\textbf{r}$. Formula (\ref{ChCoulQHD grad of int of the Green func}) works for infinite systems. If we have deal with limited in space systems we should go another way. We should explicitly consider presence of motionless ions
$$\textbf{F}_{Cl}=-e^{2}n_{e}\nabla\int
d\textbf{r}'G(\textbf{r},\textbf{r}')\biggl[n_{e}(\textbf{r}',t)-n_{0i}\biggr]$$
\begin{equation}\label{ChCoulQHD}=-e^{2}n_{e}\nabla\int
d\textbf{r}'G(\textbf{r},\textbf{r}')\biggl[n_{0e}+\delta n_{e}(\textbf{r}',t)-n_{0i}\biggr]
\end{equation}
and applying $n_{0e}=n_{0i}$ we have
\begin{equation}\label{ChCoulQHD}\textbf{F}_{Cl}=-e^{2}n_{e}\nabla\int
d\textbf{r}'G(\textbf{r},\textbf{r}')\delta n_{e}(\textbf{r}',t)
\end{equation}
used above.

Using formula (\ref{ChCoulQHD grad of int of the Green func}) we can continue our calculations presented by formula (\ref{ChCoulQHD Coulomb int term calc 1}). We see that the first two terms in the right-hand side of formula (\ref{ChCoulQHD Coulomb int term calc 1}) equal to zero. We should consider the last term of formula (\ref{ChCoulQHD Coulomb int term calc 1})
$$-e^{2}n_{0}\nabla\int G(\mid \textbf{r}-\textbf{r}'\mid)\delta n(\textbf{r}',t)d\textbf{r}'$$
$$=-e^{2}n_{0}\nabla\int G(\mid \textbf{r}-\textbf{r}'\mid)N_{A}e^{-\imath\omega t+\imath \textbf{k}\textbf{r}'}d\textbf{r}'$$
$$=-e^{2}n_{0}N_{A}e^{-\imath\omega t}\nabla\biggl(e^{i\textbf{k}\textbf{r}}\biggr)\int G(\mid\xi\mid)e^{\imath \textbf{k} \xi}d\xi$$
\begin{equation}\label{ChCoulQHD Coulomb int term calc 2}=-e^{2}n_{0}\delta n(\textbf{r},t)\imath \textbf{k}\int G(\mid\xi\mid)e^{\imath \textbf{k} \xi}d\xi,\end{equation}
we have used that the integral in the last formula is a constant. We will find that dispersion of waves depends on the integral. Using explicit form of the Green function of the Coulomb interaction (\ref{ChCoulQHD Colomb Green func in sect 2D}) we can take integral in formula (\ref{ChCoulQHD Coulomb int term calc 2}). We do it in the following way $\int G(\mid\xi\mid)e^{\imath \textbf{k} \xi}d\xi=\int \exp(\imath k \mid\xi\mid\cos\vartheta) d\vartheta d\mid\xi\mid=2\pi/k$, where we applied $G=\frac{1}{\mid\xi\mid}$.

Here we apply the equation of state of unpolarised 2DEG
\begin{equation}\label{ChCoulQHD} p_{Fe,2D}=p_{2D,unpol}=\frac{\pi\hbar^{2}n_{2D}^{2}}{2m},\end{equation}
whereas coefficient in equation of state of fully polarised 2DEG is in two times more $p_{2d, pol}=2p_{2D,unpol}=2p_{Fe,2D}$.

The force field of the Coulomb exchange interaction force
\begin{equation}\label{ChCoulQHD} \textbf{F}_{ex,2D}=\nabla\varepsilon_{C}=2^{3/2} \sqrt{2\pi}\frac{24 \textrm{arsh}1}{\pi^{2}}q_{a}^{2}\sqrt{n_{2D}}\nabla n_{2D}    \end{equation}
has been derived for fully polarised electron gas. For unpolarised electron gas the exchange force equals to zero.

Now we can write down linearised set of continuity and Euler equations for electron gas in a plane
\begin{equation}\label{ChCoulQHD} -\omega\delta n+ n_{0} \textbf{k}\delta \textbf{v}=0,\end{equation}
and
\begin{equation}\label{ChCoulQHD} -\omega mn_{0}\delta \textbf{v}+ m \textbf{k} \biggl[\frac{1}{2}v_{Fe,2D}^{2}\left(\begin{array}{c}1\\
2\\
\end{array}\right)-\frac{\beta\sqrt{2\pi}e^{2}}{\pi^{2}m_{e}}\sqrt{n_{0e}}\left(\begin{array}{c}0\\
1\\
\end{array}\right)\biggr]\delta n+\frac{\hbar^{2}k^{2}\textbf{k}}{4m}\delta n=-e^{2}n_{0} \textbf{k}\frac{2\pi}{k}\delta n, \end{equation}
where the two dimensional Fermi velocity has the following explicit  form
\begin{equation}\label{ChCoulQHD} v_{Fe,2D}=\frac{\sqrt{2\pi n_{2D}}\hbar}{m}.\end{equation}

So the spectrum of two dimensional Langmuir waves has the same structure as the three dimensional one (\ref{ChCoulQHD spectrum Langm 3D})
\begin{equation}\label{ChCoulQHD spectrum Langm 2D} \omega^{2}=\omega_{Le,2D}^2+
\left(\begin{array}{c}\frac{1}{2}v_{Fe,2D}^{2}k^2\\
v_{Fe,2D}^{2}k^2\\
\end{array}\right)-\frac{\beta\sqrt{2\pi}e^{2}}{\pi^{2}m_{e}}\sqrt{n_{0e}}k^{2}\left(\begin{array}{c}0\\
1\\
\end{array}\right)+
\frac{\hbar^2k^4}{4m^2}, \end{equation}
where $\beta=24 \textrm{arsh}1=21.153$, but with the two dimensional Langmuir frequency
\begin{equation}\label{ChCoulQHD Langmuir frq 2D}\omega_{2D}^{2}=\frac{2\pi e^2 k n_0}{m},\end{equation}
and different explicit form of the pressure and exchange interaction contribution.
The upper line in formula (\ref{ChCoulQHD spectrum Langm 2D}) corresponds unpolarised 2DEG, the lower line corresponds to the fully polarised electron gas.

The two dimensional Langmuir frequency (\ref{ChCoulQHD Langmuir frq 2D}) is a signature of low dimensional plane like objects \cite{brusov}-\cite{Tahir JPC 10}.

The Langmuir frequency in three dimensional plasmas is a constant. It does not depend on the wave vector being related to parameters of the system: an equilibrium particle concentration, particle mass and particle charge. In two dimensional plasmas situation is different. The Langmuir frequency is the function of the wave vector, so the square of the 2D Langmuir frequency grows as linear function of the wave vector module (\ref{ChCoulQHD Langmuir frq 2D}).

\subsection{\label{sec:level1} Spectrum of longitudinal waves in 1D quantum plasmas}

To support usefulness of theory of one dimensional object for real application we present a reference on a review paper on creation and application of quasi-one-dimensional structures \cite{one dimensional thing Review}. We also present a Ref. on recent achievements in work with 1D electron gas arrays \cite{Nath JAP 12}.

Let us consider equations of state $p(n)$ for one dimensional degenerate plasmas. We can single out three interesting cases:
\begin{equation}\label{ChCoulQHD} p_{Fe,1D}=\frac{\pi^{2}\hbar^{2}n_{1D}^{3}}{6m},\end{equation}
is the pressure of unpolarised 1D ideal electron gas,
\begin{equation}\label{ChCoulQHD} p_{1D}=\frac{2\pi^{2}\hbar^{2}n_{1D}^{3}}{3m},\end{equation}
is the pressure of fully spin polarised 1D ideal electron gas,
and
\begin{equation}\label{ChCoulQHD} p_{Fe,1D}=\vartheta_{1D}\frac{\pi^{2}\hbar^{2}n_{1D}^{3}}{6m},\end{equation}
is the pressure for most general case of partially polarised 1D ideal electron gas. Here we have introduced $\vartheta_{1D}$, which shows rate of polarisation of spin of electron gas. Its explicit form is
\begin{equation}\label{ChCoulQHD} \vartheta_{1D}=\frac{1}{2}[(1+\eta)^{3}+(1-\eta)^{3}],\end{equation}
where $\eta=\Delta n/n_{0}$, with $\Delta n=n_{\downarrow}-n_{\uparrow}$, $n=n_{\downarrow}+n_{\uparrow}$, with $n_{\uparrow}$ and $n_{\downarrow}$ are concentrations of spin-up and spin-down fermions.

The Euler equation for 1D plasmas
\begin{equation}\label{ChCoulQHD Euler 1D in Spectrum section} mn_{1}(\partial_{t}+v_{x}\nabla)v_{x}+\nabla_{x} p=\textbf{F}_{x Coul Int}, \end{equation}
where the force of Coulomb interaction
\begin{equation}\label{ChCoulQHD F coul int 1D} \textbf{F}_{x Coul Int}(x,t)=-en(x,t)\nabla_{x}\int \frac{n(x',t)}{\mid x-x'\mid}dx' .\end{equation}

Linearised force field has the following form
\begin{equation}\label{ChCoulQHD F coul int lin} \textbf{F}_{Coul Int(lin)}(x,t)=-en_{0}\nabla_{x}\int \frac{\delta n(x',t)}{\mid x-x'\mid}dx' .\end{equation}

NLSE for 1D quantum electron gas with the Coulomb exchange interaction can be obtained from the Euler equation (\ref{ChCoulQHD Euler 1D in Spectrum section}) (see section (\ref{sec:NLSE}) for recipes, or Refs. \cite{MaksimovTMP 1999}, \cite{Andreev PRA08}, \cite{Andreev PRB 11})
$$\imath\hbar\partial_{t}\Phi_{1D}(\textbf{r},t)=\Biggl(-\frac{\hbar^{2}\nabla^{2}}{2m}+\frac{\pi^{2}}{4}\frac{\hbar^{2}}{m}n_{1D}^{2}$$
\begin{equation}\label{ChCoulQHD NLSE int 1D with exchange} +e^{2}\int
d\textbf{r}'G(\textbf{r},\textbf{r}')\biggl[n(\textbf{r}',t)-n_{0i}\biggr]\Biggr)\Phi_{1D}(\textbf{r},t),
\end{equation}
where $n_{1D}=\mid\Phi_{1D}(\textbf{r},t)\mid^{2}$, $\int
d\textbf{r}'G(\textbf{r},\textbf{r}')\biggl[n(\textbf{r}',t)-n_{0i}\biggr]=\int
dx'G(x,x')\biggl[n(x',t)-n_{0i,1D}\biggr]$, where we have used $n(\textbf{r}',t)=n(x',t)\delta(y')\delta(z')$, $n_{0i}=n_{0i,1D}\delta(y')\delta(z')$.

To get spectrum of 1D electron gas we need to consider formula (\ref{ChCoulQHD F coul int lin}). This problem was considered in Ref. \cite{Santoyo Mexicana de Fisica}. So, we use their results.
\begin{equation}\label{ChCoulQHD F coul int lin Fourier} \textbf{F}_{Coul Int(lin)}(x,t) =-en_{0} ik_{x}\delta n(x',t) \int \frac{e^{\imath k_{x}(x-x')}}{\mid x-x'\mid}dx' .\end{equation}

Direct evaluation of integral in formula (\ref{ChCoulQHD F coul int lin Fourier}) leads to divergence. So, we cannot allow strictly 1D, when $y=z=0$. To solve this problem we appeal to physical grounds \cite{Santoyo Mexicana de Fisica}.
"The electron gas must coexist with a positive
background, thus, there should be a minimum distance, $a_{0}$, between the electrons and the
positive charge. This minimum distance would be a limit to the "size" of the cross section
of our 1D system. In a sense, this $a_{0}$ would be like a "Bohr radius" similar to that of
the hydrogen atom. Therefore we need a system which extends a macroscopic distance,
e.g., along x, and has a microscopic (finite) cross section, which should be at least $a_{0}$" \cite{Santoyo Mexicana de Fisica}. Hence we consider integral (\ref{ChCoulQHD F coul int lin Fourier}) with finite $y$ and $z$ in the Coulomb potential
$$\textbf{F}_{Coul Int(lin)}(x,t) $$
$$= -en_{0} ik_{x}\delta n(x',t) \lim_{y,z\rightarrow0}\int \frac{e^{\imath k_{x}(x-x')}}{\sqrt{(x-x')^{2}+y^{2}+z^{2}}}dx' \rightarrow$$
$$ \rightarrow -en_{0} ik_{x}\delta n(x',t) \lim_{y^{2}+z^{2}\rightarrow a_{0}^{2}}\int \frac{e^{\imath k_{x}(x-x')}}{\sqrt{(x-x')^{2}+y^{2}+z^{2}}}dx' $$
\begin{equation}\label{ChCoulQHD} =-en_{0} ik_{x}\delta n(x',t) 2K_{0}(\mid k_{x}\mid a_{0}), \end{equation}
where $K_{0}$ is the modified Bessel function of zeroth order.

On the Coulomb-type potential of the one-dimensional Schrodinger equation see also Ref. \cite{Yangqiang Ran JPA 00}.

Let us present the linearised set of QHD equations for fully polarised electrons. So it contains contribution of the exchange interaction:
\begin{equation}\label{ChCoulQHD} -\omega\delta n+ n_{0} k\delta v=0,\end{equation}
and
\begin{equation}\label{ChCoulQHD} -\omega mn_{0}\delta v+ 2 m v_{Fe,1D}^{2} k \left(\begin{array}{c}1\\
4\\
\end{array}\right)\delta n+\frac{\hbar^{2}k^{3}}{4m}\delta n=-2K_{0}(\mid k\mid a_{0}) e^{2}n_{0}k \delta n .\end{equation}

In case of unpolarised electrons the coefficient before the pressure contribution becomes 4 times smaller.

The following dispersion dependencies present
\begin{equation}\label{ChCoulQHD spectrum Langm 1D unpol} \omega^{2}=\omega_{Le,1D}^2+ 2v_{Fe,1D}^{2}k^2+
\frac{\hbar^2k^4}{4m^2}, \end{equation}
spectrum of unpolarised gas, and
\begin{equation}\label{ChCoulQHD spectrum Langm 1D pol} \omega^{2}=\omega_{Le,1D}^2+8 v_{Fe,1D}^{2}k^2+
\frac{\hbar^2k^4}{4m^2}, \end{equation}
spectrum polarised gas, where we have applied the 1D Langmuir frequency
\begin{equation}\label{ChCoulQHD Langmuir frq 1D}\omega_{1D}^{2}=\frac{ e^2 k^{2} V(k) n_{0,1D}}{m}\end{equation}
where
\begin{equation}\label{ChCoulQHD} V(k)=2K_{0}(\mid k_{x}\mid a_{0}). \end{equation}

To compare with experiment we may have a look on Ref. \cite{Hwang JP CS 07} formula (1) and text around it, where we find formula
\begin{equation}\label{ChCoulQHD omega from Hwang} \omega^{2}= \biggl(v_{Fe}^{2}+\frac{2v_{Fe}}{\pi\hbar}V(k)\biggr)k^{2},\end{equation}
in this formula $v_{Fe}=\hbar k_{F}/m$ is the 1D Fermi velocity.
Formula (\ref{ChCoulQHD omega from Hwang}) is the spectrum of plasmon excitation in 1D system obtained in terms of random-phase-approximation for non-interacting electrons and the
Luttinger liquid theory for the interacting electrons (see Refs. \cite{Li PRB 92} and \cite{Voit RPP 94}).

\subsection{\label{sec:level1} Spectrum of longitudinal waves in quantum plasmas in the external magnetic field:
Waves propagating perpendicular to external field}

\subsubsection{ 1D plasmas in magnetic field}


Here we consider string-like plasmas objects having form of the straight line. The external magnetic field is directed perpendicular to the samples.
Straightforward calculations gives us the following spectrum of the Langmuir waves
\begin{equation}\label{ChCoulQHD spectrum Langm 1D magn} \omega^{2}=\omega_{Le,1D}^2+2v_{Fe,1D}^{2}          \biggl[1+3\biggl(\frac{\Delta n_{1D}}{n_{0,1D}}\biggr)^{2}\biggr]k^2+
\frac{\hbar^2k^4}{4m^2}, \end{equation}
where $\Delta n_{1D}=n_{0,1D}\tanh\biggl(\frac{\mu B_{0}}{E_{Fe,1D}}\biggr)$, and we have assumed that ions are motionless. We have not considered the Coulomb exchange interaction in this subsubsection.

This spectrum coincides with the spectrum of the Langmuir waves when sample in the presence of external magnetic field parallel to the sample, i.e. we have "in line" magnetic field. In 2D and 3D systems similar limit cases of the Langmuir spectrum are different. In 1D system we could expect presence of the cyclotron frequency in spectrum in the perpendicular external magnetic, but it does not appear due to forces keeping electrons in the one dimension.

\subsubsection{2D plasmas in magnetic field}

Considering plane-like object, for instance two dimensional electron (hole) gas in multilayered semiconductor structure, located in an external magnetic field directed perpendicular to the plane, electron oscillation leads to the following spectrum of two dimensional Langmuir waves
$$\omega^{2}=\omega_{Le,2D}^2+\Omega^{2}+\frac{1}{2}v_{Fe,2D}^{2}\Biggl(1+\biggl(\frac{\Delta n}{n_{0,2D}}\biggr)^{2}\Biggr)k^2 $$
\begin{equation}\label{ChCoulQHD spectrum Langm 2D magn} -\biggl[\biggl(1+\frac{\Delta n}{n_{0,2D}}\biggr)^{\frac{3}{2}}-\biggl(1-\frac{\Delta n}{n_{0,2D}}\biggr)^{\frac{3}{2}}\biggr] \frac{\beta\sqrt{2\pi}e^{2}}{\pi^{2}m}\sqrt{n_{0}}k^{2}+
\frac{\hbar^2k^4}{4m^2}, \end{equation}
where $\Delta n=n_{0,2D}\tanh\biggl(\frac{\mu B_{0}}{E_{Fe,2D}}\biggr)$, $\Omega=e B_{0}/(mc)$ is the cyclotron frequency. If we consider "in plane" magnetic field (magnetic field parallel to the plane), we find that the cyclotron frequency disappears from spectrum.

\subsubsection{3D plasmas in magnetic field}

Considering plane waves in three dimensional plasmas we obtain the following spectrum of the Langmuir waves
$$\omega^{2}=\omega_{Le,3D}^2+\Omega^{2}+
\frac{1}{3}v_{Fe,3D}^{2}\biggl[1+\frac{5}{9}\biggl(\frac{\Delta n}{n_{0,3D}}\biggr)^{2}\biggr]k^2$$
\begin{equation}\label{ChCoulQHD spectrum Langm 3D magn} -\biggl[\biggl(1+\frac{\Delta n}{n_{0,3D}}\biggr)^{\frac{4}{3}}-\biggl(1-\frac{\Delta n}{n_{0,3D}}\biggr)^{\frac{4}{3}}\biggr]\sqrt[3]{\frac{3}{\pi}}\frac{e^{2}}{m}\sqrt[3]{n_{0}}k^{2} +\frac{\hbar^2k^4}{4m^2}, \end{equation}
where $\Delta n=n_{0,3D}\tanh\biggl(\frac{\mu B_{0}}{E_{Fe,3D}}\biggr)$. This spectrum is obtained under following assumptions: ions are motionless and create the positively charged background, electrons are considered as a single species of mixed spin-up and spin-down degenerate particles, system is in an external magnetic field, and wave propagate in a direction perpendicular to the external field. In wave propagate parallel to external field this formula has small change: the contribution of the cyclotron frequency disappears.

\section{\label{sec:QHD cylindr} Quantum hydrodynamics in cylindrical coordinates}

Quantum continuity and Euler equations in the cylindrical coordinates are very useful at application of QHD to cylindrical waves in three dimensional mediums \cite{Gahlot PP 13}, \cite{ul Haq JAP 10}, waves on two dimensional cylindrical objects, for instance nano-tubes, circle waves in 2DEG on a plane, and other quantum objects having cylindrical symmetry \cite{NON-Plane Cyl and Sph    CLASSIC    BEG}-\cite{NON-Plane Cyl and Sph    CLASSIC    END}. Quantum hydrodynamic description of objects has been performed \cite{NON-Plane Cyl and Sph   QUANTUM    BEG} -\cite{Spherical Quant eq state 02}, but they have not included quantum inertia force, which behave differently to their classic part depending on velocity field and thermal pressure \cite{Landau Vol 6}. Below, in this and the next sections,  we present QHD equations including quantum part of inertia force in cylindric and spherical coordinates.

Thus we starts with the cylindrical coordinates.

Cylindrical coordinates $\rho$, $\varphi$, $z$ can be introduced via the Cartesian coordinates $x$, $y$, $z$: $x=\rho\cos\varphi$, $y=\rho\sin\varphi$, $z=z$.

In cylindrical coordinates the continuity equation appears as
\begin{equation}\label{ChCoulQHD cont eq cylindr with j} \partial_{t}n+\biggl(\partial_{\rho}+\frac{1}{\rho}\biggr)j_{\rho}+\frac{1}{\rho}\partial_{\varphi}j_{\varphi}+\partial_{z}j_{z}=0,\end{equation}
with
\begin{equation}\label{ChCoulQHD} n=\langle \Psi^{*}\Psi\rangle\end{equation}
as usual, and
\begin{equation}\label{ChCoulQHD j rho cyl} j_{\rho}=   \imath\hbar\frac{1}{2m}\langle (\partial_{\rho}^{i}\Psi^{*})\Psi-\Psi^{*}\partial_{\rho}^{i}\Psi\rangle, \end{equation}
\begin{equation}\label{ChCoulQHD j varphi cyl} j_{\varphi}=\imath\hbar\frac{1}{2m}\langle \frac{1}{\rho_{i}}\biggl((\partial_{\varphi}^{i}\Psi^{*})\Psi-\Psi^{*}\partial_{\varphi}^{i}\Psi\biggr)\rangle, \end{equation}
\begin{equation}\label{ChCoulQHD j z cyl} j_{z}=      \imath\hbar\frac{1}{2m}\langle (\partial_{z}^{i}\Psi^{*})\Psi-\Psi^{*}\partial_{z}^{i}\Psi\rangle.\end{equation}
The explicit form of the quantum mechanical average $\langle ... \rangle$ is presented by formula (\ref{ChCoulQHD def density with brackets}), however, for simplicity of presentation we have drop delta function multipliers.

Applying the divergence in the cylindrical coordinates $\textmd{div} \textbf{a}\equiv \frac{1}{\rho}\partial_{\rho}(\rho a_{\rho})+\frac{1}{\rho}\partial_{\varphi}a_{\varphi}+\partial_{z}a_{z}$ we can rewrite the continuity equation (\ref{ChCoulQHD cont eq cylindr with j}) in rather usual form
\begin{equation}\label{ChCoulQHD cont eq cylindr with j via div} \partial_{t}n+\textmd{div} \textbf{j}=0.\end{equation}

The Euler equations in the cylindrical coordinates can be written as follows
\begin{equation}\label{ChCoulQHD} \partial_{t}j^{\alpha}+\biggl(\partial_{\rho}+\frac{1}{\rho}\biggr)\Pi^{\alpha\rho}+\frac{1}{\rho}\partial_{\varphi}\Pi^{\alpha\varphi}+\partial_{z}\Pi^{\alpha z}=F^{\alpha}_{inertia}+F^{\alpha}_{int},\end{equation}
where $\alpha$ stands for $\rho$, $\varphi$, and $z$. $F^{\alpha}_{inertia}$ is the inertia force, which contains quantum contribution.

From equation (\ref{ChCoulQHD DtJ}) we see that interaction appears in the Euler equation as the commutator of the operator giving the particles current $\textbf{D}_{i}/m_{i}$ and the Hamiltonian of system under consideration. The time derivative of the Hamiltonian also gives contribution in the interaction. Other terms in (\ref{ChCoulQHD DtJ}) have kinematic nature. So it was then we worked in the Cartesian coordinates. Now we have deal with the cylindrical coordinates, which are curvilinear coordinates. Thus the commutator of $\textbf{D}_{i}/m_{i}$ with the kinetic energy operator gives us some terms, which do not contain any trace of interaction. We can call them the quantum inertia force, since the inertia force appears in classic hydrodynamics in similar way.

The quantum inertia force in the cylindrical coordinates has the following form
\begin{equation}\label{ChCoulQHD F inertia_rho} F_{\rho,inertia} =\frac{\hbar^{2}}{4m^{2}}\biggl[-\frac{1}{\rho^{2}}\partial_{\rho}n-\frac{2}{\rho^{3}}\langle\Psi^{*}\partial_{\varphi (i)}^{2}\Psi+c.c.\rangle\biggr],\end{equation}
\begin{equation}\label{ChCoulQHD F inertia_varphi} F_{\varphi,inertia} =\frac{\hbar^{2}}{4m^{2}}\biggl[-\frac{1}{\rho^{3}}\partial_{\varphi}n+\frac{2}{\rho^{2}}\langle\Psi^{*}\partial_{\rho}^{i}\partial_{\varphi}^{i}\Psi+c.c.\rangle\biggr],\end{equation}
and
\begin{equation}\label{ChCoulQHD F inertia_z} F_{z,inertia}=0.\end{equation}

Tensor of the momentum current has the following structure
\begin{equation}\label{ChCoulQHD} \Pi_{\alpha\beta}=-\frac{\hbar^{2}}{4m^{2}} \langle\partial_{\beta}\partial_{\alpha}\Psi^{*}\cdot\Psi -\partial_{\alpha}\Psi^{*}\cdot\partial_{\beta}\Psi -\partial_{\beta}\Psi^{*}\cdot\partial_{\alpha}\Psi +\Psi^{*}\cdot\partial_{\beta}\partial_{\alpha}\Psi\rangle, \end{equation}
where $\partial_{\alpha}$ and $\partial_{\beta}$ stand for $\partial_{\rho}$, $\frac{1}{\rho}\partial_{\varphi}$, $\partial_{z}$. So, for instance $\Pi_{\varphi\rho}$ contains $\partial_{\rho}\biggl(\frac{1}{\rho}\partial_{\varphi}\Psi^{*}\biggr)\Psi$.

QHD in the cylindrical coordinates allows to introduce the velocity field in usual manner. Applying presentation of the many-particle wave function via its phase and amplitude $\Psi=a\exp(\imath S/\hbar)$ to the explicit form of particle current (\ref{ChCoulQHD j rho cyl})-(\ref{ChCoulQHD j z cyl}) we obtain $v_{\rho}=\hbar \langle a^{2} \partial_{\rho}^{i} S/m\rangle/n$, $v_{\varphi}=\hbar\langle a^{2} \partial_{\varphi}^{i} S/(\rho m)\rangle/n$, $v_{z}=\hbar\langle a^{2} \partial_{z}^{i} S/(m)\rangle/n$, where $S(R,t)$ is the phase of the many-particle wave function $\Psi(R,t)$.

An explicit form of the inertia force field in the cylindrical coordinates arises as
\begin{equation}\label{ChCoulQHD F inertia_rho explicit} F_{\rho,inertia}=    -\frac{1}{\rho}nv_{\varphi}^{2}-\frac{1}{\rho}p_{\varphi\varphi} +\frac{\hbar^{2}}{m^{2}}\frac{1}{\rho^{3}}\sqrt{n}\partial_{\varphi}^{2}\sqrt{n}-\frac{\hbar^{2}}{4m^{2}}\frac{1}{\rho^{2}}\partial_{\rho}n,\end{equation}
and
\begin{equation}\label{ChCoulQHD F inertia_varphi explicit} F_{\varphi,inertia}= +\frac{1}{\rho}nv_{\rho}v_{\varphi}+\frac{1}{\rho}p_{\rho\varphi} -\frac{\hbar^{2}}{m^{2}}\frac{1}{\rho^{2}}\sqrt{n}\partial_{\varphi}\partial_{\rho}\sqrt{n}-\frac{\hbar^{2}}{4m^{2}}\frac{1}{\rho^{3}}\partial_{\varphi}n.\end{equation}

We also can introduce the velocity field in the cylindrical coordinates
\begin{equation}\label{ChCoulQHD}\begin{array}{ccc} j_{\rho}= nv_{\rho}  ,& j_{\varphi}= nv_{\varphi} ,& j_{z}= nv_{z}  \end{array}.\end{equation}

The continuity equation in term of the velocity field has form
\begin{equation}\label{ChCoulQHD cont eq cylindr with v} \partial_{t}n+\biggl(\partial_{\rho}+\frac{1}{\rho}\biggr)(nv_{\rho})+\frac{1}{\rho}\partial_{\varphi}(nv_{\varphi})+\partial_{z}(nv_{z})=0. \end{equation}

General equation for evolution of $v_{\rho}$ appears as
$$n\partial_{t}v_{\rho}+n(\textbf{v}\nabla)v_{\rho}-\frac{1}{\rho}nv_{\varphi}^{2}$$
$$+\biggl(\partial_{\rho}+\frac{1}{\rho}\biggr)\biggl[p_{\rho\rho}+\frac{\hbar^{2}}{2m^{2}}\biggl((\partial_{\rho}\sqrt{n})^{2}-\sqrt{n}\partial_{\rho}^{2}\sqrt{n}\biggr)\biggr]$$
$$+\frac{1}{\rho}\partial_{\varphi}\biggl[p_{\rho\varphi}+\frac{\hbar^{2}}{2m^{2}}\frac{1}{\rho}\biggl(\partial_{\rho}\sqrt{n}\cdot\partial_{\varphi}\sqrt{n}-\sqrt{n}\partial_{\rho}\partial_{\varphi}\sqrt{n}\biggr)\biggr]$$
$$+\partial_{z}\biggl[p_{\rho z}+\frac{\hbar^{2}}{2m^{2}}\biggl(\partial_{\rho}\sqrt{n}\cdot\partial_{z}\sqrt{n}-\sqrt{n}\partial_{\rho}\partial_{z}\sqrt{n}\biggr)\biggr]$$
\begin{equation}\label{ChCoulQHD eq for v rho gen} -\frac{1}{\rho}p_{\varphi\varphi}+\frac{\hbar^{2}}{4m^{2}}\frac{1}{\rho^{2}}\partial_{\rho}n
+\frac{\hbar^{2}}{m^{2}}\frac{1}{\rho^{3}}\sqrt{n}\partial_{\varphi}^{2}\sqrt{n}=F_{\rho,int}, \end{equation}
with $F_{\rho,int}=qnE_{\rho}+q[\textbf{v},\textbf{B}]_{\rho}/c$.

General equation for evolution of $v_{\varphi}$ has the following form
$$n\partial_{t}v_{\varphi}+n(\textbf{v}\nabla)v_{\varphi}+\frac{1}{\rho}nv_{\rho}v_{\varphi}$$
$$+\biggl(\partial_{\rho}+\frac{1}{\rho}\biggr)\biggl[p_{\varphi\rho}+\frac{\hbar^{2}}{2m^{2}}\frac{1}{\rho}\biggl(\partial_{\rho}\sqrt{n}\cdot\partial_{\varphi}\sqrt{n}-\sqrt{n}\partial_{\rho}\partial_{\varphi}\sqrt{n}\biggr)+\frac{\hbar^{2}}{4m^{2}}\frac{1}{\rho^{2}}\partial_{\varphi}n\biggr]$$
$$+\frac{1}{\rho}\partial_{\varphi}\biggl[p_{\varphi\varphi}+\frac{\hbar^{2}}{2m^{2}}\frac{1}{\rho^{2}}\biggl((\partial_{\varphi}\sqrt{n})^{2}-\sqrt{n}\partial_{\varphi}^{2}\sqrt{n}\biggr)\biggr]$$
$$+\partial_{z}\biggl[p_{\varphi z}+\frac{\hbar^{2}}{2m^{2}}\frac{1}{\rho}\biggl(\partial_{\varphi}\sqrt{n}\cdot\partial_{z}\sqrt{n}-\sqrt{n}\partial_{\varphi}\partial_{z}\sqrt{n}\biggr)\biggr]$$
\begin{equation}\label{ChCoulQHD eq for v verphi gen} +\frac{1}{\rho}p_{\rho\varphi} +\frac{\hbar^{2}}{4m^{2}}\frac{1}{\rho^{3}}\partial_{\varphi}n-\frac{\hbar^{2}}{m^{2}}\frac{1}{\rho^{2}}\sqrt{n}\partial_{\varphi}\partial_{\rho}\sqrt{n}=F_{\varphi,int}, \end{equation}
where $F_{\varphi,int}=qnE_{\varphi}+q[\textbf{v},\textbf{B}]_{\varphi}/c$.

Equation for evolution of $v_{z}$ to be
$$n\partial_{t}v_{z}+n(\textbf{v}\nabla)v_{z}$$
$$+\biggl(\partial_{\rho}+\frac{1}{\rho}\biggr)\biggl[p_{z\rho}  +\frac{\hbar^{2}}{2m^{2}}\biggl(\partial_{\rho}\sqrt{n}\cdot\partial_{z}\sqrt{n}-\sqrt{n}\partial_{\rho}\partial_{z}\sqrt{n}\biggr)\biggr]$$
$$+\frac{1}{\rho}\partial_{\varphi}\biggl[p_{z\varphi}+\frac{\hbar^{2}}{2m^{2}}\frac{1}{\rho}\biggl(\partial_{z}\sqrt{n}\cdot\partial_{\varphi}\sqrt{n}-\sqrt{n}\partial_{z}\partial_{\varphi}\sqrt{n}\biggr)\biggr]$$
\begin{equation}\label{ChCoulQHD eq for v z gen} +\partial_{z}\biggl[p_{zz}+\frac{\hbar^{2}}{2m^{2}}\biggl((\partial_{z}\sqrt{n})^{2}-\sqrt{n}\partial_{z}^{2}\sqrt{n}\biggr)\biggr]=F_{z,int}, \end{equation}
where $F_{z,int}=qnE_{z}+q[\textbf{v},\textbf{B}]_{z}/c$.

The last term in the second group of terms in equation of $v_{\varphi}$ evolution (\ref{ChCoulQHD eq for v verphi gen}) $\biggl(\partial_{\rho}+\frac{1}{\rho}\biggr)\frac{\hbar^{2}}{4m^{2}}\frac{1}{\rho^{2}}\partial_{\varphi}n$ is rather similar to the second term in the last group of terms on the left-hand side of equation (\ref{ChCoulQHD eq for v verphi gen}) $\frac{\hbar^{2}}{4m^{2}}\frac{1}{\rho^{3}}\partial_{\varphi}n$. Inspite the similarity of these term we do not combine them together, since they are appeared in different ways. The last term in the second group of terms is a part of the quantum Bohm potential, since both of them have appeared from the corresponding element of the momentum current tensor $\Pi_{\varphi\rho}$. The second term in the last group of terms is the part of the inertia force field as we can see from formula (\ref{ChCoulQHD F inertia_varphi explicit}).

The electromagnetic fields obey the Maxwell equations $\textmd{div} \textbf{B}=0$, $\textmd{div} \textbf{E}=4\pi\sum_{a}q_{a}n_{a}$, $\textmd{curl} \textbf{E}=-\partial_{t}\textbf{B}$, and $\textmd{curl} \textbf{B}=4\pi \sum_{a}q_{a}n_{a}\textbf{v}_{a}/c+\partial_{t}\textbf{E}$.

\subsection{\label{sec:level1} Dispersion of cylindrical waves}

In literature on fundamental properties of quantum plasmas most attention is focused on plane linear and nonlinear perturbations. So we decided to consider quantum cylindrical waves in 3D electron gas.

In absence of angle dependence $\partial_{\varphi}=0$ we have the continuity equation
\begin{equation}\label{ChCoulQHD cont eq cylindr with v simp} \partial_{t}n+\biggl(\partial_{\rho}+\frac{1}{\rho}\biggr)(nv_{\rho})+\partial_{z}(nv_{z})=0,\end{equation}
and the Euler equation
$$n\partial_{t}v_{\rho}+n(\textbf{v}\nabla)v_{\rho}-\frac{1}{\rho}nv_{\varphi}^{2}-\frac{1}{\rho}p_{\varphi\varphi}+\frac{\hbar^{2}}{4m^{2}}\frac{1}{\rho^{2}}\partial_{\rho}n$$
$$+\biggl(\partial_{\rho}+\frac{1}{\rho}\biggr)p_{\rho\rho}
+\frac{\hbar^{2}}{4m^{2}}\biggl(\partial_{\rho}+\frac{1}{\rho}\biggr)\biggl(\frac{(\partial_{\rho}n)^{2}}{n}-\partial_{\rho}^{2}n\biggr)$$
\begin{equation}\label{ChCoulQHD eq for v rho simp} +\frac{\hbar^{2}}{4m^{2}}\partial_{z} \biggl(\frac{\partial_{\rho}n\cdot\partial_{z}n}{n}-\partial_{\rho}\partial_{z}n\biggr)=F_{\rho,int}, \end{equation}

\begin{equation}\label{ChCoulQHD eq for v verphi simp} n\partial_{t}v_{\varphi}+n(\textbf{v}\nabla)v_{\varphi}+\frac{1}{\rho}nv_{\rho}v_{\varphi}=F_{\varphi,int}, \end{equation}
and
$$n\partial_{t}v_{z}+n(\textbf{v}\nabla)v_{z}$$
$$+ \frac{\hbar^{2}}{4m^{2}} \biggl(\partial_{\rho}+\frac{1}{\rho}\biggr)\biggl(\frac{\partial_{\rho}n\cdot\partial_{z}n}{n}-\partial_{\rho}\partial_{z}n\biggr)$$
\begin{equation}\label{ChCoulQHD eq for v z simp} +\partial_{z}p_{zz}+\frac{\hbar^{2}}{4m^{2}}\partial_{z}\biggl(\frac{(\partial_{z}n)^{2}}{n}-\partial_{z}^{2}n\biggr)=F_{z,int}\end{equation}
Independence of the hydrodynamic variables on the angle $\varphi$ gave significant simplification of QHD equations. In equations (\ref{ChCoulQHD eq for v rho simp})-(\ref{ChCoulQHD eq for v z simp}) we have also assumed that pressure is isotropic. For isotropic pressure $p_{\rho\rho}=p_{\varphi\varphi}=p_{zz}\equiv p$ and nondiagonal elements of the pressure tensor equal to zero \cite{Landau Vol 6}.

In cylindrical waves all physical depend on the radial coordinate $\rho$ only. Assuming that we can present further simplification of the QHD equations. Hence we have
\begin{equation}\label{ChCoulQHD cont eq cylindr with v simp II} \partial_{t}n+\biggl(\partial_{\rho}+\frac{1}{\rho}\biggr)(nv_{\rho})=0,\end{equation}

$$n\partial_{t}v_{\rho}+nv_{\rho}\partial_{\rho}v_{\rho}-\frac{1}{\rho}nv_{\varphi}^{2}+\frac{\hbar^{2}}{4m^{2}}\frac{1}{\rho^{2}}\partial_{\rho}n$$
\begin{equation}\label{ChCoulQHD eq for v rho simp II}+\partial_{\rho}p_{\rho\rho}
+\frac{\hbar^{2}}{4m^{2}}\biggl(\partial_{\rho}+\frac{1}{\rho}\biggr)\biggl(\frac{(\partial_{\rho}n)^{2}}{n}-\partial_{\rho}^{2}n\biggr)=F_{\rho,int},\end{equation}

\begin{equation}\label{ChCoulQHD eq for v verphi simp II} n\partial_{t}v_{\varphi}+nv_{\rho}\partial_{\rho}v_{\varphi}+\frac{1}{\rho}nv_{\rho}v_{\varphi}=F_{\varphi,int}, \end{equation}
and
$$n\partial_{t}v_{z}+nv_{\rho}\partial_{\rho}v_{z}$$
\begin{equation}\label{ChCoulQHD eq for v z simp} +\partial_{z}p_{zz}+\frac{\hbar^{2}}{4m^{2}}\partial_{z}\biggl(\frac{(\partial_{z}n)^{2}}{n}-\partial_{z}^{2}n\biggr)=F_{z,int}.\end{equation}

Next we consider linearised set of QHD equations for longitudinal waves, assuming $n=n_{0}+\delta n$, $\textbf{v}=0+\delta \textbf{v}$, $\phi=\delta\phi$, with $E_{\rho}=-\partial_{\rho}\phi$
\begin{equation}\label{ChCoulQHD cont eq cylindr with v simp II lin} \partial_{t}\delta n+n_{0}\biggl(\partial_{\rho}+\frac{1}{\rho}\biggr)\delta v_{\rho}=0,\end{equation}

$$n_{0}\partial_{t}\delta v_{\rho}+\frac{\hbar^{2}}{4m^{2}}\frac{1}{\rho^{2}}\partial_{\rho}n$$
\begin{equation}\label{ChCoulQHD eq for v rho simp II lin}+\partial_{\rho}p_{\rho\rho}
-\frac{\hbar^{2}}{4m^{2}}\biggl(\partial_{\rho}+\frac{1}{\rho}\biggr)\biggl(\partial_{\rho}^{2}n\biggr)=-\frac{e}{m}n_{0}\partial_{\rho}\delta\phi+\frac{eB_{0}}{mc}n_{0}\delta v_{\varphi},\end{equation}

\begin{equation}\label{ChCoulQHD eq for v verphi simp II lin} n_{0}\partial_{t}v_{\varphi}=-\frac{eB_{0}}{mc}\delta v_{\rho}, \end{equation}
and
$\partial_{t}v_{z}=0$.

Poisson equation
\begin{equation}\label{ChCoulQHD} \frac{1}{\rho}\partial_{\rho}(\rho\partial_{\rho}\phi)=-4\pi e\delta n \end{equation}
Together with the continuity equation it gives $\partial_{\rho}\phi=4\pi\imath e n_{0}\delta v_{\rho}/\omega$.

\begin{equation}\label{ChCoulQHD} (\omega^{2}-\Omega^{2}-\omega_{Le,3D}^{2})\delta n+U^{2}\triangle_{\rho}\delta n-\frac{\hbar^{2}}{4m^{2}}\triangle_{\rho}^{2}\delta n=0, \end{equation}
where $\Omega=eB_{0}/(mc)$ is the electron cyclotron frequency, and $\triangle_{\rho}=\biggl(\partial_{\rho}+1/\rho\biggr)\partial_{\rho}$ is the radial part of the Laplacian.

Consideration of cylindrical waves in the quantum plasmas gives same spectrum as three dimensional plane waves
\begin{equation}\label{ChCoulQHD Spectrum Langm 3D cylindric} \omega^{2}_{cyl}= \omega_{Le,3D}^{2}+\Omega^{2}+U^{2}k^{2}+\frac{\hbar^{2}k^{4}}{4m^{2}}.\end{equation}

with the wave vector $k$ satisfying the wave equation for cylindrical waves
\begin{equation}\label{ChCoulQHD} \frac{1}{\rho}\partial_{\rho}\biggl(\rho \partial_{\rho}n\biggr)+k^{2}n=0.\end{equation}
Hence, for the traveling wave, we have the following solution
\begin{equation}\label{ChCoulQHD wave cylindric H} n=N_{0}e^{-\imath\omega t} H_{0}^{(1)}(k\rho),\end{equation}
where $H_{0}^{(1)}(k\rho)$ is the Hankel function.

For large $k\rho\gg1$ one finds simplification of solution (\ref{ChCoulQHD wave cylindric H}) via exponent
\begin{equation}\label{ChCoulQHD} n=N_{0}\sqrt{\frac{2}{\pi}}\frac{e^{-\imath\omega t +\imath k \rho-\imath\pi/4}}{\sqrt{k\rho}}.\end{equation}

\section{\label{sec:QHD spheric} Quantum hydrodynamics in spherical coordinates}

At consideration of spherical waves in 3D plasmas \cite{ul Haq JAP 10}, waves in 2DEG on spherical surface, and other spherically symmetric systems we need to have the MPQHD equations in spherical coordinates.

Let us present the Cartesian coordinates via the spherical coordinates
$$ x=r\sin\theta\cos\varphi,$$
$$ y=r\sin\theta\sin\varphi,$$
$$ z=r\cos\theta,$$
where $r\in[0,\infty)$, $0\leq\theta\leq\pi$, and $0\leq\varphi\leq 2\pi$.

In spherical coordinates the continuity equation appears as
\begin{equation}\label{ChCoulQHD cont eq sph with j} \partial_{t}n+\frac{1}{r^{2}}\partial_{r}(r^{2}j_{r})
+\frac{1}{r\sin\theta}\partial_{\theta}(\sin\theta j_{\theta})+\frac{1}{r\sin\theta}\partial_{\varphi}j_{\varphi}=0,\end{equation}
with traditional definition of the particle concentration
\begin{equation}\label{ChCoulQHD} n=\langle \Psi^{*}\Psi\rangle,\end{equation}
and components of the particle current vector
\begin{equation}\label{ChCoulQHD j r sph} j_{r}=   \imath\hbar\frac{1}{2m}\langle (\partial_{r}^{i}\Psi^{*})\Psi-\Psi^{*}\partial_{r}^{i}\Psi\rangle,\end{equation}
\begin{equation}\label{ChCoulQHD j theta sph} j_{\theta}=\imath\hbar\frac{1}{2m}\langle \frac{1}{r_{i}}\biggl((\partial_{\theta}^{i}\Psi^{*})\Psi-\Psi^{*}\partial_{\theta}^{i}\Psi\biggr)\rangle,\end{equation}
\begin{equation}\label{ChCoulQHD j varphi sph} j_{\varphi}=      \imath\hbar\frac{1}{2m}\langle \frac{1}{r_{i}\sin\theta_{i}}\biggl( (\partial_{\varphi}^{i}\Psi^{*})\Psi-\Psi^{*}\partial_{\varphi}^{i}\Psi\biggr)\rangle.\end{equation}
Definition of angular brackets $\langle ...\rangle$ is presented by formula (\ref{ChCoulQHD def density with brackets}). In this section we neglect the delta functions existing in definitions of the many-particle hydrodynamic quantities, since we want to give simple presentation of our formulas and be focus on their application.

Applying the divergence in the spherical coordinates $\textmd{div} \textbf{a}\equiv \frac{1}{r^{2}}\partial_{r}(r^{2} a_{r})+\frac{1}{r\sin\theta}\partial_{\theta}(\sin\theta a_{\theta})+\frac{1}{r\sin\theta}\partial_{\varphi}a_{\varphi}$ we can rewrite the continuity equation (\ref{ChCoulQHD cont eq cylindr with j}) in rather usual form
\begin{equation}\label{ChCoulQHD cont eq sph with j via div} \partial_{t}n+\textmd{div} \textbf{j}=0.\end{equation}

The Euler equations in the spherical coordinates can be written as follows
\begin{equation}\label{ChCoulQHD} \partial_{t}j^{\alpha}+\biggl(\partial_{r}+\frac{2}{r}\biggr)\Pi^{\alpha r} +\biggl(\frac{1}{r}\partial_{\theta}+\frac{1}{r}ctg\theta\biggr)\Pi^{\alpha\theta}+\frac{1}{r\sin\theta}\partial_{\varphi}\Pi^{\alpha \varphi} =F^{\alpha}_{inertia}+F^{\alpha}_{int},\end{equation}
where $\alpha$ stands for $r$, $\theta$, and $\varphi$. $F^{\alpha}_{inertia}$ is the inertia force, which contains quantum contribution.

From equation (\ref{ChCoulQHD DtJ}) we see that interaction appears in the Euler equation as the commutator of the operator giving the particles current $\textbf{D}_{i}/m_{i}$ and the Hamiltonian of system under consideration. The time derivative of the Hamiltonian also gives contribution in the interaction. Other terms in (\ref{ChCoulQHD DtJ}) have kinematic nature. So it was then we worked in the Cartesian coordinates. Now we have deal with the spherical coordinates, which are curvilinear coordinates. Thus the commutator of $\textbf{D}_{i}/m_{i}$ with the kinetic energy operator gives us some terms, which do not contain any trace of interaction. We can call them the quantum inertia force, since the inertia force appears in classic hydrodynamics in similar way.

The quantum inertia force in the spherical coordinates has the following form
$$F_{r,inertia}=-\frac{\hbar^{2}}{2m^{2}}\biggl[\frac{1}{r^{2}}\partial_{r}n+\frac{1}{r^{3}}cth\theta\partial_{\theta}n$$
\begin{equation}\label{ChCoulQHD F inertia_r   sph} +\langle\frac{1}{r_{i}^{3}}(\Psi^{*}\partial_{\theta (i)}^{2}\Psi+c.c.)\rangle+\langle\frac{1}{r_{i}^{3}\sin^{2}\theta_{i}}(\Psi^{*}\partial_{\varphi (i)}^{2}\Psi+c.c.)\rangle\biggr],\end{equation}
$$F_{\theta,inertia}=\frac{\hbar^{2}}{4m^{2}}\biggl[-\frac{1}{r^{3}}\biggl(1+\frac{1}{\sin^{2}\theta}\biggr)\partial_{\theta}n$$
\begin{equation}\label{ChCoulQHD F inertia_theta sph} -2\langle\frac{1}{r_{i}^{3}}\frac{\cos\theta_{i}}{\sin^{3}\theta_{i}}(\Psi^{*}\partial_{\varphi (i)}^{2}\Psi+c.c.)\rangle +2\langle\frac{1}{r_{i}^{2}}(\Psi^{*}\partial_{r}^{i}\partial_{\theta}^{i}\Psi+c.c.)\rangle\biggr],\end{equation}
and
$$F_{\varphi,inertia}=\frac{\hbar^{2}}{4m^{2}}\biggl[-\frac{1}{\sin^{3}\theta}\frac{1}{r^{3}}\partial_{\varphi}n$$
\begin{equation}\label{ChCoulQHD F inertia_varphi sph} +2\langle\frac{1}{r_{i}^{2}\sin\theta_{i}}(\Psi^{*}\partial_{r}^{i}\partial_{\varphi}^{i}\Psi+c.c.)\rangle +2\langle\frac{1}{r_{i}^{3}}\frac{\cos\theta_{i}}{\sin^{2}\theta_{i}}(\Psi^{*}\partial_{\varphi}^{i}\partial_{\theta}^{i}\Psi+c.c.)\rangle\biggr].\end{equation}

Tensor of the momentum current has the following structure
\begin{equation}\label{ChCoulQHD} \Pi_{\alpha\beta}=-\frac{\hbar^{2}}{4m^{2}} \langle\partial_{\beta}^{i}\partial_{\alpha}^{i}\Psi^{*}\cdot\Psi -\partial_{\alpha}^{i}\Psi^{*}\cdot\partial_{\beta}^{i}\Psi -\partial_{\beta}^{i}\Psi^{*}\cdot\partial_{\alpha}^{i}\Psi +\Psi^{*}\cdot\partial_{\beta}^{i}\partial_{\alpha}^{i}\Psi\rangle, \end{equation}
where $\partial_{\alpha}^{i}$ and $\partial_{\beta}^{i}$ stand for $\partial_{\rho}^{i}$, $\frac{1}{\rho_{i}}\partial_{\varphi}^{i}$, $\partial_{z}^{i}$ for $i$th particle. So, for instance $\Pi_{\theta r}$ contains $\partial_{r}\biggl(\frac{1}{r}\partial_{\theta}\Psi^{*}\biggr)\Psi$, $\Pi_{\varphi r}$ contains $\partial_{r}\biggl(\frac{1}{r\sin\theta}\partial_{\varphi}\Psi^{*}\biggr)\Psi$,  and  $\Pi_{\varphi \theta}$ contains $\frac{1}{r}\partial_{\theta}\biggl(\frac{1}{r\sin\theta}\partial_{\varphi}\Psi^{*}\biggr)\Psi$.

An explicit form of the inertia force field in the spherical coordinates arises as
$$F_{r,inertia}=\frac{n}{r}(v_{\theta}^{2}+v_{\varphi}^{2})+\frac{1}{r}(p_{\theta\theta}+p_{\varphi\varphi})-\frac{\hbar^{2}}{2m^{2}}\biggl[\frac{1}{r^{2}}\partial_{r}n+\frac{1}{r^{3}}cth\theta\partial_{\theta}n\biggr]$$
\begin{equation}\label{ChCoulQHD F inertia_r   sph explicit}  -\frac{\hbar^{2}}{m^{2}}\frac{1}{r^{3}}\biggl(\sqrt{n}\partial_{\theta}^{2}\sqrt{n} +\frac{1}{\sin^{2}\theta}\sqrt{n}\partial_{\varphi}^{2}\sqrt{n}\biggr),\end{equation}

$$F_{\theta,inertia}=\frac{n}{r}\biggl(cth\theta v_{\varphi}^{2}-v_{r}v_{\theta}\biggr)+\frac{1}{r}\biggl(cth\theta p_{\varphi\varphi}-p_{r\theta}\biggr) -\frac{\hbar^{2}}{4m^{2}}\frac{1}{r^{3}}\biggl(1+\frac{1}{\sin^{2}\theta}\biggr)\partial_{\theta}n$$
\begin{equation}\label{ChCoulQHD F inertia_theta sph explicit} +\frac{\hbar^{2}}{m^{2}}\biggl(\frac{1}{r^{2}}\sqrt{n}\partial_{r}\partial_{\theta}\sqrt{n} -\frac{\cos\theta}{\sin^{3}\theta}\frac{1}{r^{3}}\sqrt{n}\partial_{\varphi}^{2}\sqrt{n}\biggr) ,\end{equation}
and
$$F_{\varphi,inertia}=-\frac{n}{r}\biggl(v_{r}v_{\varphi}+cth\theta v_{\varphi}v_{\theta}\biggr)-\frac{1}{r}\biggl(p_{r\varphi}+cth\theta p_{\varphi\theta}\biggr) -\frac{\hbar^{2}}{4m^{2}}\frac{1}{\sin^{3}\theta}\frac{1}{r^{3}}\partial_{\varphi}n$$
\begin{equation}\label{ChCoulQHD F inertia_varphi sph explicit} +\frac{\hbar^{2}}{m^{2}}\biggl(\frac{1}{\sin\theta}\frac{1}{r^{2}}\sqrt{n}\partial_{r}\partial_{\varphi}\sqrt{n} +\frac{\cos\theta}{\sin^{2}\theta}\frac{1}{r^{3}}\sqrt{n}\partial_{\varphi}\partial_{\theta}\sqrt{n}\biggr) .\end{equation}

Similarly to cylindrical case, the QHD in the spherical coordinates allows to introduce the velocity field. Applying presentation of the many-particle wave function via its phase and amplitude $\Psi=a\exp(\imath S/\hbar)$ to the explicit form of particle current (\ref{ChCoulQHD j r sph})-(\ref{ChCoulQHD j varphi sph}) we obtain $v_{r}=\hbar \langle a^{2} \partial_{r}^{i} S/m\rangle/n$, $v_{\theta}=\hbar\langle a^{2} \partial_{\theta} S/(rm)\rangle/n$, $v_{\varphi}=\hbar\langle a^{2} \partial_{\varphi} S/(r\sin\theta m)\rangle/n$, where $S(R,t)$ is the phase of the many-particle wave function $\Psi(R,t)$.

We also can introduce the velocity field in the spherical coordinates
\begin{equation}\label{ChCoulQHD}\begin{array}{ccc} j_{r}= nv_{r}  ,& j_{\theta}= nv_{\theta} ,& j_{\varphi}= nv_{\varphi}  \end{array}.\end{equation}

The continuity equation in term of the velocity field has form
\begin{equation}\label{ChCoulQHD cont eq sph with v} \partial_{t}n+\biggl(\partial_{r}+\frac{2}{r}\biggr)(nv_{r}) +\frac{1}{r}\biggl(\partial_{\theta}+ctg\theta\biggr)(nv_{\theta}) +\frac{1}{r\sin\theta}\partial_{\varphi}(nv_{\varphi})=0. \end{equation}

General equation for evolution of $v_{r}$ appears as
$$n\partial_{t}v_{r}+n(\textbf{v}\nabla)v_{r}-\frac{n}{r}(v_{\theta}^{2}+v_{\varphi}^{2})-\frac{1}{r}(p_{\theta\theta}+p_{\varphi\varphi})$$
$$+\biggl(\partial_{r}+\frac{2}{r}\biggr)\biggl[p_{rr}+\frac{\hbar^{2}}{2m^{2}}\biggl((\partial_{r}\sqrt{n})^{2}-\sqrt{n}\partial_{r}^{2}\sqrt{n}\biggr)\biggr]$$
$$+\frac{1}{r}\biggl(\partial_{\theta}+ctg\theta\biggr)\biggl[p_{r\theta}  +\frac{\hbar^{2}}{2m^{2}}\frac{1}{r}\biggl(\partial_{r}\sqrt{n}\cdot\partial_{\theta}\sqrt{n}-\sqrt{n}\partial_{\theta}\partial_{r}\sqrt{n}\biggr)\biggr]$$
$$+\frac{1}{r\sin\theta}\partial_{\varphi}\biggl[p_{r\varphi}  +\frac{\hbar^{2}}{2m^{2}}\frac{1}{r\sin\theta}\biggl(\partial_{\varphi}\sqrt{n}\cdot\partial_{r}\sqrt{n}-\sqrt{n}\partial_{\varphi}\partial_{r}\sqrt{n}\biggr)\biggr]$$
\begin{equation}\label{ChCoulQHD Euler sph general   r} +\frac{\hbar^{2}}{2m^{2}}\biggl[\frac{1}{r^{2}}\partial_{r}n+\frac{1}{r^{3}}cth\theta\partial_{\theta}n\biggr] +\frac{\hbar^{2}}{m^{2}}\frac{1}{r^{3}}\biggl(\sqrt{n}\partial_{\theta}^{2}\sqrt{n} +\frac{1}{\sin^{2}\theta}\sqrt{n}\partial_{\varphi}^{2}\sqrt{n}\biggr)=F_{r,int}, \end{equation}
where $F_{r,int}=qnE_{r}+q[\textbf{v},\textbf{B}]_{r}/c$.

General equation for evolution of $v_{\theta}$ arises as
$$n\partial_{t}v_{\theta}+n(\textbf{v}\nabla)v_{\theta}
-\frac{n}{r}\biggl(cth\theta v_{\varphi}^{2}-v_{r}v_{\theta}\biggr)-\frac{1}{r}\biggl(cth\theta p_{\varphi\varphi}-p_{r\theta}\biggr)$$
$$+\biggl(\partial_{r}+\frac{2}{r}\biggr)\biggl[p_{\theta r}  +\frac{\hbar^{2}}{2m^{2}}\frac{1}{r}\biggl(\partial_{r}\sqrt{n}\cdot\partial_{\theta}\sqrt{n}-\sqrt{n}\partial_{r}\partial_{\theta}\sqrt{n}\biggr)+\frac{\hbar^{2}}{4m^{2}}\frac{1}{r^{2}}\partial_{\theta}n\biggr]$$
$$+\frac{1}{r}\biggl(\partial_{\theta}+ctg\theta\biggr)\biggl[p_{\theta\theta}+\frac{\hbar^{2}}{2m^{2}}\frac{1}{r^{2}}\biggl((\partial_{\theta}\sqrt{n})^{2}-\sqrt{n}\partial_{\theta}^{2}\sqrt{n}\biggr)\biggr]$$
$$+\frac{1}{r\sin\theta}\partial_{\varphi}\biggl[p_{\theta\varphi}  +\frac{\hbar^{2}}{2m^{2}}\frac{1}{r^{2}\sin\theta}\biggl(\partial_{\varphi}\sqrt{n}\cdot\partial_{\theta}\sqrt{n}-\sqrt{n}\partial_{\varphi}\partial_{\theta}\sqrt{n}\biggr)\biggr]$$
\begin{equation}\label{ChCoulQHD Euler sph general   theta} +\frac{\hbar^{2}}{4m^{2}}\frac{1}{r^{3}}\biggl(1+\frac{1}{\sin^{2}\theta}\biggr) \partial_{\theta}n
-\frac{\hbar^{2}}{m^{2}}\biggl(\frac{1}{r^{2}}\sqrt{n}\partial_{r}\partial_{\theta}\sqrt{n} -\frac{\cos\theta}{\sin^{3}\theta}\frac{1}{r^{3}}\sqrt{n}\partial_{\varphi}^{2}\sqrt{n}\biggr)=F_{\theta,int}, \end{equation}
where $F_{\theta,int}=qnE_{\theta}+q[\textbf{v},\textbf{B}]_{\theta}/c$.

General equation for evolution of $v_{\varphi}$ arises as
$$n\partial_{t}v_{\varphi}+n(\textbf{v}\nabla)v_{\varphi} +\frac{n}{r}\biggl(v_{r}v_{\varphi}+cth\theta v_{\varphi}v_{\theta}\biggr)+\frac{1}{r}\biggl(p_{r\varphi}+cth\theta p_{\varphi\theta}\biggr)$$
$$+\biggl(\partial_{r}+\frac{2}{r}\biggr)\biggl[p_{\varphi r}  +\frac{\hbar^{2}}{2m^{2}}\frac{1}{r\sin\theta}\biggl(\partial_{r}\sqrt{n}\cdot\partial_{\varphi}\sqrt{n}-\sqrt{n}\partial_{r}\partial_{\varphi}\sqrt{n}\biggr)+\frac{\hbar^{2}}{4m^{2}}\frac{1}{r^{2}\sin\theta}\partial_{\varphi}n\biggr]$$
$$+\frac{1}{r}\biggl(\partial_{\theta}+ctg\theta\biggr)\biggl[p_{\varphi\theta}  +\frac{\hbar^{2}}{2m^{2}}\frac{1}{r^{2}\sin\theta}\biggl(\partial_{\theta}\sqrt{n}\cdot\partial_{\varphi}\sqrt{n}-\sqrt{n}\partial_{\theta}\partial_{\varphi}\sqrt{n}\biggr)+\frac{\hbar^{2}}{4m^{2}}\frac{1}{r^{2}}\frac{\cos\theta}{\sin^{2}\theta}\partial_{\varphi}n\biggr]$$
$$+\frac{1}{r\sin\theta}\partial_{\varphi}\biggl[p_{\varphi\varphi}+\frac{\hbar^{2}}{2m^{2}}\frac{1}{r^{2}\sin^{2}\theta}\biggl((\partial_{\varphi}\sqrt{n})^{2}-\sqrt{n}\partial_{\varphi}^{2}\sqrt{n}\biggr)\biggr]$$
\begin{equation}\label{ChCoulQHD Euler sph general   varphi} +\frac{\hbar^{2}}{4m^{2}}\frac{1}{\sin^{3}\theta}\frac{1}{r^{3}}\partial_{\varphi}n
-\frac{\hbar^{2}}{m^{2}}\biggl(\frac{1}{\sin\theta}\frac{1}{r^{2}}\sqrt{n}\partial_{r}\partial_{\varphi}\sqrt{n} +\frac{\cos\theta}{\sin^{2}\theta}\frac{1}{r^{3}}\sqrt{n}\partial_{\varphi}\partial_{\theta}\sqrt{n}\biggr)=F_{\varphi,int}, \end{equation}
where $F_{\varphi,int}=qnE_{\varphi}+q[\textbf{v},\textbf{B}]_{\varphi}/c$.

\subsection{\label{sec:level1} Dispersion of the quantum Langmuir waves propagating on spherical surface}

As an application of the quantum hydrodynamics in spherical coordinates derived in this section we consider a specific example of the 2DEG, it is electron gas located on a spherical surface. Fullerene like structure open possibilities for creation of new kinds of 2D objects. Nanotubes are also belong to family of fullerenes, but they have the cylindric geometry.

At fixed $r=R$ the set of QHD equations (\ref{ChCoulQHD cont eq sph with v}), (\ref{ChCoulQHD Euler sph general   r}), (\ref{ChCoulQHD Euler sph general   theta}), and (\ref{ChCoulQHD Euler sph general   varphi}) takes the following form
\begin{equation}\label{ChCoulQHD cont eq sph with v  rR} \partial_{t}n +\frac{1}{R}\biggl(\partial_{\theta}+ctg\theta\biggr)(nv_{\theta}) +\frac{1}{R\sin\theta}\partial_{\varphi}(nv_{\varphi})=0, \end{equation}

$$n\partial_{t}v_{\theta}+n(\textbf{v}\nabla)v_{\theta}
-\frac{n}{R}cth\theta v_{\varphi}^{2}-\frac{1}{R}cth\theta p_{\varphi\varphi}+\frac{\hbar^{2}}{2m^{2}}\frac{1}{R^{3}}\partial_{\theta}n$$
$$+\frac{1}{R}\biggl(\partial_{\theta}+ctg\theta\biggr)\biggl[p_{\theta\theta}+\frac{\hbar^{2}}{2m^{2}}\frac{1}{R^{2}}\biggl((\partial_{\theta}\sqrt{n})^{2}-\sqrt{n}\partial_{\theta}^{2}\sqrt{n}\biggr)\biggr]$$
$$+\frac{1}{R\sin\theta}\partial_{\varphi}\biggl[ \frac{\hbar^{2}}{2m^{2}}\frac{1}{R^{2}\sin\theta}\biggl(\partial_{\varphi}\sqrt{n}\cdot\partial_{\theta}\sqrt{n}-\sqrt{n}\partial_{\varphi}\partial_{\theta}\sqrt{n}\biggr)\biggr]$$
\begin{equation}\label{ChCoulQHD Euler v theta with v  rR} +\frac{\hbar^{2}}{4m^{2}}\frac{1}{R^{3}}\biggl(1+\frac{1}{\sin^{2}\theta}\biggr)\partial_{\theta}n
+\frac{\hbar^{2}}{m^{2}}\frac{\cos\theta}{\sin^{3}\theta}\frac{1}{R^{3}}\sqrt{n}\partial_{\varphi}^{2}\sqrt{n}=F_{\theta,int}, \end{equation}
where $F_{\theta,int}=qnE_{\theta}=-qn\frac{1}{R}\partial_{\theta}\widetilde{\phi}$,
and
$$n\partial_{t}v_{\varphi}+n(\textbf{v}\nabla)v_{\varphi} +\frac{n}{R}cth\theta v_{\varphi}v_{\theta} +\frac{\hbar^{2}}{2m^{2}}\frac{1}{R^{3}\sin\theta}\partial_{\varphi}n$$
$$+\frac{1}{R}\biggl(\partial_{\theta}+ctg\theta\biggr)\biggl[  \frac{\hbar^{2}}{2m^{2}}\frac{1}{R^{2}\sin\theta}\biggl(\partial_{\theta}\sqrt{n}\cdot\partial_{\varphi}\sqrt{n}-\sqrt{n}\partial_{\theta}\partial_{\varphi}\sqrt{n}\biggr)+\frac{\hbar^{2}}{4m^{2}}\frac{1}{R^{2}}\frac{\cos\theta}{\sin^{2}\theta}\partial_{\varphi}n\biggr]$$
$$+\frac{1}{R\sin\theta}\partial_{\varphi}\biggl[p_{\varphi\varphi}+\frac{\hbar^{2}}{2m^{2}}\frac{1}{R^{2}\sin^{2}\theta}\biggl((\partial_{\varphi}\sqrt{n})^{2}-\sqrt{n}\partial_{\varphi}^{2}\sqrt{n}\biggr)\biggr]$$
\begin{equation}\label{ChCoulQHD Euler v varphi with v  rR} +\frac{\hbar^{2}}{4m^{2}}\frac{1}{\sin^{3}\theta}\frac{1}{R^{3}}\partial_{\varphi}n
-\frac{\hbar^{2}}{m^{2}}\frac{\cos\theta}{\sin^{2}\theta}\frac{1}{R^{3}}\sqrt{n}\partial_{\varphi}\partial_{\theta}\sqrt{n}=F_{\varphi,int}, \end{equation}
where $F_{\varphi,int}=qnE_{\varphi}=-qn\frac{1}{R\sin\theta}\partial_{\varphi}\widetilde{\phi}$. We apply $M$ for mass of electrons in this subsection to distinguish it from the quantum number $m$.

Set of equations (\ref{ChCoulQHD cont eq sph with v  rR})-(\ref{ChCoulQHD Euler v varphi with v  rR}) allows us to consider dispersion of longitudinal waves in electron gas located on a spherical surface.
In this subsection we also assume that thermal pressure tensor is isotropic, hence diagonal elements are equal each other and nondiagonal elements equal to zero.

Let us present linearised form of equations (\ref{ChCoulQHD cont eq sph with v  rR}), (\ref{ChCoulQHD Euler v theta with v  rR}) and (\ref{ChCoulQHD Euler v varphi with v  rR})
\begin{equation}\label{ChCoulQHD cont eq sph with v  rR lin} \partial_{t}\delta n +\frac{n_{0}}{R}\biggl(\partial_{\theta}+ctg\theta\biggr)\delta v_{\theta} +\frac{n_{0}}{R\sin\theta}\partial_{\varphi}\delta v_{\varphi}=0, \end{equation}

$$n\partial_{t}\delta v_{\theta}+\frac{1}{R}\partial_{\theta}\delta p
+\frac{\hbar^{2}}{2m^{2}}\frac{1}{R^{3}}\partial_{\theta}\delta n$$
$$ -\frac{\hbar^{2}}{4m^{2}}\frac{1}{R^{3}}\biggl(\partial_{\theta}+ctg\theta\biggr)\partial_{\theta}^{2}\delta n -\frac{\hbar^{2}}{4m^{2}}\frac{1}{R^{3}\sin^{2}\theta} \partial_{\varphi}^{2}\partial_{\theta}\delta n$$
\begin{equation}\label{ChCoulQHD Euler v theta with v  rR lin} +\frac{\hbar^{2}}{4m^{2}}\frac{1}{R^{3}}\biggl(1+\frac{1}{\sin^{2}\theta}\biggr)\partial_{\theta}\delta n
+\frac{\hbar^{2}}{2m^{2}}\frac{\cos\theta}{\sin^{3}\theta}\frac{1}{R^{3}}\partial_{\varphi}^{2}\delta n=\delta F_{\theta,int}, \end{equation}

and

$$n\partial_{t}\delta v_{\varphi}+\frac{1}{R\sin\theta}\partial_{\varphi}\delta p +\frac{\hbar^{2}}{2m^{2}}\frac{1}{R^{3}\sin\theta}\partial_{\varphi}\delta n$$
$$+\frac{\hbar^{2}}{4m^{2}}\frac{1}{R^{3}}\biggl(\partial_{\theta}+ctg\theta\biggr)\biggl[  -\frac{1}{\sin\theta}\partial_{\theta}\partial_{\varphi}\delta n +\frac{\cos\theta}{\sin^{2}\theta}\partial_{\varphi}\delta n\biggr]$$
\begin{equation}\label{ChCoulQHD Euler v varphi with v  rR lin} -\frac{\hbar^{2}}{4m^{2}}\frac{1}{R^{3}\sin^{3}\theta}\partial_{\varphi}^{3}\delta n +\frac{\hbar^{2}}{4m^{2}}\frac{1}{\sin^{3}\theta}\frac{1}{R^{3}}\partial_{\varphi}\delta n
-\frac{\hbar^{2}}{2m^{2}}\frac{\cos\theta}{\sin^{2}\theta}\frac{1}{R^{3}}\partial_{\varphi}\partial_{\theta}\delta n= F_{\varphi,int}, \end{equation}
where perturbation of the force fields are related to perturbation of the electrostatic potential caused by the perturbations of the particle concentration
\begin{equation}\label{ChCoulQHD pot int in sph section} \widetilde{\phi}(\textbf{r},t)=q\int d\textbf{r}'\frac{\delta n(\textbf{r}',t)}{\mid \textbf{r}-\textbf{r}'\mid}, \end{equation}
where $\textbf{r}$ and $\textbf{r}'$ are point of space, which belong to the sphere of radius $R$.

We consider perturbations in electron gas located on a sphere of radius $R$. hence all physical quantities can be presented via expansion on spherical harmonics
\begin{equation}\label{ChCoulQHD} g(\theta,\varphi)=\sum_{l=0}^{+\infty}\sum_{m=-l}^{l}A_{lm}(R)\mathrm{Y}_{lm}(\theta,\varphi), \end{equation}
where
\begin{equation}\label{ChCoulQHD spherical harmonics} \mathrm{Y}_{lm}(\theta,\varphi)=\frac{1}{\sqrt{2\pi}}e^{\imath m\varphi}\Theta_{lm}(\cos\theta), \end{equation}
is the spherical harmonics, with
\begin{equation}\label{ChCoulQHD} \Theta_{lm}(\cos\theta)=\sqrt{\frac{2l+1}{2}\frac{(l-m)!}{(l+m)!}}P_{l}^{m}(\cos\theta), \end{equation}
and
\begin{equation}\label{ChCoulQHD} P_{l}^{m}(x)=(1-x^{2})^{m/2}\frac{d^{m}P_{l}(x)}{dx^{m}}, \end{equation}
where $-l\leq m\leq l$, and $P_{l}(\cos\theta)$ are the Legendre polynomials.

Let us present several examples of the spherical harmonics
\begin{equation}\label{ChCoulQHD} \mathrm{Y}_{l0}(\theta,\varphi)=\sqrt{\frac{2l+1}{4\pi}}P_{l}(\cos\theta), \end{equation}
\begin{equation}\label{ChCoulQHD} \mathrm{Y}_{10}(\theta,\varphi)=\sqrt{\frac{3}{4\pi}}\cos\theta, \end{equation}
and
\begin{equation}\label{ChCoulQHD} \mathrm{Y}_{1,\pm1}(\theta,\varphi)=\pm\sqrt{\frac{3}{8\pi}}\cos\theta e^{\pm\imath\varphi}. \end{equation}

Potential of the Coulomb interaction in spherical coordinates can be presented in the following form
\begin{equation}\label{ChCoulQHD} \frac{1}{\mid \textbf{r}-\textbf{r}'\mid} =\frac{1}{R}\sum_{l=0}^{\infty}\sum_{m=-l}^{l}\frac{4\pi}{2l+1}\mathrm{Y}_{lm}(\theta,\varphi)\mathrm{Y}_{lm}^{*}(\theta',\varphi'). \end{equation}

Simple case of the wave equation in the spherical coordinates at $r=R$ is
\begin{equation}\label{ChCoulQHD} \Delta_{\theta,\varphi}f+l(l+1)f=0, \end{equation}
this is the equation of wave on a spherical surface, where
\begin{equation}\label{ChCoulQHD} \Delta_{\theta,\varphi}f \equiv\frac{1}{\sin\theta}\partial_{\theta}(\sin\theta\partial_{\theta}f)+\frac{1}{\sin^{2}\theta}\partial_{\varphi}^{2}. \end{equation}

In our calculation of spectrum we have applied the following formula of differentiation of the spherical harmonics
\begin{equation}\label{ChCoulQHD sph form appl diff} \frac{d\Theta_{lm}(x)}{dx}=-\frac{mx}{1-x^{2}}\Theta_{lm}+\sqrt{\frac{l(l+1)-m(m+1)}{1-x^{2}}}\Theta_{l,m+1} , \end{equation}
and
\begin{equation}\label{ChCoulQHD} \partial_{\theta} \mathrm{Y}_{lm}(\theta,\varphi)=-\sin\theta\frac{e^{im\varphi}}{\sqrt{2\pi}}\frac{d\Theta_{lm}(x)}{dx}|_{x=\cos\theta}. \end{equation}
In formulas of differentiation of $\Theta_{lm}(x)$ we should put $\Theta_{lm}(x)=0$ at $m=\pm(l+1)$.

We also apply the following relation between spherical harmonics with different $m$
\begin{equation}\label{ChCoulQHD sph form appl rel of m} \frac{2mx}{\sqrt{1-x^{2}}}\Theta_{lm}= \sqrt{l(l+1)-m(m+1)}\Theta_{l,m+1} -\sqrt{l(l+1)-m(m-1)}\Theta_{l,m-1} . \end{equation}

From explicit form of spherical harmonics (\ref{ChCoulQHD spherical harmonics}) we find the formula of differentiation of the spherical harmonics on angle $\varphi$
\begin{equation}\label{ChCoulQHD} \partial_{\varphi} \mathrm{Y}_{lm}(\theta,\varphi)=im \mathrm{Y}_{lm}(\theta,\varphi). \end{equation}


\subsubsection{\label{sec:level1} Spectrum on Langmuir waves on sphere}

Considering equations (\ref{ChCoulQHD cont eq sph with v  rR lin}), (\ref{ChCoulQHD Euler v theta with v  rR lin}), and (\ref{ChCoulQHD Euler v varphi with v  rR lin}), after straightforward calculations we obtain
$$\omega^{2}R^{2}\delta n -l(l+1)\frac{en_{0}}{m}\delta\widetilde{\phi}-\frac{l(l+1)}{m}\delta p -l(l+1)\frac{3}{4}\frac{\hbar^{2}}{m^{2}R^{2}} \delta n -\frac{\hbar^{2}}{4m^{2}}\frac{[l(l+1)]^{2}}{R^{2}}\delta n$$
$$-\frac{\hbar^{2}}{2m^{2}}\frac{1}{R^{2}}(1+m^{2})\frac{ctg\theta}{\sin^{2}\theta}\partial_{\theta}\delta n+\frac{\hbar^{2}}{m^{2}}\frac{1}{R^{2}}(\frac{3}{4}+ctg^{2}\theta)\frac{m^{2}}{\sin^{2}\theta}\delta n$$
\begin{equation}\label{ChCoulQHD disp eq nonExpl on sph} +\frac{\hbar^{2}}{4m^{2}}\frac{1}{R^{2}} \biggl(\frac{m^{2}}{\sin^{4}\theta}-\frac{l(l+1)}{\sin^{2}\theta}\biggr)\delta n=0. \end{equation}
The five first terms in equation (\ref{ChCoulQHD disp eq nonExpl on sph}) have constant coefficients, then the other terms contain coefficients explicitly depending on angle $\theta$ and one of them contains derivative on angle $\theta$. Angle depending terms arise from the inertia force $F^{\theta}_{inertia}$. These terms leads to interinfluence of different modes $\delta n_{lm}$. In absence of the angle depending terms we would have infinite set of independent equations for $\delta n_{lm}$. We can get rid of angle dependence of terms in equations (\ref{ChCoulQHD disp eq nonExpl on sph}) applying formulae (\ref{ChCoulQHD sph form appl diff}), (\ref{ChCoulQHD sph form appl rel of m}). Hence we obtain mix of $\delta n_{lm}$ with $\delta n_{lm\pm2}$, $\delta n_{lm\pm4}$ in one equation. Thus we have coupled set of algebraic equation for $\delta n_{lm}$ at different $m\in[-l,l]$. We do not have any mix at $l=0$ and $l=1$, since $m\pm2$ and $m\pm4$ do not exist in this case. Quantities $l=2$ and $l=3$ open possibility for mix with $m\pm2$ states. At larger $l$ we have possibility of mix of $\delta n_{lm}$ with $\delta n_{lm\pm2}$, $\delta n_{lm\pm4}$. Nevertheless we neglect below intermode influence at all quantum numbers $l$.

\begin{equation}\label{ChCoulQHD} \int \mathrm{Y}_{lm}(\theta',\varphi')\mathrm{Y}_{l'm'}^{*}(\theta',\varphi') d\Omega'=\delta_{ll'}\delta_{mm'}, \end{equation}
where $d\Omega'=\sin\theta' d\theta' d\varphi'$ is an element of spherical angle.

Working on a sphere surface we should present expansion of the Coulomb potential having spherical symmetry
\begin{equation}\label{ChCoulQHD} \int\frac{\delta n}{\mid \textbf{r}-\textbf{r}'\mid}d\textbf{r}'=\sum_{l=0}^{\infty}\sum_{m=-l}^{l}\frac{4\pi R}{2l+1} \delta n_{lm} \mathrm{Y}_{lm}(\theta,\varphi) .\end{equation}

\begin{equation}\label{ChCoulQHD} \left(\begin{array}{c}\delta n\\
\delta p\\
\widetilde{\phi} \\
\end{array}\right) = \sum_{l=0}^{\infty}\sum_{m=-l}^{l} \left(\begin{array}{c} n_{lm}\\
p_{lm}\\
\widetilde{\phi}_{lm} \\
\end{array}\right) \mathrm{Y}_{lm}(\theta,\varphi). \end{equation}

Since we consider electron gas on a sphere, we need an equation of state $p=p(n)$ to get closed set of QHD equations. Hence we derived the following pressure for ideal gas of degenerate particles on a sphere with radius $R$:
\begin{equation}\label{ChCoulQHD} P_{Fe,Sph}=\frac{\pi^{2}R^{2}\hbar^{2}}{4m}n_{2D}^{3}, \end{equation}
we also obtain the Fermi velocity of particles on sphere $v_{Fe,Sph}=\pi R\hbar n_{2D}/(2m)$, and corresponding Fermi energy $E_{Fe,Sph}=\pi^{2}R^{2}\hbar^{2}n_{2D}^{2}/(8m)$.

In Refs. \cite{Spherical Quant eq state 01}, \cite{Spherical Quant eq state 02} authors use usual 3D Fermi pressure $p=(3n/8\pi)^{2/3} (4\pi^{2}\hbar^{2}/3m) n$ for particles on a sphere. It was also used for particles on a cylinder (see Refs. \cite{Sph}, \cite{on cylindre  quantum no inertia END 2}).

First of all we present spectrum of collective excitations related to the Coulomb interaction and the fermi pressure neglecting by the quantum Bohm potential and quantum part of inertia forces
\begin{equation}\label{ChCoulQHD Langmuir zer ord spectr on Sph} \omega^{2}=\frac{l(l+1)}{2l+1}\omega_{Le,Sp}^{2}+l(l+1)\frac{3\pi\hbar^{2}n_{0,2D}^{2}}{4m^{2}}, \end{equation}
where
\begin{equation}\label{ChCoulQHD Langmuir freq on sph} \omega_{Le,Sp}^{2}=\frac{4\pi e^{2}n_{0,2D}}{m}\frac{1}{R} \end{equation}
is the Langmuir frequency for two-dimensional electron gas on the sphere of radius $R$.

Let us consider the first term in spectrum (\ref{ChCoulQHD Langmuir zer ord spectr on Sph}). At large $l$ it increases as $\frac{l(l+1)}{2l+1}\rightarrow l/2$. Including presence of $1/r$ in the Langmuir frequency (\ref{ChCoulQHD Langmuir freq on sph}) we see that the first term in formula (\ref{ChCoulQHD Langmuir zer ord spectr on Sph}) is proportional to $k_{eff}=l/R$ similarly to spectrum of the plane 2DEG (\ref{ChCoulQHD Langmuir frq 2D}).

We also see that the second term describing contribution of the Fermi pressure is proportional to $k_{eff}^{2}$ as we usually have for the pressure contribution.

Including contribution of the theta-inertia force we find
\begin{equation}\label{ChCoulQHD} \omega^{2}=\frac{l(l+1)}{2l+1}\omega_{Le,Sp}^{2}+l(l+1)\frac{3\pi\hbar^{2}n_{0,2D}^{2}}{4m^{2}}+\frac{\hbar^{2}}{4m^{2}}\frac{1}{R^{4}} l(l+1)(3+l(l+1))+\widetilde{\Omega}^{2},  \end{equation}
where $\widetilde{\Omega}^{2}$ is caused by the $F^{\theta}_{inertia}$ has the following rather huge explicit form
$$\widetilde{\Omega}^{2}= \frac{\hbar^{2}}{M^{2}R^{4}}\biggl[\frac{1}{8}(Z_{m,m+1}^{l}+Z_{m,m-1}^{l})\biggl(m(1-9m/2+m^{2})+l(l+1)/2\biggr)$$
$$-\frac{1}{32}m(1-5m/2+m^{2})\biggl(Z_{m,m+1}^{l}Z^{l}_{m+1,m+2}+(Z^{l}_{m,m+1}+Z^{l}_{m,m-1})^{2}+Z^{l}_{m,m-1}Z^{l}_{m-1,m-2}\biggr)$$
\begin{equation}\label{ChCoulQHD} m^{2}-\frac{1}{4}l(l+1)  +\frac{1}{4}(m+1)(m^{2}+1)Z_{m,m+1}^{l}\biggl(\frac{1}{4}(Z^{l}_{m+1,m+2}+Z^{l}_{m,m+1}+Z^{l}_{m,m-1})-1\biggr) \biggr],\end{equation}
with
\begin{equation}\label{ChCoulQHD} Z_{m,m+1}^{l}=\frac{l(l+1)-m(m+1)}{m(m+1)} .\end{equation}

\section{\label{sec:Concl} Conclusions and further generalizations}

This paper has described basic principles of the MPQHD on simples example: the Coulomb interaction. The method of MPQHD is based on using of explicit form of the Schrodinger equation and Hamiltonian of particles. So it can be used up to the semi-relativistic approximation only, since we do not have a convenient many-particle Dirac equation. Thus, we have, in literature, application of the MPQHD to the semi-relativistic effects \cite{pavel}, \cite{Andreev IJMP B 12}, \cite{Ivanov arxiv 2012}, \cite{Ivanov IJMP B 14}.

Nevertheless, there is another way of generalization of described scheme. There is a wide spread opinion that hydrodynamic equations appear via a kinetic equation. When we get equations for moments of the distribution function, which are product of the momentum of different degree on the distribution function integrated over the momentum. However we have shows that the QHD equations can be derived from the quantum mechanics directly, with no use of the kinetics. It is possible to derive the classic hydrodynamics directly from the Newton equations, with no use of kinetic equations using proper definition of the particle concentration \cite{Kuz'menkov 12}.

Nevertheless kinetics allows to obtained detailed information about thermal properties of the system, which affects many collective properties of plasmas. Consequently it is worthwhile to have quantum kinetics. And it is interesting to know: can we derive quantum kinetic and QHD from the same principles, as it was done  in classic physics \cite{Kuz'menkov 91}, \cite{Kuz'menkov 12}? We can give positive answer on this question \cite{Andreev kinetics 12}, \cite{Andreev kinetics 13}, \cite{Andreev annigilation}. The principles described in this paper can be used to derive the quantum kinetics. In this case we should start from definition of the quantum distribution function operator
\begin{equation}\label{ChCoulQHD distrib func operator} \hat{f}=\sum_{i}\delta(\textbf{r}-\widehat{\textbf{r}}_{i})\delta(\textbf{p}-\widehat{\textbf{p}}_{i}).\end{equation}
Quantum mechanical averaging of the operator of many-particle distribution function gives us the microscopic distribution function for system of spinning particles \cite{Andreev kinetics 12}, \cite{Andreev kinetics 13}
$$f(\textbf{r}, \textbf{p},t)=\frac{1}{4}\int \Biggl(\psi^{*}(R,t)\sum_{i}\biggl(\delta(\textbf{r}-\textbf{r}_{i})\delta(\textbf{p}-\widehat{\textbf{p}}_{i})$$
\begin{equation}\label{ChCoulQHD distribution function def} +\delta(\textbf{p}-\widehat{\textbf{p}}_{i})\delta(\textbf{r}-\textbf{r}_{i})\biggr)\psi(R,t)+h.c.\Biggr)dR.\end{equation}

The method of MPQHD is a representation of many-particle quantum mechanics derived rigorously from the many-particle Schrodinger equation. It is useful up to the semi-relativistic approximation of quantum hydrodynamics and quantum kinetics of various physical systems.


\begin{thebibliography}{99}


\bibitem{Madelung ZP 26} E. Madelung, Z. Phys. \textbf{40}, 332 (1926).

\bibitem{Takabayasi PTP 55} T. Takabayasi, Progr. Theor. Phys. \textbf{14}, 283 (1955).


\bibitem{Recami PRA 98} E. Recami and G. Salesi, Phys. Rev. À \textbf{54}, 98 (1998).
\bibitem{Wong JMP 76} C. X. Wong, J. Math. Phys. \textbf{17}, 1008 (1976).
\bibitem{Bialynicki PRD 71} I. Bialynicki-Birula and Z. Bialynicki-Birula, Phys. Rev. D \textbf{3}, 2410 (1971).
\bibitem{Sanin OS 96} A. L. Sanin, Opt. Spektrosk. \textbf{80}, 540 (1996).

\bibitem{Rand PF 64} S. Rand, Phys. Fluids \textbf{7}, 64 (1964).

\bibitem{Bohm PR 52} David Bohm, Phys. Rev. \textbf{85}, 166 (1952).


\bibitem{MaksimovTMP 1999} L. S. Kuz'menkov and S. G. Maksimov,
Theoretical and Mathematical Physics \textbf{118}, 227 (1999).

\bibitem{Klimontovich book} Yu. L. Klimontovich, \emph{Statistical Physics} [in Russian], Nauka, Moscow (1982); English transl., Harwood, New
York (1986).



\bibitem{Weinberg book} S. Weinberg, \emph{Gravitation and Cosmology} (John Wiley and Sons, Inc., New York, 1972).

\bibitem{Drofa1996} M. A. Drofa, L. S. Kuzmenkov, Theor. Math. Phys. \textbf{108}, N1, 849 (1996).


\bibitem{Kuz'menkov 12} L. S. Kuz'menkov, P. A. Andreev, PIERS Proceedings, p.158, August 19-23, Moscow, Russia (2012).


\bibitem{Kuz'menkov 91} L. S. Kuz'menkov, Theoretical and Mathematical
Physics \textbf{86}, 159 (1991).




\bibitem{Shukla RMP 11} P. K. Shukla, B. Eliasson,
Rev. Mod. Phys. \textbf{83},  885 (2011).

\bibitem{Shukla UFN 10} P. K. Shukla, B. Eliasson, Phys. Usp. \textbf{53} 51
(2010) [Uspehi Fizihceskih Nauk \textbf{180}, 55 (2010)].



\bibitem{MaksimovTMP 2001} L. S. Kuz'menkov, S. G. Maksimov, and V. V. Fedoseev, Theor.
Math. Fiz. \textbf{126} 136 (2001) [Theoretical and Mathematical
Physics, \textbf{126} 110 (2001)].


\bibitem{MaksimovTMP 2001 b} L. S. Kuz'menkov, S. G. Maksimov, and V. V. Fedoseev, Theor.
Math. Fiz. \textbf{126} 258 (2001) [Theoretical and Mathematical
Physics, \textbf{126} 212 (2001)].


\bibitem{Andreev Asenjo 13} P. A. Andreev,
F. A. Asenjo, and S. M. Mahajan, arXiv: 1304.5780.


\bibitem{Trukhanova PTEP 13} M. Iv. Trukhanova,  Progr. Theor. Exp. Phys. \textbf{2013}, 111I01 (2013).

\bibitem{pavel} P. A. Andreev and L. S. Kuz'menkov, Russian Phys. Jour. {\bf 50}, 1251 (2007).

\bibitem{Andreev IJMP B 12} P. A. Andreev, L. S. Kuz'menkov,  Int. J. Mod. Phys. B \textbf{26},  1250186 (2012);
Andreev P. A., Kuz'menkov L. S., The quantum hydrodynamics equation with
the spin-current and spin-orbit interaction, Dynamic of complex systems
(Dinamika slojnykh sistem), Vol. \textbf{3}, N. 3, pp. 3-30, 2009 (in Russian).

\bibitem{Andreev annigilation} Pavel A. Andreev, arXiv:1404.4899.

\bibitem{Ivanov arxiv 2012} A. Yu. Ivanov, P. A. Andreev,
Russian Physics Journal, \textbf{56}, 325 (2013);
A. Yu. Ivanov, P. A. Andreev, and L. S. Kuzmenkov, arXiv:1209.6124.

\bibitem{Ivanov IJMP B 14} A. Yu. Ivanov, P. A. Andreev, and L. S. Kuzmenkov, Int. J. Mod. Phys. B \textbf{28}, 1450132 (2014).



\bibitem{Andreev PRA08} P. A. Andreev, L. S. Kuz'menkov, Phys. Rev. A \textbf{78}, 053624 (2008).
\bibitem{Andreev IJMP B 13} P. A. Andreev, Int. J. Mod. Phys. B \textbf{27}, 1350017 (2013).

\bibitem{Andreev PRB 11} P. A. Andreev, L. S. Kuz'menkov, M. I. Trukhanova, Phys. Rev. B \textbf{84}, 245401 (2011).

\bibitem{Andreev 2013 non-int GP} P. A. Andreev, Mod. Phys. Lett. B \textbf{27}, 1350096 (2013).

\bibitem{Andreev 2013 non-int GP} P. A. Andreev, Mod. Phys. Lett. B \textbf{27}, 1350096 (2013).


\bibitem{Andreev EPJ D Pol} P. A. Andreev and L. S. Kuz'menkov, Eur. Phys. J. D \textbf{67}, 216 (2013).


\bibitem{Uzdensky arxiv review 14} D. A. Uzdensky and S. Rightley, Reports on Progress in
Physics, \textbf{77}, Issue 3, 036902 (2014).


\bibitem{Koide PRC 13} T. Koide, Phys. Rev. C \textbf{87}, 034902 (2013).




\bibitem{Kuzmenkov RPJ 04} L. S. Kuz'menkov and D. E. Kharabadze, Russ. Phys. J., No. 4, 437 (2004).


\bibitem{Polyakov 79} P. A. Polyakov, Soviet Physics Jour. {\bf 22}, 310 (1979);
D. V. Vagin, N. E. Kim, P. A. Polyakov, A. E. Rusakov,
Izvestiya RAN (Proceedings of Russian Academy of science) \textbf{70}, 443 (2006).

\bibitem{Andreev VestnMSU 2007} P. A. Andreev, L. S. Kuz'menkov,
Moscow University Physics Bulletin \textbf{62}, N.5, 271 (2007).

\bibitem{Marklund PRL 07} M. Marklund and G. Brodin, Phys. Rev. Lett. \textbf{98}, 025001 (2007).
\bibitem{brodinMHD} G. Brodin and M. Marklund, New J. Phys. \textbf{9}, 277 (2007).








\bibitem{Brodin PRL 08 a} G. Brodin, M. Marklund, and G. Manfredi, Phys. Rev. Lett. \textbf{100}, 175001 (2008).


\bibitem{Misra JPP 10} A. P. Misra, G. Brodin, M. Marklund and P. K. Shukla, J. Plasma Physics
\textbf{76}, 857 (2010).

\bibitem{Mahajan PRL 11} S. M. Mahajan and F. A. Asenjo, Phys. Rev. Lett. \textbf{107},
195003 (2011).
\bibitem{Mahajan PL A 13} S. M. Mahajan, F. A. Asenjo, Phys. Lett. A \textbf{377}, 1430 (2013).


\bibitem{Andreev spin-up and spin-down 1405} P. A. Andreev, arXiv:1405.0719.

\bibitem{Asenjo kinetic 12} F. A. Asenjo, J. Zamanian, M. Marklund, G. Brodin,
and P. Johansson, New J. Phys. \textbf{14}, 073042 (2012).



\bibitem{Haas PRE 00} F. Haas, G. Manfredi, M. Feix, Phys. Rev. E \textbf{62},
2763(2000).


\bibitem{Wigner PR 84} M. Hillery, R. F. O'Connell, M. O. Scully, E. P. Wigner, Physics Reports, \textbf{106}, 121  (1984).



\bibitem{Krishnaswami PRL and arXiv 14} G. S. Krishnaswami, R. Nityananda, A. Sen, and
A. Thyagaraja, Phys. Rev. Lett. \textbf{112}, 129501 (2014);
G. S. Krishnaswami, R. Nityananda, A. Sen, and A. Thyagaraja, arXiv:1407.6865.



\bibitem{Landau Vol 3} L. D. Landau, E. M. Lifshitz,
\emph{Quantum Mechanics: Non-Relativistic Theory}. Vol. 3 (3rd ed.). Pergamon Press,  (1977).



\bibitem{Andreev 1401 finite ions} P. A. Andreev and L. S. Kuz'menkov, arXiv:1401.3224.


\bibitem{Kroll book} N. A. Krall and A. W. Trivelpiece, {\it Principles of Plasma Physics} (McGraw-Hill, Inc. 1973).



\bibitem{L.P.Pitaevskii RMP 99} F. Dalfovo, S. Giorgini, L. P. Pitaevskii,
and S. Stringari,  Rev. Mod. Phys. \textbf{71}, 463 (1999).
\bibitem{Giorgini RMP 08} S. Giorgini, L. P. Pitaevskii,
and S. Stringari,  Rev. Mod. Phys. \textbf{80}, 1215 (2008).

\bibitem{Yi PRA 00} S. Yi and L. You, Phys. Rev. A, \textbf{61}, 041604(R) (2000).
\bibitem{Goral PRA 00} K. Goral, K. Rzazewski, and T.
Pfau, Phys. Rev. A \textbf{61}, 051601(R) (2000).
\bibitem{Santos PRL 00} L. Santos, G. V. Shlyapnikov, P. Zoller, and M.
Lewenstein, Phys. Rev. Lett. \textbf{85}, 1791 (2000).
\bibitem{Yi PRA 01} S. Yi and L. You, Phys. Rev. A, \textbf{63}, 053607 (2001).

\bibitem{Quemener CR 12} Goulven Quemener, and Paul S. Julienne, Chem. Rev. \textbf{112}, 4949 (2012). \textbf{???}
\bibitem{Ho PRL 98} T. L. Ho, Phys. Rev. Lett. \textbf{81}, 742 (1998).
\bibitem{Machida JPSJ 98} T. Ohmi and K. Machida, J. Phys. Soc. Jpn. \textbf{67},
1822 (1998).
\bibitem{Kawaguchi Ph Rep 12} Yuki Kawaguchi, Masahito Ueda, Physics Reports \textbf{520},  253 (2012).
\bibitem{Stamper-Kurn RMP 13} Dan M. Stamper-Kurn, Masahito Ueda, Rev. Mod. Phys. \textbf{85}, 1191 (2013).



\bibitem{Landau Vol 10} L. D. Landau, E. M. Lifshitz, \emph{Physical kinetics}. Vol. 10. Pergamon Press,  (1977).

\bibitem{Rukhadze book 9}  A. F. Alexandrov, L. S. Bogdankevich,
and A. A. Rukhadze, \textit{Principles of Plasma Electrodynamics} (Vysshaya Shkola, Moscow, 1978; Springer-Verlag, Berlin, 1984).


\bibitem{Khan JAP 14} S. A. Khan and Sunia Hassan, Journal of Applied Physics \textbf{115}, 204304 (2014). 


\bibitem{Landau v5 eq st} L. Landau and E. M. Lifshitz, Statistical Physics (Pergamon, New York,
1980).  
\bibitem{Mushtaq PP 12} A. Mushtaq, R. Maroof, Zulfiaqr Ahmad, and A. Qamar, Phys. Plasmas \textbf{19}, 052101
(2012).
\bibitem{Mushtaq PP EPJD 11} A. Mushtaq, and S. V. Vladimirov, Eur. Phys. J. D \textbf{64}, 419 (2011).
\bibitem{Asenjo PL A 12} F. A. Asenjo, Phys. Lett. A \textbf{376}, 2496 (2012).
\bibitem{Masood PP 11} W. Masood, H. Rizvi, and M. Siddiq
Phys. Plasmas \textbf{18}, 102316 (2011).
\bibitem{Mushtaq PP 10} A. Mushtaq, and S. V. Vladimirov, Phys. Plasmas \textbf{17}, 102310 (2010).



\bibitem{Andreev spin-up and spin-down p2} P. A. Andreev, L. S. Kuz'menkov, arXiv:1406.6252.


\bibitem{Haas PP 03} F. Haas, L. G. Garcia, J. Goedert, and G. Manfredi, Phys. Plasmas  \textbf{10}, 3858 (2003).





\bibitem{unmodified pressure beg}
S. A. Khan and H. Saleem, Physics of Plasmas  \textbf{16}, 052109 (2009).

\bibitem{}
W. Masood, S. Karim and H. A. Shah, Phys. Scr. \textbf{82}, 045503 (2010);
W. Masood,
Physics of Plasmas  \textbf{17}, 052312 (2010).

\bibitem{}
Shi-Sen Ruan, ShanWu, Majid Raissan, Ze Cheng, Astrophys Space Sci \textbf{346}, 431 (2013).

\bibitem{}
S. Hussain, S. Mahmood, A. Mushtaq, Astrophys Space Sci \textbf{346}:359 (2013);
Arshad M. Mirza and W. Masood, Physics of Plasmas \textbf{18}, 122701 (2011).

\bibitem{}
A. E. Dubinov, and A. A. Dubinova, Plasma Physics Reports \textbf{34}, 403 (2008).

\bibitem{unmodified pressure end}
Arshad M. Mirza, W. Masood, Tufail A. Khan, Astrophys Space Sci. \textbf{346}, 279 (2013).











\bibitem{represented Fermi 3D beg}
A. Mushtaq and S. A. Khan,
Physics of Plasmas  \textbf{14}, 052307 (2007).

\bibitem{}
C. Bhowmik, A. P. Misra, and P. K. Shukla,
Physics of Plasmas  \textbf{14}, 122107 (2007).

\bibitem{}
Q. Haque, S. Mahmood, and A. Mushtaq,
Physics of Plasmas  \textbf{15}, 082315 (2008);
Q. Haque and S. Mahmood,
Physics of Plasmas  \textbf{15}, 034501 (2008);
Q. Haque, Physics of Plasmas  \textbf{15}, 094502 (2008).

\bibitem{}
A. P. Misra and S. Samanta,
Physics of Plasmas \textbf{15}, 122307 (2008).

\bibitem{}
M. Sadiq, S. Ali, and R. Sabry,
Physics of Plasmas  \textbf{16}, 013706 (2009).

\bibitem{}
W. Masood, Shahid M. Mirza, and Arshad M. Mirza,
Physics of Plasmas  \textbf{16}, 084503 (2009).

\bibitem{}
A. S. Bains, A. P. Misra, N. S. Saini, and T. S. Gill, Physics of Plasmas  \textbf{17}, 012103 (2010).

\bibitem{}
S. Mahmood, S. A. Khan, and H. Ur-Rehman, Physics of Plasmas  \textbf{17}, 112312 (2010);
S. Mahmood, N. Akhtar, and H. Ur-Rehman, Phys. Scr. \textbf{83}, 035505 (2011).


\bibitem{}
S. Hussain and S. Mahmood, Physics of Plasmas  \textbf{18}, 082109 (2011);
N. Akhtar and S. Mahmood, Physics of Plasmas \textbf{18}, 112506 (2011);
S. Mahmood, Q. Haque, Physics Letters A \textbf{374}, 872 (2010).

\bibitem{}
Utpal Kumar Samanta, Asit Saha, Prasanta Chatterjee, Astrophys Space Sci. \textbf{347}, 293 (2013).

\bibitem{}
S. Mahmood, N. Akhtar, and S. A. Khan, J. Plasma Physics \textbf{78},  3 (2012)

\bibitem{}
Amar P. Misra, S. Samanta, And A. R. Chowdhury, J. Plasma Physics, \textbf{74}, 197 (2008).

\bibitem{represented Fermi 3D end}
A. P. Misra, N. K. Ghosh, Astrophys Space Sci. \textbf{331}, 605 (2011)









\bibitem{relativistic   pressure beg}
M. Akbari-Moghanjoughi, Physics of Plasmas  \textbf{18}, 072702 (2011).

\bibitem{}
M. Akbari-Moghanjoughi, Physics of Plasmas  \textbf{18}, 082706 (2011).

\bibitem{Rel exch 1}
M. Akbari-Moghanjoughi, Physics of Plasmas  \textbf{19}, 042701 (2012).

\bibitem{Rel exch 2}
M. Akbari-Moghanjoughi, Physics of Plasmas \textbf{20}, 042706 (2013).

\bibitem{}
Swarniv Chandra, Basudev Ghosh, Astrophys Space Sci. \textbf{342}, 417 (2012).

\bibitem{}
S. A. Khan, Indian J. Phys. \textbf{88}, 433 (2014).

\bibitem{relativistic   pressure end}
M. Akbari-Moghanjoughi, and P. K. Shukla, Physical Review E \textbf{86}, 066401 (2012).




\bibitem{mixed eq state 1}
Prerana Sharma and R. K. Chhajlani, Physics of Plasmas  \textbf{20}, 092101 (2013).

\bibitem{mixed eq state 2}
Haijun Ren, Zhengwei Wu, Jintao Cao, and Paul K. Chu, J. Phys. A: Math. Theor. \textbf{41}, 115501 (2008)

\bibitem{Sharma PP 14} Prerana Sharma and R. K. Chhajlani, Phys. Plasmas \textbf{21}, 032101 (2014).














\bibitem{mod over 3 beg}
P. K. Shukla and S. Ali,
Physics of Plasmas \textbf{12}, 114502 (2005).

\bibitem{}
S. Ali and P. K. Shukla,
Physics of Plasmas \textbf{13}, 022313 (2006).

\bibitem{}
Amar P. Misra and A. Roy Chowdhury,
Physics of Plasmas \textbf{13}, 072305 (2006).

\bibitem{}
S. Ali and P. K. Shukla,
Physics of Plasmas \textbf{13}, 102112 (2006).

\bibitem{}
W. F. El-Taibany and Miki Wadati,
Physics of Plasmas \textbf{14}, 042302 (2007).

\bibitem{}
W. M. Moslem, P. K. Shukla, S. Ali, and R. Schlickeiser,
Physics of Plasmas \textbf{14}, 042107 (2007).

\bibitem{}
S. Ali, W. M. Moslem, P. K. Shukla, and R. Schlickeiser,
Physics of Plasmas \textbf{14}, 082307 (2007).

\bibitem{}
Amar P. Misra and P. K. Shukla,
Physics of Plasmas \textbf{14}, 082312 (2007).

\bibitem{}
Biswajit Sahu and Rajkumar Roychoudhury,
Physics of Plasmas  \textbf{14}, 072310 (2007).

\bibitem{}
S. A. Khan and A. Mushtaq,
Physics of Plasmas  \textbf{14}, 083703 (2007).


\bibitem{}
W. M. Moslem, S. Ali, P. K. Shukla, X. Y. Tang, and G. Rowlands,
Physics of Plasmas \textbf{14}, 082308 (2007).

\bibitem{}
W. Masood, A. Mushtaq, and R. Khan,
Physics of Plasmas  \textbf{14}, 123702 (2007).

\bibitem{}
S. Mahmood, Physics of Plasmas \textbf{15}, 014502 (2008).

\bibitem{}
W. Masood and A. Mushtaq, Physics of Plasmas \textbf{15}, 022306 (2008).

\bibitem{}
S. A. Khan and W. Masood, Physics of Plasmas 15, 062301 (2008).

\bibitem{}
Q. Haque and H. Saleem, Physics of Plasmas  \textbf{15}, 064504 (2008).

\bibitem{}
S. A. Khan, S. Mahmood, and H. Saleem,
Physics of Plasmas  \textbf{15}, 082303 (2008).

\bibitem{}
H. Ur-Rehman, S. A. Khan, W. Masood, and M. Siddiq,
Physics of Plasmas  \textbf{15}, 124501 (2008).

\bibitem{}
S. Mahmood and W. Masood,
Physics of Plasmas  \textbf{15}, 122302 (2008).


\bibitem{}
R. Sabry, W. M. Moslem, F. Haas, S. Ali, and P. K. Shukla,
Physics of Plasmas \textbf{15}, 122308 (2008).

\bibitem{}
Ch. Uzma, I. Zeba, H. A. Shah, and M. Salimullah,
Journal of Applied Physics \textbf{105}, 013307 (2009).

\bibitem{}
S. A. Khan, W. Masood, and M. Siddiq,
Physics of Plasmas \textbf{16}, 013701 (2009).

\bibitem{}
W. Masood, M. Siddiq, Shahida Nargis, and Arshad M. Mirza
Physics of Plasmas  \textbf{16}, 013705 (2009).

\bibitem{}
Om Prakash Sah, Physics of Plasmas  \textbf{16}, 012105 (2009).

\bibitem{}
O. P. Sah and J. Manta, Physics of Plasmas  \textbf{16}, 032304 (2009).

\bibitem{}
M. Salimullah, M. Jamil, I. Zeba, Ch. Uzma, and H. A. Shah,
Physics of Plasmas  \textbf{16}, 034503 (2009).

\bibitem{}
M. Salimullah, I. Zeba, Ch. Uzma, and M. Jamil,
Physics of Plasmas \textbf{16}, 033703 (2009);
S. A. Khan, S. Mahmood, and S. Ali,
Physics of Plasmas \textbf{16}, 044505 (2009).

\bibitem{}
Prasanta Chatterjee, Kaushik Roy, Sithi V. Muniandy, S. L. Yap, and C. S. Wong,
Physics of Plasmas \textbf{16}, 042311 (2009).

\bibitem{}
A. P. Misra, C. Bhowmik, and P. K. Shukla,
Physics of Plasmas 16, 072116 (2009).

\bibitem{}
Prasanta Chatterjee, Kaushik Roy, Ganesh Mondal, S. V. Muniandy, S. L. Yap, and C. S. Wong,
Physics of Plasmas \textbf{16}, 122112 (2009).

\bibitem{}
M. Akbari-Moghanjoughi, Physics of Plasmas  \textbf{17}, 052302 (2010);
S. K. El-Labany, N. M. El-Siragy, W. F. El-Taibany, E. F. El-Shamy, and E. E. Behery,
Physics of Plasmas \textbf{17}, 053705 (2010).

\bibitem{}
Prasanta Chatterjee, Brindaban Das, Ganesh Mondal, S. V. Muniandy, and C. S. Wong,
Physics of Plasmas \textbf{17}, 103705 (2010).

\bibitem{}
Basudev Ghosh, Swarniv Chandra, and S. N. Paul,
Physics of Plasmas \textbf{18}, 012106 (2011).

\bibitem{}
Prasanta Chatterjee, Malay kr. Ghorui, and C. S. Wong,
Physics of Plasmas  \textbf{18}, 103710 (2011).

\bibitem{}
S. Tasnim, S. Islam, and A. A. Mamun,
Physics of Plasmas  \textbf{19}, 033706 (2012).

\bibitem{}
Biswajit Sahu, Swarup Poria, Uday Narayan Ghosh, and Rajkumar Roychoudhury
Physics of Plasmas \textbf{19}, 052306 (2012).


\bibitem{}
S. Ghosh, Swati Dubey, R. Vanshpal,
Physics Letters A \textbf{375}, 43 (2010).

\bibitem{}
S. Ali, W.M. Moslem, P.K. Shukla, I. Kourakis, Physics Letters A \textbf{366}, 606 (2007).

\bibitem{}
M. Salimullah, I. Zebaa, Ch. Uzma, H. A. Shaha,
G. Murtaza, Physics Letters A \textbf{372}, 2291 (2008).

\bibitem{}
S. Mahmood, A. Mushtaq, Physics Letters A \textbf{372}, 3467 (2008).

\bibitem{}
A. P. Misra, N. K. Ghosh, Physics Letters A \textbf{372}, 6412 (2008).

\bibitem{}
M. Salimullah, A. Hussain, I. Sara, G. Murtaza, H. A. Shah,
Physics Letters A \textbf{373}, 2577 (2009).

\bibitem{}
S. K. El-Labany, E. F. El-Shamy, W. F. El-Taibany, P. K. Shukla
Physics Letters A \textbf{374}, 960 (2010).

\bibitem{}
S. K. El-Labany, E. F. El-Shamy, W. F. El-Taibany, P. K. Shukla,
Physics Letters A \textbf{374}, 960 (2010);
S. K. EL-Labany, E. F. EL-Shamy, and M. G. El-Mahgoub,
Physics of Plasmas \textbf{19}, 062105 (2012).

\bibitem{}
Yunliang Wang, Yushan Dong, B. Eliasson
Physics Letters A \textbf{377}, 2604 (2013).

\bibitem{}
S. A. Khan, Q. Haque, Chin. Phys. Lett. \textbf{25}, 4329 (2008).

\bibitem{}
Samiran Ghosh, EPL \textbf{99}, 36002 (2012).

\bibitem{}
M. Salimullah, M. Ayub, H. A. Shah, and W. Masood, Phys. Scr. \textbf{76}, 655 (2007).

\bibitem{}
T. S. Gill, A. S. Bains and C. Bedi, Journal of Physics: Conference Series \textbf{208}, 012040 (2010).

\bibitem{}
Punit Kumar, and Chhaya Tiwari, Journal of Physics: Conference Series \textbf{208}, 012051 (2010).

\bibitem{}
R. Vanshpal, S. Dubey, S. Ghosh, Journal of Physics: Conference Series \textbf{365}, 012046 (2012).

\bibitem{}
M. R. Rouhani, Z. Mohammadi, A. Akbarian, Astrophys Space Sci. \textbf{349}, 265 (2014);
P. K. Shukla, J. Plasma Physics, \textbf{74}, 107 (2008).

\bibitem{}
M. Jamil, Ch. Uzma, K. Zubia, I. Zeba, H. M. Rafique,
And M. Salimullah, J. Plasma Physics, \textbf{78}, 589 (2012).

\bibitem{}
B. Sahu and R. Roychoudhury, Indian J. Phys. \textbf{86}, 401 (2012).

\bibitem{}
Prasanta Chatterjee, Malay Kumar Ghorui, and Rajkumar Roychoudhury, PRAMANA — journal of physics, \textbf{80}, 519 (2013).

\bibitem{}
Malay Kumar Ghorui, Utpal Kumar Samanta, Prasanta Chatterjee, Astrophys Space Sci. \textbf{345}, 273 (2013);
Malay Kumar Ghorui, Ganesh Mondal, Prasanta Chatterjee,
Astrophys Space Sci. \textbf{346}, 191 (2013)


\bibitem{}
M. Akbari-Moghanjoughi, Indian J. Phys. \textbf{86}, 413 (2012).

\bibitem{}
M. K. Ghorui, P. Chatterjee, and R. Roychoudhury, Indian J. Phys. \textbf{87}, 77 (2013).

\bibitem{}
K. Roy, and P. Chatterjee, Indian J. Phys. \textbf{85}, 1653 (2011).

\bibitem{}
Nikhil Chakrabarti, Janaki Sita Mylavarapu, Manjistha Dutta, and Manoranjan Khan, Physical Review E \textbf{83}, 016404 (2011).

\bibitem{}
Samiran Ghosh, and Nikhil Chakrabarti, Physical Review E \textbf{87}, 033102 (2013).

\bibitem{}
Kamel Ourabah, and Mouloud Tribeche, Physical Review E \textbf{88}, 045101 (2013).

\bibitem{}
S. K. EL-Labany, E. F. EL-Shamy, M. G. El-Mahgoub, Astrophys Space Sci. \textbf{339}, 195 (2012);
W. Masood, M. Salimullah, H. A. Shah, Physics Letters A \textbf{372}, 6757 (2008);
Biswajit Sahu,
PRAMANA — journal physics \textbf{76},  933 (2011).


\bibitem{mod over 3 end}
S. K. El-Labany, E. F. El-Shamy, S. K. El-Sherbeny, Astrophys Space Sci. \textbf{340}, 93 (2012).






\bibitem{mod over 3  For dust}
Qi Xue-Hong, Duan Wen-Shan, Chen Jian-Min, and Wang Shan-Jin, Chin. Phys. B \textbf{20}, 025203 (2011).


%

\bibitem{mod over 3 in Rel plasmas}

Basudev Ghosh, Swarniv Chandra, and Sailendra Nath Paul, PRAMANA journal of physics \textbf{78}, 779 (2012).







\bibitem{good modif beg}
L. A. Rios and P. K. Shukla,
Physics of Plasmas \textbf{15}, 074501 (2008);
Q. Haque, Phys. Scr. \textbf{80}, 055501 (2009).

\bibitem{}
Zhengwei Wu, Haijun Ren, Jintao Cao, and Paul K. Chu,
Physics of Plasmas \textbf{15}, 082103 (2008);
Hai Jun Ren, Jintao Cao, and Zhengwei Wu,
Physics of Plasmas \textbf{15}, 102108 (2008);
Haijun Ren, Zhengwei Wu, Jintao Cao, and Paul K. Chu,
Physics of Plasmas \textbf{16}, 103705 (2009).

\bibitem{}
M. Shahid and G. Murtaza, Physics of Plasmas  \textbf{20}, 082124 (2013).

\bibitem{good modif end}
M. Shahid, A. Hussain, and G. Murtaza,
Physics of Plasmas  \textbf{20}, 092121 (2013).











\bibitem{Eq of state Strange diff D}
P. K. Shukla and B. Eliasson, PRL \textbf{96}, 245001 (2006);
Bengt Eliasson And Padma K. Shukla, Plasma and Fusion Research: Review Articles Volume \textbf{4}, 032 (2009).


\bibitem{Manfredi PRB 2001} G. Manfredi, and F. Haas, Physical Review B, \textbf{64}, 075316 (2001).




\bibitem{Khan PP 08 DIAW}
S. A. Khan, A. Mushtaq, and W. Masood, Physics of Plasmas  \textbf{15}, 013701 (2008).

\bibitem{Akbari Moghanjoughi PP 10 RaOC}
M. Akbari-Moghanjoughi, Physics of Plasmas \textbf{17}, 082317 (2010).

\bibitem{Akbari Moghanjoughi PRAMANA 11}
M. Akbari-Moghanjoughi, and N. Ahmadzadeh-Khosroshahi, PRAMANA journal of physics \textbf{77}, 369 (2011).

\bibitem{Fathalian PP 10}
Ali Fathalian and Shahram Nikjo,
Physics of Plasmas \textbf{17}, 103710 (2010).

\bibitem{Wang PLA 08 SaCS}
Yue-yue Wang, Jie-fang Zhang, Physics Letters A \textbf{372}, 6509 (2008).







\bibitem{Ali NJP 08 PSoNE}
S. Ali, W. M. Moslem, I. Kourakis, and P. K. Shukla, New Journal of Physics \textbf{10}, 023007 (2008).

\bibitem{Sahu ASS 13 KPsiqp}
Biswajit Sahu, Naba Kumar Ghosh, Astrophys Space Sci. \textbf{343}, 289 (2013).





\bibitem{Nozieres PR 58} P. Nozieres and D. Pines, Phys. Rev. \textbf{111}, 442 (1958).
\bibitem{Kanazawa PTP 60} H. Kanazawa, S. Misawa and K. Fujita, Progr. Theoret. Phys. (Kyoto) \textbf{23}, 426 (1960).


\bibitem{Schweber} S. Schweber, \emph{An Introdution to Relativistic Quantum Field
Theory} (Evantson, Peterson, Elmsford, NY, 1961).

\bibitem{Andreev 1403 exchange} P. A. Andreev, arXiv:1403.6075 (accepted for publication in Ann. Phys.).




\bibitem{DuBois AnnP 59} D. F. DuBois, Ann. of Phys. \textbf{8}, 24 (1959).
\bibitem{Burt PR 62} P. Burt and Hugo D. Wahlquist, Phys. Rev. \textbf{125}, 1785 (1962).
\bibitem{Roos PR 61} O. von Roos and J. Zmuidzinas, Phys. Rev. \textbf{121}, 941 (1961).
\bibitem{Ter Haar RPP 61} D. Ter Haar, Rep. Progr. Phys. \textbf{24}, 304 (1961).
\bibitem{Hedin JP C 71} L. Hedin, and B. I. Lundqvist,  J. Phys. C: Solid St. Phys., \textbf{4}, 2064 (1971).
\bibitem{Brey PRB 90} L. Brey, Jed Dempsey, N. F. Johnson, and B. I. Halperin, Phys. Rev. B \textbf{42}, 1240 (1990).
\bibitem{Datta JAP 83} S. Datta and R. L. Gunshor, Journal of Applied Physics \textbf{54}, 4453 (1983). 



\bibitem{Trukhanova 1405 exchange} Mariya Iv. Trukhanova and Pavel A. Andreev, arXiv:1405.6294.

\bibitem{Zamanian PRE 13}
J. Zamanian, M. Marklund, and G. Brodin, Physical Review E \textbf{88}, 063105 (2013).



\bibitem{Akbari PP 14} M. Akbari-Moghanjoughi, Phys. Plasmas \textbf{21}, 032110 (2014). 




\bibitem{Krasheninnikov JETP 80} M. V. Krasheninnikov and A. V. Chaplik, Sov. Phys.
JETP \textbf{52}, 279 (1980).


\bibitem{Andreev kinetics 12} P. A. Andreev, arXiv:1212.0099.

\bibitem{Andreev kinetics 13} P. A. Andreev, arXiv:1308.3715.


\bibitem{brusov} P. N. Brusov, J. Soviet Math.  {\bf 23}, 2389 (1983).  ????
\bibitem{krash} M. V. Krasheninnikov and A. V. Chaplik, Sov. Phys. JETP {\bf 52}, 279 (1980). ????
\bibitem{Oji PRB 86} H. C. A. Oji and A. H. MacDonald, Phys. Rev. B  \textbf{33}, 3810 (1986).
\bibitem{Batke PRB 86} E. Batke, D. Heitmann, C.W.
Tu, Phys. Rev. B \textbf{34}, 6951 (1986).
\bibitem{Xu PRB 06} L. J. Xu, X. G. Wu,
Phys. Rev. B \textbf{74}, 165315 (2006).
\bibitem{Burkov PRB 04} A. A. Burkov, Alvaro S. Nunez and
A. H. MacDonald, Phys. Rev. B \textbf{70}, 155308 (2004).
\bibitem{Tahir JPC 10} M. Tahir, K. Sabeeh and A. MacKinnon, J. Phys.: Condens. Matter \textbf{22}, 015801 (2010).



\bibitem{one dimensional thing Review}
Jia Grace Lu, Paichun Chang, Zhiyong Fan, Materials Science and Engineering R \textbf{52}, 49 (2006).


\bibitem{Nath JAP 12} Digbijoy N. Nath, Pil Sung Park, Michele Esposto, David Brown, Stacia Keller, Umesh K. Mishra, and Siddharth
Rajan, Journal of Applied Physics \textbf{111}, 043715 (2012).






\bibitem{Santoyo Mexicana de Fisica} Bernardo Mendoza Santoyo, Marcelo Del Castillo-Mussot,
Revista Mexicana de Fisica \textbf{39}, No. 4, 640 (1993). 

\bibitem{Yangqiang Ran JPA 00} Yangqiang Ran, Lihui Xue, Sizhu Hu,
and Ru-Keng Su, J. Phys. A: Math. Gen. \textbf{33} 9265 (2000).


\bibitem{Hwang JP CS 07}  C. G. Hwang, S. Y. Shin, and J. W. Chung,
Journal of Physics: Conference Series \textbf{61}, 454 (2007).

\bibitem{Li PRB 92} Q. P. Li, S. Sarma and R. Joynt, Phys. Rev. B \textbf{45}, 13713 (1992).
\bibitem{Voit RPP 94} J. Voit, Rep. Prog. Phys. \textbf{57} 977 (1994).












\bibitem{Gahlot PP 13} Ajay Gahlot, Ritu Walia, Jyotsna Sharma, Suresh C. Sharma, and
Rinku Sharma, Phys. Plasmas \textbf{20}, 013706 (2013). 



\bibitem{ul Haq JAP 10} Muhammad Noaman ul Haq, R. Saeed, and Asif Shah, Journal Of Applied Physics \textbf{108}, 043301 (2010).



























\bibitem{NON-Plane Cyl and Sph    CLASSIC    BEG}
M. M. Hossain, A. A. Mamun, and K. S. Ashrafi, Physics of Plasmas \textbf{18}, 103704 (2011).

\bibitem{}
M. S. Alama, M. M. Masudb, and A. A. Mamun, Plasma Physics Reports, \textbf{39},  1011 (2013).

\bibitem{}
J. Plasma Physics, \textbf{79}, 37 (2013).
Deb Kumar Ghosh, Uday Narayan Ghosh, And Prasanta Chatterjee

\bibitem{}
Deb Kumar Ghosh, Prasanta Chatterjee, and Uday Narayan Ghosh,
Physics of Plasmas \textbf{19}, 033703 (2012).

\bibitem{}
Uday Narayan Ghosh, and Prasanta Chatterjee, J. Plasma Physics,  \textbf{79},  789 (2013).

\bibitem{}
S. A. Khan, and Arshad M. Mirza, Commun. Theor. Phys. \textbf{55}, 151 (2011).

\bibitem{}
T. Akhter, M. M. Hossain, and A. A. Mamun, Commun. Theor. Phys. \textbf{59}, 745 (2013).

\bibitem{}
A. A. Mamun, and P. K. Shukla, EPL \textbf{94}, 65002 (2011).

\bibitem{on cylindre  Classic   Single}
Constantine Yannouleas, Eduard N. Bogachek, and Uzi Landman,
Physical Review B \textbf{53}, 225 (1996).


\bibitem{NON-Plane Cyl and Sph    CLASSIC    END}
Deb Kumar Ghosh, Uday Narayan Ghosh, Prasanta Chatterjee, and C. S. Wong,
PRAMANA journal of physics \textbf{80}, 665 (2013).


\bibitem{NON-Plane Cyl and Sph   QUANTUM    BEG}
Biswajit Sahu and Rajkumar Roychoudhury
Physics of Plasmas  \textbf{14}, 012304 (2007).

\bibitem{}
A. P. Misra, P. K. Shukla, and C. Bhowmik
Physics of Plasmas \textbf{14}, 082309 (2007).

\bibitem{}
A. Mushtaq, Physics of Plasmas \textbf{14}, 113701 (2007).

\bibitem{}
R. Sabry, S. K. El-Labany, and P. K. Shukla,
Physics of Plasmas \textbf{15}, 122310 (2008).

\bibitem{}
Amar P. Misra, Chandan Bhowmik, Physics Letters A \textbf{369}, 90 (2007).


\bibitem{}
S. A. Khan, S. Mahmood, Arshad M. Mirza, Physics Letters A \textbf{372}, 148 (2008).


\bibitem{}
S. A. Khan, S. Mahmood, Arshad M. Mirza, Physics Letters A \textbf{372}, 148 (2008).

\bibitem{}
Yue-yue Wang, Jie-fang Zhang, Physics Letters A \textbf{372}, 3707 (2008).

\bibitem{}
Shalini Bagchi, Kasturi Roychowdhury, A. P. Mishra,
A. Roy Chowdhury, Commun Nonlinear Sci. Numer. Simulat. \textbf{15}, 275 (2010).

\bibitem{}
S. Ali, W. M. Moslem, I. Kourakis, and P. K. Shukla,
New Journal of Physics \textbf{10}, 023007 (2008).



\bibitem{}
M. M. Hossain, and A. A. Mamun, J. Phys. A: Math. Theor. \textbf{45}, 125501 (2012).


\bibitem{}
Jiu-Ning Han, Jun-Hua Luo, Jun-Xiu Li, Astrophys Space Sci. \textbf{349}, 305 (2014).

\bibitem{NON-Plane Cyl and Sph   QUANTUM     END}
M. Hasan, M. M. Hossain, A. A. Mamun, Astrophys Space Sci. \textbf{345}, 113 (2013).




\bibitem{on cylindre  quantum no inertia   BEG}
Alireza Abdikian and Mehran Bagheri,
Physics of Plasmas \textbf{20}, 102103 (2013).



\bibitem{}
Li Wei, You-Nian Wang, Physical Review B \textbf{75}, 193407 (2007).

\bibitem{}
H. Terças, J. T. Mendonça, and P. K. Shukla,
Physics of Plasmas \textbf{15}, 072109 (2008).

\bibitem{on cylindre  quantum no inertia    END}
Afshin Moradi, J. Phys.: Condens. Matter \textbf{21}, 045303 (2009).


\bibitem{on cylindre  quantum no inertia  BEG 2}
Ali Fathalian and Shahram Nikjo, Physics of Plasmas \textbf{17}, 103710 (2010).

\bibitem{}
Alireza Abdikian and Mehran Bagheri, Physics of Plasmas  \textbf{20}, 102103 (2013).

\bibitem{Sph}
Ali Fathalian, Shahram Nikjo, Solid State Communications \textbf{150}, 1062 (2010).

\bibitem{}
Afshin Moradi, Physica E \textbf{42}, 43 (2009).

\bibitem{}
K. A. Vijayalakshmi, And T. P. Nafeesa Baby, PRAMANA — journal of physics, \textbf{80}, 289 (2013).

\bibitem{}
Afshin Moradi, J. Phys.: Condens. Matter \textbf{21}, 045303 (2009).

\bibitem{on cylindre  quantum no inertia END 2}
Ali Fathaliana, and Shahram Nikjo, Chin. Phys. B \textbf{21}, 057306 (2012).








\bibitem{Spherical Quant eq state 01}
Ali Fathalian, Shahram Nikjo, Optics Communications \textbf{283}, 5051 (2010).


\bibitem{Spherical Quant eq state 02}
Ali Fathalian, Shahram Nikjo, Optics Communications \textbf{284}, 2236 (2011).


\bibitem{Landau Vol 6} L. D. Landau, E. M. Lifshitz, \emph{Hydrodynamics}. Vol. 6. Pergamon Press,  (1977).



\end{thebibliography}
\end{document}